\documentclass[%
  reprint,
  amsmath,amssymb,
  aps,
  prc,
  nofootinbib,
  floatfix,
 superscriptaddress, longbibliography
]{revtex4-2}
\usepackage[utf8]{inputenc}
\usepackage{graphicx}
\usepackage{dcolumn}
\usepackage{bm}
\usepackage{xspace}
\usepackage{tcolorbox}
\usepackage{tabularx}
\usepackage{enumitem}
\newcommand{\jpsi}{\ensuremath{{J}\hspace{-.08em}/\hspace{-.14em}\psi}\xspace}

\newcommand{\pp}{{\ensuremath{p+p}}\xspace}
\newcommand{\pA}{\ensuremath{p+A}\xspace}

\newcommand{\hA}{\ensuremath{h+A}\xspace}
\newcommand{\eA}{\ensuremath{e+A}\xspace}
\newcommand{\AAn}{\ensuremath{A+A}\xspace}

\newcommand{\rpa}{\ensuremath{R_{pA}}\xspace}

\newcommand{\rha}{\ensuremath{R_{hA}}\xspace}

\newcommand{\gev}{\ifmmode \xspace
\ensuremath{\textnormal{GeV}\xspace} \else GeV \fi}

\newcommand{\Tr}{\mathrm{Tr}}

\newcommand{\pt}{\ensuremath{p_T}\xspace}

\newcommand{\tf}{t_{\textnormal{f}}}

\usepackage[colorlinks=true, linkcolor=blue, citecolor=blue, urlcolor=blue]{hyperref}

\begin{document}

\title{Nuclear Cold QCD: Review and Future Strategy}

\author{F. Arleo}
\affiliation{Subatech, IMT Atlantique, Université de Nantes, IN2P3/CNRS, Nantes, France}

\author{P. Caucal}
\affiliation{Subatech, IMT Atlantique, Université de Nantes, IN2P3/CNRS, Nantes, France}

\author{A. Deshpande}
\affiliation{Center for Frontiers in Nuclear Science (CFNS), Department of Physics and Astronomy, Stony Brook University, Stony Brook, NY 11794, USA}
\affiliation{Physics Department, Brookhaven National Laboratory, Upton, NY 11973, USA}

\author{J. M. Durham}
\affiliation{Los Alamos National Laboratory, Physics Division, Los Alamos, NM 87545, USA}

\author{G. M. Innocenti}
\affiliation{Massachusetts Institute of Technology, Cambridge, MA 02139, USA}

\author{J. Jalilian-Marian}
\affiliation{Department of Natural Sciences, Baruch College, CUNY, New York, NY, USA}
\affiliation{The Graduate Center, CUNY, New York, NY 10016, USA}

\author{A. Kusina}
\affiliation{Institute of Nuclear Physics, Polish Academy of Sciences, Krakow, PL-31342, Poland}

\author{M. X. Liu}
\affiliation{Los Alamos National Laboratory, Physics Division, Los Alamos, NM 87545, USA}

\author{Y. Mehtar-Tani}
\affiliation{Physics Department, Brookhaven National Laboratory, Upton, NY 11973, USA}

\author{C.-J. Naïm}
\email{Corresponding author: charlesjoseph.naim@stonybrook.edu}
\affiliation{Center for Frontiers in Nuclear Science (CFNS), Department of Physics and Astronomy, Stony Brook University, Stony Brook, NY 11794, USA}

\author{H. Paukkunen}
\affiliation{Department of Physics, University of Jyväskylä, P.O. Box 35, FI-40014 University of Jyväskylä, Finland}
\affiliation{Helsinki Institute of Physics, P.O. Box 64, FI-00014 University of Helsinki, Finland}

\author{S. Platchkov}
\affiliation{IRFU, CEA, Université Paris-Saclay, 91191 Gif-sur-Yvette, France}

\author{F. Salazar}
\affiliation{Physics Department, Brookhaven National Laboratory, Upton, NY 11973, USA}
\affiliation{Department of Physics, Temple University, Philadelphia, PA 19122, USA}
\affiliation{RIKEN-BNL Research Center, Brookhaven National Laboratory, Upton, NY 11973, USA}

\author{I. Vitev}
\affiliation{Los Alamos National Laboratory, Theoretical Division, Los Alamos, NM 87545, USA}

\author{R. Vogt}
\affiliation{Nuclear and Chemical Sciences Division, Lawrence Livermore National Laboratory, Livermore, CA 94551, USA}
\affiliation{Department of Physics and Astronomy, University of California, Davis, Davis, CA 95616, USA}

\begin{abstract}
This review examines data from hadron-nucleus collisions, primarily focusing on hard processes like Drell-Yan, heavy flavor and quarkonium production. It highlights observed modifications of particle yields as functions of momentum and rapidity, aiming to clarify the underlying QCD effects on cold nuclear matter. It outlines strategies for future experiments, including the Electron-Ion Collider, to distinguish between these effects. Key questions address the universality of suppression mechanisms and the role of nonperturbative physics, providing a road map for upcoming measurements of hadrons on nuclei.
\end{abstract}

\maketitle


\section{Introduction}
\label{sec:introduction}
Perturbative Quantum Chromodynamics (pQCD) has proven remarkably successful in describing high energy processes, particularly in proton-proton collisions, where its predictions closely match the experimental results. However, extending pQCD to hadron-nucleus collisions presents significant challenges due to the additional complexity arising from the effects related to the nuclear environment. The corresponding cold nuclear matter (CNM) effects alter the parton dynamics in the initial and the final state and modify the particle production yields. Despite decades of research, no coherent and universally accepted framework has yet emerged to consistently describe the data in such systems. 

The root cause of this unsatisfactory situation lies in the multitude of nuclear effects that can influence the nuclear cross sections. The well known EMC effect~\cite{Aubert:1983xm} provided evidence that the nuclear parton distributions functions (nPDFs) in bound nucleons differ from those in free nucleons, modifying the effective parton densities probed in the collision \cite{deFlorian:2011fp,Eskola:2016oht,Eskola:2021nhw, Khanpour:2020zyu, Kovarik:2015cma,Duwentaster:2022kpv, AbdulKhalek:2022fyi, Helenius:2021tof}. Most hadron-nucleus experiments primarily probe the leading-twist shadowing region of the target nucleus at low momentum fraction $x$, leading to a reduction of the particle yields relative to an incoherent sum over all the nucleons. 

Another effect, relevant at high center-of-mass energies, is due to gluon saturation. Saturation prevents an unlimited increase of the gluon density at low $x$ values, which would otherwise violate unitarity constraints.  This saturation has an important impact on particle production, resulting in additional suppression at very forward rapidities. This suppression is similar to the suppression resulting from the purely nonperturbative leading-twist shadowing. Although saturation and shadowing originate from two different QCD mechanisms, differentiating between them remains a difficult task.  

In addition, the incoming \textit{or} outgoing parton may lose some of its energy by medium-induced gluon radiation when traversing the nucleus, known as Landau-Pomeranchuk-Migdal (LPM) energy loss ~\cite{Gyulassy:2000fs,Zakharov:2000iz}. Similarly, the incoming \textit{and} outgoing partons could undergo energy loss, which is summed coherently and is known as Fully Coherent Energy Loss (FCEL) ~\cite{Arleo:2010rb,Arleo:2012rs}. 
A related, but distinct, phenomenon arises from multiple soft scatterings of the parton traversing the nuclear medium. Each consecutive scattering contributes to a slight modification of the transverse momentum of the parton as well as the broadening of the particle production yields as a function of transverse momentum, $p_T$~\cite{Kopeliovich:2010aa, Arleo:2020rbm, Mueller:2016gko, Johnson:2000dm, Gyulassy:2002yv,Baier:1996sk}. Since the broadening scales with the number of scatterings, it typically grows linearly with the nuclear size. 

Furthermore, heavy mesons produced in the final state are subject to inelastic interactions with the nuclear medium. When their formation length  is shorter than the size of the nucleus, the newly formed mesons may be dissociated via scattering with partons in a high-density environment \cite{Vogt:2004dh, Vogt:2010aa}. In addition, both the initially produced $Q \overline Q$ pair and the nascent heavy meson can interact with the surrounding nucleons. The corresponding absorption cross section can vary significantly depending on the energy and production mechanism. Moreover, the quarkonium state may undergo further interactions with comoving hadrons or partons, referred to comovers \cite{Vogt:1990us, Elena}. These interactions, which result in the break-up of the bound state, are affected by the local density of the medium. The effects of collisional dissociation, nuclear absorption, and comover interactions are strongly dependent on the path length of the state through the nuclear medium and all lead to attenuation of the observed heavy meson yields.
In a typical high-energy hadron-nucleus experiment, several CNM effects may be present simultaneously. Their relative importance, however, depends strongly on the process considered, the kinematic regime, and the specific observables under scrutiny. The difficulties in quantitatively establishing the relative  contributions of these effects adds to the challenge of forming a coherent theoretical framework.

This work aims to address these questions by reviewing the current state of experimental data and theoretical understanding, with an emphasis on particle production. We summarize the present state of knowledge, address known limitations, and examine discrepancies between theoretical predictions and experimental data. Our aim is to outline strategies for improving the theoretical description of CNM phenomena in the coming decade. Importantly, we do not seek to challenge the validity of any specific theoretical model. Rather, through an experimental lens, we aim to establish a set of key observables that can help distinguish between cold nuclear matter effects on various processes.

The upcoming Electron-Ion Collider (EIC) \cite{AbdulKhalek:2021gbh} and future fixed-target experiments \cite{Adams:2676885, SeaQuest:2017kjt} will offer unprecedented opportunities to probe CNM with greater precision and across a broad kinematic range. Leveraging these new experimental capabilities, we seek to narrow the gap between theoretical predictions and experimental observations, paving the way for a unified description of cold nuclear matter effects. 

We identify three key questions that future data should address:
\begin{enumerate}[left=0.1cm]
\item \textbf{What are the relative contributions of perturbative (gluon saturation) and nonperturbative (nPDF) QCD dynamics to the modification of nuclear structure functions and particle production spectra at small $x$?}
\item \textbf{How do parton energy loss mechanisms affect particle production in $h+A$ and $e+A$ collisions?}
\item \textbf{What other effects modify particle production and how do we distinguish among final-state interactions?}
\end{enumerate}

By addressing these questions, we aim to provide a road map for future research and contribute to a deeper understanding of cold nuclear matter and its implications for the interpretation of data from $\eA$, $\pA$, and $\AAn$ collisions.

This paper is organized in eight sections. The most relevant fixed-target and collider data are reviewed in Sec.~\ref{sec:data}. Section~\ref{sec:nPDF_saturation} discusses the present status of nPDFs and gluon saturation at small $x$.  Radiative energy loss effects are reviewed in Sec.~\ref{sec:eloss}. Final-state interactions including collisional dissociation, absorption, and comovers are discussed in Sec.~\ref{sec:finalstate}. Additional effects due to impact-parameter dependence and intrinsic charm are presented in Sec~\ref{sec:other_effects}. Section~\ref{sec:future_experience} explores the experimental opportunities opened by the newest physics facilities and, particularly, by the  EIC.  A short conclusion is drawn in Sec.~\ref{sec:conclusion}.   
\section{Drell-Yan and heavy flavor data on nuclear targets}
\label{sec:data}
Hadron-nucleus collisions serve as a powerful tool to probe the properties of cold nuclear matter, offering unique insights into the dynamics of partonic interactions in the nuclear medium. Assuming the validity of collinear factorization, they provide a useful context for extracting and studying nPDFs.  
Processes such as Drell-Yan and single inclusive hadron production are particularly valuable probes of these effects. Fixed-target experiments, which benefit from a large experimental rapidity, $y$, coverage due to the beam boost, primarily explore nuclear effects at higher values of the parton momentum fraction $x$, $x > 10^{-3}$. In contrast, data from the Large Hadron Collider (LHC) extend these studies into the low $x$ regime, even though they cannnot cover the full rapidity range available, reaching $x$ values as small as $\sim 10^{-5}$, where gluon saturation is expected to become significant. The nuclear modification factor, $\rha$, quantifies the impact of the nuclear medium on particle production by comparing cross sections in hadron-nucleus, $\hA$, to hadron-proton, $h+p$, collisions. Studied as a function of rapidity or Feynman $x$ ($x_F$) and transverse momentum, $p_T$, $\rha$ provides insight into the evolution of nuclear effects across kinematic regimes. 

Despite extensive measurements at fixed-target facilities and colliders, the mechanisms driving the nuclear effects remain ill-determined. It does not appear possible to precisely determine the contributions from all proposed cold nuclear matter effects based on only one type of measurement. 
Gaining a more quantitative insight into these effects requires a combined analysis of various processes. The current status of the main processes used to investigate cold nuclear matter effects---namely Drell-Yan, quarkonium and heavy-flavor production in hadronic collisions, as well as particle production in ultraperipheral collisions---is summarized below.

\subsection{Drell-Yan Data}

At leading order (LO), the electromagnetic Drell-Yan process describes the production of a virtual photon, $\gamma^{\star}$, via quark-antiquark annihilation  
and its decay into a dilepton (lepton-antilepton pair), $\gamma^{\star} \to \ell^+\ell^-$. Along with deep-inelastic scattering (DIS), this process provides complementary information about the partonic structure of both the beam particle and the target proton or nucleus. 
The Drell-Yan cross section is well understood and can be calculated using fixed-order perturbative methods \cite{Melnikov:2006di, Baur:2001ze, Catani:2007vq} or employing advanced approaches such as Collins-Soper-Sterman (CSS) resummation and transverse-momentum-dependent (TMD) factorization \cite{Camarda:2019zyx}, which accurately describes the low $p_T$ portion of the dilepton spectrum. The nuclear dependence of the Drell-Yan process is quantified by the ratio $\rha^{\text{DY}}$, which compares the yields in $\hA$ collisions to those in $h+p$ or, occasionally, in $h+ {\rm d}$ collisions.  

Most of the Drell-Yan data come from fixed-target experiments, primarily performed at CERN and Fermilab, spanning the center-of-mass energy range $15 \leq \sqrt{s} \leq38.7$~GeV. The earliest experiments~\cite{Branson:1977cj, Anderson:1979tt, Frisch:1982hp}, which used proton and pion beams at 225\,GeV and measured only total cross sections, showed no evidence of nuclear effects. This conclusion was based on comparisons of $p+p$ and $\pA$ cross sections using the relation $\sigma_{pA} = A^\alpha \sigma_{pp}$, where the exponent $\alpha$ was found to be consistent with unity.
More detailed comparisons between heavy and light targets were performed in the beginning of the 1980s using  pion beams at CERN \cite{Badier:1981ci} and proton beams at Fermilab~\cite{Ito:1980ev}.  In spite of the relatively large statistics, both experiments reported  the absence of nuclear effects. A clear variation of $\rha^{\text{DY}}$  from unity was observed \cite{Bordalo:1987cs} a few years later by the NA10 experiment at CERN, when comparing the data collected on hydrogen and tungsten targets.\footnote{The nuclear modification factor $R_{hA}$ can be related to $\alpha$ on the cross section level by $R_{hA} = \sigma_{pA}/(A \sigma_{pp}) = A^{\alpha - 1}$.}  The measured $\rha^{\rm DY}$ as a function of the target parton momentum fraction, $x_2$, was found to agree with the recently discovered EMC effect~\cite{Aubert:1983xm}. A pronounced increase of $\rha^{\text{DY}}$ around $p_T=2$ GeV (the ``Cronin peak"~\cite{Cronin:1974zm}) was also observed \cite{Bordalo:1987cr}. The NA10 results were soon qualitatively confirmed by extensive measurements by the E772 experiment at Fermilab~\cite{Alde:1990im}. Taking data with a 800 GeV proton beam, the experiment compared results from a deuterium target with four other nuclear targets (C, Ca, Fe and W) in the $x_F$ region between 0.1 and 0.3.\footnote{Note that Feynman $x$, $x_F$, can be defined as $x_F = x_1 - x_2$ the difference between the projectile momentum fraction, $x_1$ and the target momentum fraction, $x_2$, with a range $-1 \leq x_F \leq 1$. Another definition is $x_F = p_z / p_z^{\text{max}}$, where $p_z$ is the momentum of the dilepton along the beam direction, and $p_z^{\text{max}} = \sqrt{s}/2$.} The data were found to be consistent with unity, except at low $x_2$, where the suppression increased with the mass of the target, $A$, an effect fairly compatible with the leading-twist nuclear shadowing measured in DIS.  An increase of  $\rha^{\text{DY}}$ for $p_T \gtrsim 1$ GeV with $A$ was also clearly observed.  
A much larger $x_F$ region, up to $x_F=0.95$, was covered by the E866 collaboration~\cite{Vasilev:1999fa} using three targets: Be, Fe and W. These data thus extended to larger values of $x_1$ and thus to smaller values of $x_2$, providing increased sensitivity to nuclear effects in these two domains. At low $x_2$, the experimental Fe/Be and W/Be ratios are compatible with the leading-twist nuclear shadowing observed in DIS while for $x_1 \gtrsim 0.7$ the observed suppression provided the first evidence for parton energy loss in the projectile in the Drell-Yan process~\cite{Neufeld:2010dz, Arleo:2018zjw, Eskola:2021nhw}.

High statistics Drell-Yan data employing a 450 GeV proton beam was obtained by the NA50 collaboration at CERN \cite{NA50:2003fvu}. When corrected for nuclear isospin, the resulting cross sections for a number of nuclei from Be to W perfectly scale with the number of nucleon-nucleon collisions, in line with previous data on proton and deuteron targets. More recently, preliminary data \cite{Lin:2017eoc} from the E906 experiment at Fermilab (SeaQuest) using a proton beam at 120 GeV became available. The measured nuclear modification factor $\rpa^{\text{DY}}$ was found to be smaller than unity as a function of $x_F$, with a stronger suppression at large $x_F \sim 0.8$.  The modification appears to be more pronounced for W/C ratio than for Fe/C. 

Measurements of the dilepton continuum have been extended by several collider experiments. The PHENIX collaboration at RHIC \cite{Leung:2018tql} reported data for $p+p$ and $p+\mathrm{Au}$ collisions at $\sqrt{s_{NN}} = 200$ GeV, focusing on the dilepton mass region below 10 GeV and $p_T \lesssim M$. A slight enhancement in the $p_T$ spectra was observed at forward rapidity around $p_T = 2$~GeV. Due to the low statistics, these data do not exhibit any clear nuclear effects. 

More recently, the CMS collaboration at the LHC~\cite{CMS:2021ynu} presented extensive dimuon data from $\sqrt{s_{NN}} = 8.16$~TeV $p + {\rm Pb}$ collisions, covering the mass range $15 \leq M \leq600$~GeV with $15  \leq p_T \leq 120$~GeV. The measured $\rpa^{\text{DY}}$ clearly favors calculations incorporating nuclear PDFs. The presence of nuclear effects is also necessary to account for the forward ($y>0$) to backward ($y<0$) rapidity ratios. 

The study of the Drell-Yan continuum is complemented by $Z$ and $W^\pm$ vector boson measurements made by the LHC experiments, see recent results from $p + {\rm Pb}$ collisions by CMS \cite{CMS:2015ehw,CMS:2015zlj,CMS:2019leu,CMS:2021ynu}, ALICE \cite{ALICE:2016rzo,ALICE:2020jff} and LHCb \cite{LHCb:2014jgh, LHCb:2022kph} at 5.02 and 8.16 TeV. Within uncertainties, the ALICE data show only a small difference between the $\pp$ and $\pA$ cross sections. In contrast, the LHCb data indicate a slight suppression of the nuclear cross section in the forward region and an enhancement in the backward region compared to data in $\pp$ collisions. Both data sets are compatible with the nPDF calculations. This difference may partly arise from the distinct rapidity coverages: ALICE data probe $-4.46 < y < -2.96$ and $2.03 < y < 3.53$, while LHCb data access slightly different rapidity regions with $-4.0 < y < -2.5$ and $1.5 < y < 4.0$. The different kinematic ranges could influence the observed nuclear effects.  

The Drell-Yan, $Z$ and $W^\pm$ vector boson data from the fixed-target and collider experiments discussed above are summarized in Table~\ref{tab:table_DY}.
\begin{table*}[tbp]
  \begin{center}
    \begin{tabular}{ccccccc}
      \hline 
      \hline 
      Exp. & Beam  & $\sqrt{s_{NN}}$ (GeV) & Final State & Target ($A$) & Observable & Ref. \\
      \hline 
      E906 & $p$ & 15  & DY & Fe, W &  $x_F$ &  \cite{Lin:2017eoc} \\ 
      \hline 
      COMPASS & $\pi^{-}$  & 18.9  & DY & NH$_3$, Al, W & $x_F$, \pt & \cite{Aghasyan:2017jop} \\ 
      \hline 
      NA3 & $\pi^{-}$ & 16.7, 19.4, 22.9 & DY & Pt & $x_F$, \pt & \cite{Badier:1981ci,Badier:1982zb} \\
      \hline 
      NA10 & $\pi^{-}$ & 16.2, 23.1  & DY & W  & $x_F$, \pt &  \cite{Bordalo:1987cr,Bordalo:1987cs} \\ 
      \hline 
      E615 & $\pi^{\pm}$ & 21.7  & DY & W & $x_F$ & \cite{Heinrich:1989cp} \\
       \hline 
      E288 & $p$ & 27.4  & DY & Be, Cu, Pt & $M$, \pt & \cite{Ito:1980ev} \\
      \hline 
      E772 & $p$ & 38.7  & DY & D, Ca, Fe, W & $x_F$, \pt & \cite{Alde:1990im} \\
      \hline 
      E866 & $p$ & 38.7  & DY & Be, Fe, W & $x_F$, \pt & \cite{Vasilev:1999fa} \\
      \hline 
      \hline  
       PHENIX & $p$ & 200   & DY & Au & \pt & \cite{Leung:2018tql} \\
      \hline 
      CMS & $p$ & 8016 &  DY & Pb &  $y$, \pt & \cite{CMS:2021ynu}  \\
       & $p$ & 5020 &  $W$, $Z$ & Pb &  $y$ & \cite{CMS:2015ehw,CMS:2015zlj}  \\
        & $p$ & 8016 &  $W$, $Z$ & Pb &  $y$  & \cite{CMS:2019leu,CMS:2021ynu}  \\
      \hline 
      ALICE & $p$ & 5020 &  $W$, $Z$ & Pb &  $y$ & \cite{ALICE:2016rzo,ALICE:2020jff}  \\
            & $p$ & 8016 &  $Z$ & Pb &  $y$ & \cite{ALICE:2020jff}  \\
      \hline 
     LHCb & $p$ & 5000 &  $Z$ & Pb &  $y$ & \cite{LHCb:2014jgh}  \\
      
       & $p$ & 8016 &  $Z$ & Pb &  $y$, \pt & \cite{LHCb:2022kph}  \\
      \hline
      \hline 
    \end{tabular}
    \caption{List of data sets for Drell-Yan (DY) and $Z$, $W^\pm$ vector bosons production on nuclear targets.}
    \label{tab:table_DY}
    \end{center}
\end{table*} 

\subsection{Quarkonium Data}  
Quarkonium production involves the formation of a heavy quark-antiquark bound state through hadronization, $h + A \to [Q\bar{Q}] + X$. Several approaches have been used to describe this process, including the Color Evaporation Model (CEM) \cite{Fritzsch:1977ay, Halzen:1977im, Gluck:1977zm,Gavai:1994in}, the Improved Color Evaporation Model (ICEM) \cite{Ma:2016exq}, the Color Singlet Model (CSM) \cite{Chang:1979nn}, and Non-Relativistic Quantum Chromodynamics (NRQCD) \cite{Bodwin:1994jh}. The  CEM and CSM have not be rigorously applied beyond hadroproduction.  The ICEM has been successfully been applied to production and polarization of $J/\psi$ and $\Upsilon$ in $p+p$ and $J/\psi$ in $e+p$ collisions \cite{Cheung:2017loo,Cheung:2017osx,Cheung:2018tvq,Cheung:2018upe,Cheung:2021epq,Cheung:2024bnt}.  Work is ongoing to apply it to other quarkonium states and production mechanisms.  
Although theoretically sound and phenomenologically successful, the NRQCD approach appears incomplete, as outlined in the next paragraph.

In the NRQCD framework, the short distance cross section for quark-antiquark production can be calculated perturbatively. However, the subsequent evolution of the quark-antiquark pair into a bound quarkonium state is governed by nonperturbative dynamics. The associated long-distance matrix 
elements (LDMEs) must therefore be determined from fits to data. A further complication arises in the low $p_T$ region, $p_T \lesssim M_{Q}$, where pQCD is less reliable and nonperturbative effects may dominate. Fortunately, most of the nonperturbative effects cancel in cross section ratios measured under identical kinematical conditions, allowing for more direct comparison with theory. 

The earliest exploratory experiments \cite{Branson:1977ci, Antipov:1977ss, Anderson:1979tt}, primarily focused on deviations of the exponent $\alpha$, extracted from the data, from unity. Measurements at CERN \cite{Corden:1982zs, Badier:1983dg} and Fermilab \cite{Katsanevas:1987pt, Kartik:1990it} demonstrated that a constant $\alpha$  could not adequately describe the observed $x_F$ and $p_T$ spectra. Detailed proton-induced measurements on several nuclear targets conducted by the Fermilab E772 \cite{Alde:1990wa} and E866 \cite{Leitch:1995yc} collaborations confirmed the strong variation of  $\alpha$ in both $\jpsi$ and $\psi'$ production.
  
Furthermore, the suppression of the first three bottonium mesons \cite{Alde:1991sw} was also investigated. By comparing $\Upsilon$(1S) and $\Upsilon$(2S+3S) production on several nuclear targets with data on deuterium, the E772 experiment found that $\alpha$ changes significantly with $x_F$ and $p_T$. Within the statistical uncertainties, the values of $\alpha$ obtained for the 1S and summed (2S+3S) states were identical. The $A$ dependence of $\rpa^\Upsilon$ was found to be nearly half as strong as that measured for charmonium, suggesting a weaker nuclear dependence for the heavier bottomonium mesons.

\begin{table*}
  \begin{center}
    \begin{tabular}{ccccccr}
      \hline 
      \hline 
      Exp. & Beam  & $\sqrt{s_{NN}}$ (GeV) & Final State & Target ($A$) & Observable & Ref. \\
      \hline 
      E444    & $\pi^{-}$ & 20.5  & $\jpsi$ & C, Cu, W & $x_F$, \pt & \cite{Anderson:1979tt} \\
                     & $p$ & 20.5  & $\jpsi$ & C, Cu, W & $x_F$, \pt & \\
                     \hline
      WA39  & $\pi^{-}$ & 8.6  & $\jpsi$ & H, W & $x_F$, \pt & \cite{Corden:1982zs} \\
                 & $\pi^{+}$ & 8.6  & $\jpsi$ & H, W & $x_F$, \pt & \\
      \hline 
      NA3  &  $p$ & 19.4  &  $\jpsi$ & H, Pt  & $x_F$, \pt & \cite{Badier:1983dg} \\
           & $\pi^{+}$ & 19.4 & $\jpsi$ & H, Pt & $x_F$, \pt & \\
           & $\pi^{-}$ & 16.7 & $\jpsi$ & H, Pt & $x_F$, \pt &  \\
           & $\pi^{-}$ & 19.4 & $\jpsi$ & H, Pt & $x_F$, \pt &  \\
           & $\pi^{-}$ & 22.9 & $\jpsi$ & H, Pt & $x_F$, \pt &  \\
      \hline 
      E537    & $\pi^{-}$ & 15.3  & $\jpsi$ & Be, Cu, W & $x_F$, \pt & \cite{Katsanevas:1987pt} \\
                     &  $\bar{p}$ & 15.3  & $\jpsi$ & Be, Cu, W & $x_F$, \pt & \\
      \hline 
      E672    & $\pi^{-}$ & 31.5  & $\jpsi$ & C, Al, Cu, Pb & $x_F$, \pt & \cite{Kartik:1990it} \\
      \hline 
      E772      & $p$  &  38.7    & $\jpsi$, $\psi'$ &  D, Ca, Fe, W & \pt & \cite{Alde:1990wa} \\
                   & $p$ &  38.7    & $\Upsilon$ &  D, Ca, Fe, W & \pt & \cite{Alde:1991sw} \\
      \hline 
      E789      & $p$  & 38.7  & $\jpsi$ &  Be, Cu &  $x_F$ & \cite{Kowitt:1993ns} \\
                & $p$ &    38.7     & $\jpsi$ &  Be, Cu, W &  $x_F$, \pt & \cite{Leitch:1995yc} \\
      \hline 
      E866   
          &$p$& 38.7 & $\jpsi$, $\psi'$ &  Be, Fe, W  & $x_F$, \pt & \cite{Leitch:1999ea} \\
      \hline 
      NA38  & $p$ & 29.1  & $\jpsi$, $\psi'$  & C, Al, Cu, W & $x_F$, \pt & \cite{NA38:1998lyg} \\
      \hline
      NA50  & $p$ & 29.1  & $\jpsi$, $\psi'$  & Be, Al, Cu, Ag, W & $x_F$, \pt  & \cite{NA50:2003fvu} \\
      & $p$ & 27.4  & $\jpsi$, $\psi'$  & Be, Al, Cu, Ag, W & $x_F$ &\cite{NA50:2006rdp} \\
      \hline
       HERA-B    & $p $& 41.6  & $\jpsi$, $\psi'$ & C, Ti, W & $x_F$, \pt & \cite{HERA-B:2006bhy} \\
      & $p$ & 41.6  & $\jpsi$              & C, Ti, W & $x_F$, \pt & \cite{HERA-B:2008ymp} \\
      \hline
      NA60  & $p$ & 17.2  & $\jpsi$ & Be, Al, Cu, W, Pb, U & $x_F$ & \cite{NA60:2010wey} \\
     & $p$ &  27.4  & $\jpsi$ & Be, Al, Cu, W, Pb, U & $x_F$ & \\
      \hline 
         SMOG  & $p$ & 68.5 &  $J/\psi$ & Ne & $y$, \pt & \cite{LHCb:2022sxs} \\
                    & $p$ & 86.6 &  $J/\psi$ & He, Ar & $y$, \pt & \cite{LHCb:2018jry} \\
                    & $p$ & 110.4 &  $J/\psi$ & He, Ar & $y$, \pt & \cite{LHCb:2018jry} \\
      \hline 
      \hline  
    \end{tabular}
    \caption{Quarkonia production data available in fixed-target experiments with nuclear targets.}
    \label{tab:table_Quarkonia_FT}
    \end{center}
\end{table*}

\begin{table*}
  \begin{center}
    \begin{tabular}{ccccccr}
      \hline 
      \hline 
      Exp. & Beam  & $\sqrt{s_{NN}}$ (GeV) & Final State & Target ($A$) & Observable & Ref. \\
      \hline 
 PHENIX & d & 200   & $\jpsi$ & Au & $y$, \pt & \cite{Adare:2012qf} \\
     & d &  & $\jpsi$, $\psi'$ & Au & $y$, \pt & \cite{PHENIX:2013pmn} \\
      & $p$, $^3$He &    & $\jpsi$ & Al, Au & $y$, \pt & \cite{PHENIX:2019brm} \\
      & $p$ &    & $\psi'$ & Al, Au & $y$, \pt & \cite{PHENIX:2022nrm} \\ 
     \hline 
     STAR & d & 200   & $\jpsi$ & Au & \pt & \cite{STAR:2021zvb} \\
      & d &    & $\Upsilon$ & Au & $x_{F}$ & \cite{STAR:2013kwk} \\
      \hline 
      ALICE & $p$ &  5020 & $\jpsi$ & Pb & $y$, \pt & \cite{Adam:2015jsa} \\
        & $p$ &  & $\psi'$ & Pb & $y$, \pt & \cite{ALICE:2014cgk} \\
         & $p$ &  & $\Upsilon$ & Pb & $y$, \pt & \cite{ALICE:2014ict} \\
            & $p$ &  8160 & $\jpsi$ & Pb & $y$, \pt & \cite{ALICE:2022zig} \\
            & $p$ &   & $\psi'$ & Pb & $y$, \pt & \cite{ALICE:2020vjy} \\
          & $p$ &   & $\Upsilon$ & Pb & $y$, \pt &  \cite{ALICE:2019qie} \\
      \hline 
      LHCb & $p$ & 5020 &  $\jpsi$ & Pb & $y$, \pt & \cite{LHCb:2013gmv} \\
           & $p$ &  &  $\psi'$ & Pb & $y$, \pt & \cite{LHCb:2016vqr} \\
           & $p$ &  &  $\Upsilon$ & Pb & $y$, \pt & \cite{LHCb:2014rku} \\
           & $p$ & 8160 &  $\jpsi$ & Pb & $y$, \pt & \cite{LHCb:2017ygo} \\
           & $p$ &  &  $\psi'$ & Pb & $y$, \pt & \cite{LHCb:2024taa} \\
           & $p$ &  &  $\Upsilon$ & Pb & $y$, \pt & \cite{LHCb:2018psc} \\
      \hline 
       CMS & $p$ & 5020 &  $\jpsi$ & Pb & $y$, \pt & \cite{CMS:2017exb} \\
             & $p$ &  &  $\psi'$ & Pb & $y$, \pt & \cite{CMS:2018gbb} \\
             & $p$ &  &  $\Upsilon$ & Pb & $y$, \pt & \cite{CMS:2018bbk, CMS:2022wfi} \\
      \hline 
      \hline  
    \end{tabular}
    \caption{Quarkonia production data available from collider experiments with hadrons on nuclei.}
    \label{tab:table_Quarkonia_C}
    
    \end{center}
\end{table*}

A few years later, the E866 collaboration  \cite{Leitch:1999ea} repeated the E772 studies by significantly extending the kinematic coverage of the data. A substantial variation of $J/\psi$ suppression as a functions of both $x_F$ and $p_T$ was observed. Moreover, the data showed distinct suppression patterns for $J/\psi$ and $\psi'$, for all positive values of $x_F$. This suppression becomes more pronounced at $x_F \gtrsim 0.3$.  Complementary high-statistics measurements by the NA50 collaboration at CERN confirmed \cite{NA50:2003fvu} that the loosely bound $\psi'$ is more strongly suppressed than the $J/\psi$. Further investigation using a higher-intensity proton beam showed \cite{NA50:2006rdp} that the ratio of $\psi'$ to $J/\psi$ production significantly decreases with increasing $A$, also indicating stronger $\psi'$suppression. 

The HERA-B collaboration also measured $\rpa^{\psi'}$ for three nuclear targets \cite{HERA-B:2006bhy}, finding results consistent with those of E866 \cite{Leitch:1999ea}. Moreover, their results  confirmed the modification of the measured $p_T$ spectra with $A$, providing support for explanations involving multiple parton interactions. By extending the measurements into the negative $x_F$ region, HERA-B also observed \cite{HERA-B:2008ymp} an increase of $\alpha$ with decreasing $x_F$, with $\alpha > 1$ for $x_F\lesssim-0.15$. In fixed-target interactions, such negative values of $x_F$ correspond to the antishadowing region in the target $x_2$ range. 

The NA50 collaboration at CERN investigated $J/\psi$ suppression using seven nuclear targets 158 and 450 GeV proton beams. Their analysis \cite{NA60:2010wey},  presented as a function of the thickness of the nuclear target traversed by the $J/\psi$, $L$,  revealed a significant $\sqrt{s_{NN}}$ dependence that became  more pronounced for larger $L$. This energy dependence remained when the data were presented as a function of $x_2$. 

More recently, heavy flavor production measurements in fixed-target mode with the LHC proton beams of 2.5, 4.0 and 6.5 GeV ($\sqrt{s}/2$ in collider mode) were reported. Using a dedicated internal gas target (SMOG), the LHCb collaboration was able to fill gaps in the fixed-target energy reach from $\sqrt{s_{NN}} = 69$ to 110.4 GeV \cite{LHCb:2018jry,LHCb:2022sxs}.  These first, promising, measurements are in good agreement with previous data at lower energies.  

Subsequent measurements at the LHC have confirmed the suppression patterns observed in fixed-target experiments at lower beam energies. In $\pA$ collisions, the nuclear modification factor $\rpa$ for $J/\psi$ exhibits stronger suppression at forward rapidity compared to backward rapidity, indicating a rapidity-dependent nuclear effect. This behavior is consistent across measurements by ALICE~\cite{Adam:2015jsa, ALICE:2014cgk, ALICE:2014ict, ALICE:2022zig, ALICE:2020vjy, ALICE:2019qie}, LHCb~\cite{LHCb:2013gmv, LHCb:2016vqr, LHCb:2017ygo, LHCb:2024taa, LHCb:2018psc, LHCb:2014rku}, and CMS~\cite{CMS:2017exb, CMS:2018gbb, CMS:2018bbk, CMS:2022wfi}. Furthermore, the suppression of different quarkonium states varies with the strength of their binding energies. The hierarchy of suppression observed follows the trend $\rpa^{\psi({\rm 2S})} < \rpa^{J/\psi} < \rpa^{\Upsilon({\rm 1S})}$, where the $\psi$(2S) states are more suppressed and the $\Upsilon$(1S) states being least suppressed  \cite{Arnaldi:2014kta, Paul:2014vta, Moon:2023lgc}. 
States with lower binding energies, like the $\psi$(2S), are thus more easily dissociated.  This hierarchy is more pronounced at backward rapidity than at forward rapidity.  In addition, the suppression of $\Upsilon$ states follows the hierarchy $\rpa^{\Upsilon({\rm 3S})} < \rpa^{\Upsilon({\rm 2S})} <\rpa^{\Upsilon({\rm 1S})}$ \cite{CMS:2022wfi}. The relative suppression among the $\Upsilon$ excited states is also more pronounced at backward than at forward rapidity. In summary, the suppression patterns of quarkonium states in $\pA$ collisions at the LHC are rapidity-dependent and reflect the interplay between CNM effects, quarkonium masses and binding energies. 

These quarkonium data are summarized in Tables~\ref{tab:table_Quarkonia_FT} and \ref{tab:table_Quarkonia_C}.

\subsection{Open heavy-flavor data}

Open heavy-flavor production in $\hA$ collisions offers complementary insights into CNM effects relative to quarkonium production. At LO, heavy quarks are predominantly produced through $2 \to 2$ scatterings. Following production, they undergo hadronization via fragmentation into various charged and neutral states, including $D^0$, $D^\pm$, and $D_s$ mesons for charm quarks, and $B^0$, $B^\pm$, $B_c$ and $B_s$ mesons for bottom quarks. The corresponding nuclear modification factors have been investigated both at fixed-target and collider facilities.  

The first experiments were with pion beams. WA82 at CERN \cite{Adamovich:1992fx} and E769 at Fermilab \cite{Alves:1992ux} investigated the $A$ dependence of $D^{\pm}$, $D^0$ and $\bar D^0$ mesons. The resulting cross sections were assumed to follow the usual $A^{\alpha}$ dependence. In both experiments, the extracted values of $\alpha$ were found to be compatible with unity. 
Moreover, no statistically significant dependence on $p_T$ or $x_F$ was observed. A subsequent analysis of $D^{*}$ mesons by E769~\cite{Alves:1993igi} corroborated these findings. An absence of nuclear effects was also reported \cite{Adamovich:1996xf} by the WA92 experiment at CERN for $D^0$, $D^+$, and $D_{s}^{+}$ mesons and their charge conjugates. Similarly, no nuclear dependence was observed in the central rapidity region (near $x_F=0$) by the E789 collaboration \cite{E789:1994nhc}, using an 800 GeV proton beam. More recently, these results  were confirmed and their precision improved by the HERA-B collaboration for $D^0$, $D^+$, $D_s^+$, and $D^{*+}$ mesons in $\pA$ collisions at 920 GeV \cite{HERA-B:2007rfd}.

Somewhat different results were obtained \cite{BlancoCovarrubias:2009py} by the SELEX collaboration, using $p$, $\pi^+$, $\pi^-$ and $\Sigma^-$ beams. The production of $D^0$, $D^+$, $D_s^+$, and $D^{*+}$ mesons and their antiparticles was found to be independent of the beam used and, within the statistical uncertainties, independent of $x_F$ in the interval $0.1<x_F<1.0$. However, the average value of $\alpha$ in their measurements is $\alpha=0.85\pm 0.03$, which is significantly smaller than unity. The discrepancy between the SELEX results and the other measurements has not yet been fully clarified.  

While no significant $A$ dependence was observed for $D$ mesons, a marked asymmetry was observed between $D^+$ and $D^-$ meson distributions as a function of $x_F$ and $p_T$.  With a $\pi^-$ beam, the $D^-$, which shares a valence $d$ quark with the beam, is designated as leading charm while the $D^+$, which does not, is considered nonleading.  The E791 collaboration \cite{E791:1996htn} compared their $D$ meson asymmetries with those from E769 \cite{Alves:1992ux} and WA82 \cite{Adamovich:1992fx} and found that, even though the energies were different, the asymmetries were similar. More recently, SMOG@LHCb has demonstrated a similar asymmetry in $p+{\rm Ne}$ interactions \cite{LHCb:2022cul}.  These data cannot be easily explained within convention production in pQCD but could be interpreted as being due to effects such as intrinsic charm or color connections.

Open charm data were also collected by the LHCb collaboration in fixed-target mode \cite{LHCb:2022sxs}, using the LHC proton beam at 2.5 TeV and a gaseous neon target. A comparison between $J/\psi$ and $D^0$ production shows a cross section ratio that more than doubles for 1 $\lesssim p_T \lesssim$ 4 GeV, while it remains nearly constant as a function of $y$. 

Measurements  in collider mode have also been made at the LHC by the ALICE \cite{ALICE:2019fhe, Tarhini:2017xtb} and LHCb collaborations \cite{LHCb:2017yua, LHCb:2023kqs} at $\sqrt{s_{NN}} = 5.02$ TeV, with additional $D^0$ and $B$ meson data from LHCb at $\sqrt{s_{NN}} = 8.16$ TeV \cite{LHCb:2019avm,LHCb:2022dmh}. The LHCb measurements reveal distinct rapidity-dependent effects. In $p+\mathrm{Pb}$ collisions, $\rpa^{D}$ exhibits a weak $p_T$ dependence at backward rapidity, while strong suppression is observed at forward rapidity with low $p_T$. Similarly, $B^+$ meson production at $\sqrt{s_{NN}} = 8.16$ TeV shows a more pronounced suppression at forward rapidity compared to backward rapidity. Low $p_T$ suppression is observed in the forward region while there is no significant suppression at backward rapidity.  

In addition, the LHCb collaboration has reported the first observation of strangeness enhancement in charm hadronization in $p+\mathrm{Pb}$ collisions at $\sqrt{s_{NN}} = 8.16$~TeV~\cite{LHCb:2023rpm} by measuring the $D_s^+$ to $D^+$ ratio as a function of charged-particle multiplicity in both the backward ($-5.0 < y < -2.5$) and forward ($1.5 < y < 4.0$) rapidity regions. A significant enhancement of $D_s^+$ over $D^+$ production is observed in high-multiplicity events, particularly at low $p_T$ and backward rapidity.

Recently, the ratio of the 
$\Lambda_c^+$ charm baryon relative to $D^0$ production in $p+{\rm Pb}$ collisions at $\sqrt{s} = 5.02$~TeV was published by the CMS collaboration as a function of $p_T$ for $-1.46 < y < 0.54$. An $A$ dependence of charm baryon relative to charm meson production is observed \cite{CMS:2023frs,CMS:2024tjo}. A rapidity-dependent analysis could further elucidate the underlying interaction dynamics.

\begin{table*}
  \begin{center}
    \begin{tabular}{ccccccc}
      \hline 
      \hline 
      Exp. & Beam  & $\sqrt{s_{NN}}$ (GeV) & Final State & Target ($A$) & Observable & Ref. \\
      \hline 
      E769    & $\pi^{+}$,  & 21.7  & $D^{\pm}$, $D^0$, $\bar D^0$ & Be, Al, Cu, W & $x_F$, \pt & \cite{Alves:1992ux} \\
      \hline
      WA82 & $\pi^{-}$ & 25.3 & $D^0$, $D^{\pm}$ & Si, Cu, W & $x_F$ & \cite{Adamovich:1992fx} \\
      \hline 
      E791 & $\pi^{-}$ & 30.6 & $D^{\pm}$ & Cu, Pt & $x_F$, $p_{T}$ & \cite{E791:1996htn} \\
      \hline 
      E781 & $\pi^{+}$, $p$ & 31.8 & $D^{\pm}$, $D^0$ & C, Cu & $x_F$, \pt & \cite{BlancoCovarrubias:2009py} \\      
           & $\pi^{-}$, $\Sigma^{- }$ & 33.5 & $D^{\pm}$, $D^0$ & C, Cu & $x_F$ & \cite{BlancoCovarrubias:2009py}  \\
      \hline 
      E789  & $p$ & 38.7  & $D^0$, $\bar D^0$ & Be, Au & $p_{T}$ & \cite{E789:1994nhc} \\
      \hline 
      HERA-B & $p$ & 41.5 & $D^0$, $D^{+}$,  $D^{*}$  &  C, Ti, W & $x_F$, \pt & \cite{HERA-B:2007rfd} \\    
      \hline 
      SMOG & $p$ & 68.5 & $D^0$, $\bar D^0$ & Ne & $y$, \pt & \cite{LHCb:2022cul} \\
           & $p$ & 86.6, 110.4 & $D^0$ & He, Ar & $y$, \pt & \cite{LHCb:2018jry} \\
      \hline 
      \hline 
      ALICE & $p$ & 5020 & $D^{+}$, $D^0$, $D^{*+}$, $D^{+}_{\text{S}}$& Pb & $y$, \pt & \cite{ALICE:2019fhe} \\
      \hline 
      LHCb & $p$ & 5020 & $D^0$, $D^{+}$, $D^{+}_{s}$ & Pb & $y$, \pt & \cite{LHCb:2017yua, LHCb:2023kqs} \\
           & $p$ & 8016 & $D^0$ & Pb & $y$, \pt & \cite{LHCb:2022dmh} \\ 
           & $p$ &  & $D^{+}_s $ & Pb & $y$, \pt & \cite{LHCb:2023rpm} \\
           & $p$  & & $B^{+}$, $B^0$ & Pb & $y$, \pt & \cite{LHCb:2019avm} \\
        \hline
       CMS & $p$ & 5020 & $B^+$, $B^0$, $B_s^0$ & Pb & $y$, $p_T$ & \cite{CMS:2014tfa} \\
         & $p$ & 8160 & $B^+$ & Pb & $p_T$ & \cite{CMS:2024nrk} \\
      \hline 
      \hline 
    \end{tabular}%
    \caption{Open heavy flavor production data at fixed-target and collider energies.}
    \label{tab:table_OpenHeavy}
  \end{center}
\end{table*}

\subsection{Ultraperipheral collisions in ultrarelativistic heavy-ion collisions}

\begin{figure}[tbp]
  \begin{center}
   \includegraphics[width=\linewidth,clip]{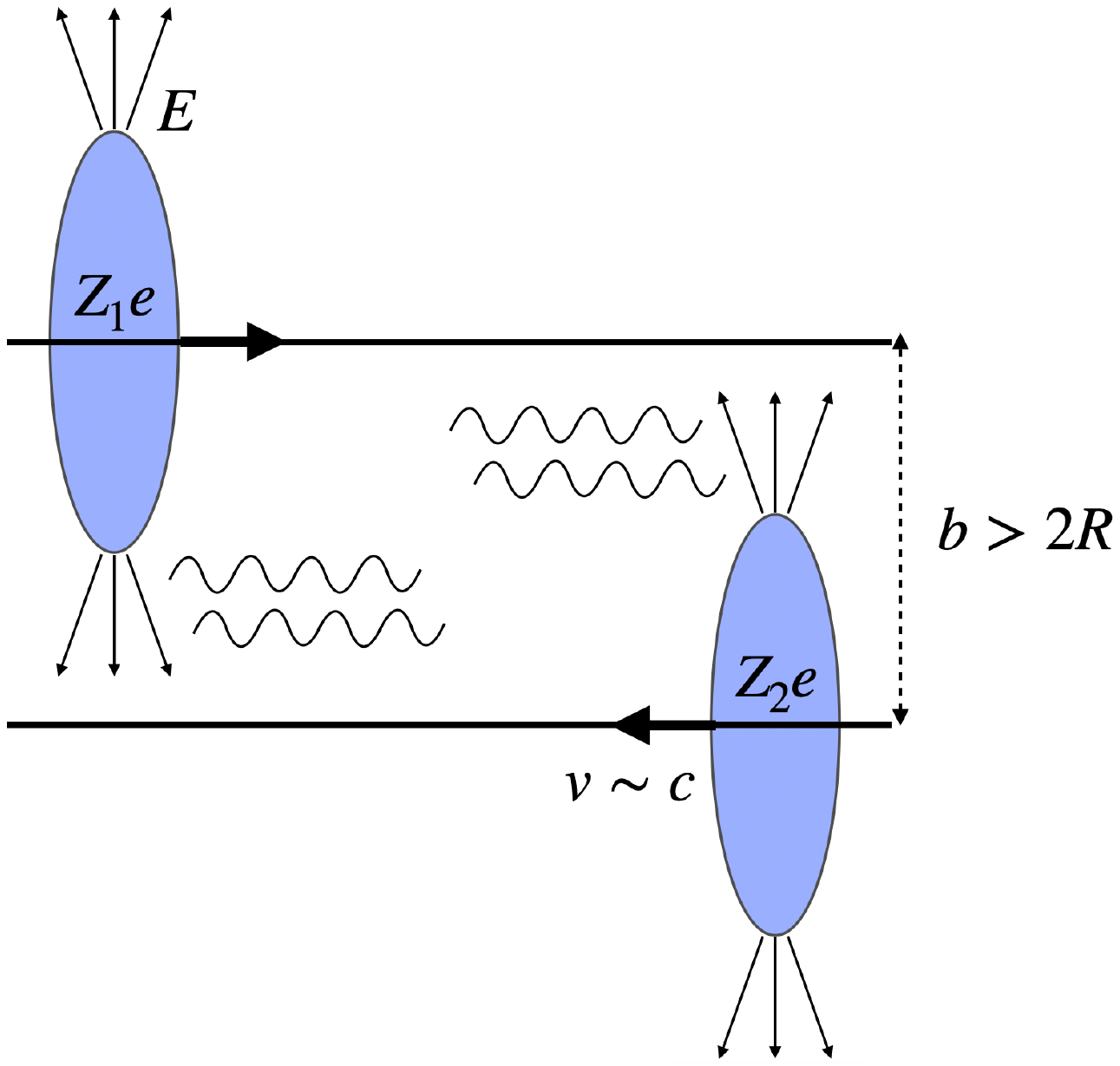}
    \caption{Sketch of an ultraperipheral heavy-ion collision.}
   \label{fig:LHCUPCs}
  \end{center}
\end{figure}

Ultraperipheral heavy-ion collisions (UPCs) provide a unique experimental tool to generate large fluxes of high-energy photons for $\gamma +A$ collisions to provide further characterization the nature of nuclear matter at low $x$~\cite{Strikman:2005yv}. As shown in Fig.~\ref{fig:LHCUPCs}, UPCs are characterized by an impact parameter that exceeds the sum of the two colliding nuclear radii. These interactions are dominated by scatterings of quasi-real photons from the strong electromagnetic fields accompanying ultrarelativistic heavy nuclei.\footnote{In such collisions, the photon flux from each nucleus is proportional to the square of the nuclear charge, $Z^2$.} The study of heavy-quark production in UPCs, in particular, relies on perturbative calculations to model the partonic scattering.

Measurements of coherent $\jpsi$ photoproduction in UPCs by STAR \cite{STAR:2019yox} at RHIC and  ALICE \cite{ALICE:2021gpt, ALICE:2019tqa, ALICE:2024midy} and CMS  \cite{CMS:2016itn} at the LHC could constrain the gluon nPDF in nuclei for $Q^2 \sim M^2_{c\overline{c}}$ as low as $x \approx 10^{-5}$, confirming strong suppression of low $x$ gluons~\cite{CMS:2023snh,ALICE:2023jgu}. More recently, the first measurement of incoherent $\jpsi$ production in UPCs~\cite{CMSPAS-HIN-23-009}  provided insights into event-by-event gluon density fluctuations. CMS results show significant nuclear suppression, comparable to coherent photoproduction, for $x<10^{-4}$.  The suppression decreases with increasing $x$. 

UPC measurements, summarized in Table~\ref{tab:UPC_Jpsi}, show that coherent $J/\psi$ photoproduction cross sections are suppressed relative to the impulse approximation\footnote{The impulse approximation assumes that the photoproduction process occurs independently of the surrounding nuclear environment, treating the nucleus as a sum of free nucleons.} for $x$ values in the range $10^{-5} <  x< 10^{-2}$. The STAR collaboration, at $\sqrt{s_{NN}} = 200$ GeV, observed a nuclear suppression factor of $\approx 0.90 \pm 0.10$ at $x \sim 10^{-2}$, consistent with leading-twist shadowing predictions \cite{STAR:2019yox}. At $\sqrt{s_{NN}} = 5.02$ TeV, ALICE measurements at forward rapidity indicate 40–60\% suppression of coherent $J/\psi$ production at $x \sim 10^{-5}$  \cite{ALICE:2019tqa, ALICE:2021gpt, ALICE:2024midy}. The CMS results at similar $x$ values are compatible with those from ALICE, within experimental uncertainties \cite{CMS:2016itn}.  LHCb measurements \cite{LHCb:2022ahs}  in peripheral Pb+Pb collisions at $\sqrt{s_{NN}} = 5.02$ TeV further confirm the suppression at $x < 10^{-4}$, although there are some tensions with  midrapidity results from other experiments.  CMS has also  reported the first measurement of $D^0$ production in UPCs. Photonuclear events, selected via zero-degree calorimeters and events with large rapidity gaps, confirm the importance of such studies. The results offer a $Q^2$-dependent characterization of nuclear matter at low $x$~\cite{CMSPAS-HIN-24-003,LHCb:2022ahs}.

\begin{table*}
  \begin{center}
    \begin{tabular}{cccccc}
      \hline
      \hline
      Experiment & System & $\sqrt{s_{NN}}$ (GeV) & Final State & Observable & Reference \\
      \hline
      STAR & Au+Au & 200 & $J/\psi$ & $y$ & \cite{STAR:2023nos, STAR:2023vvb} \\
        &  &  & $\psi'$ & $y$ & \cite{STAR:2023vvb} \\
            \hline
       PHENIX & Au+Au & 200  & $J/\psi$ & $y$ & \cite{PHENIX:2009xtn} \\
      \hline
      ALICE & Pb+Pb & 2760 & $J/\psi$ & $y$ & \cite{ALICE:2012yye} \\
        & & 5020 & $J/\psi$,  $\psi'$ &  
      $y$ & \cite{ALICE:2021gpt} \\
            &  &  & $J/\psi$ & $y$ & 
            \cite{ALICE:2019tqa} \\
            &  &  & $J/\psi$ & 
            $y$ & \cite{ALICE:2024midy} \\
      \hline
       CMS & Pb+Pb & 2760 & $J/\psi$ & $y$
      & \cite{CMS:2016itn} \\
       &  & 5020 & $J/\psi$ & $y$ 
      & \cite{CMS:2016itn} \\
      \hline
      LHCb & Pb+Pb & 5020 & $J/\psi$ &  
      $y$ & \cite{LHCb:2022ahs} \\
      \hline
      \hline
    \end{tabular}
    \caption{Recent measurements of exclusive $J/\psi$ and $\psi'$ photoproduction in UPCs.}
    \label{tab:UPC_Jpsi}
  \end{center}
\end{table*}

\subsection{Discussion}

We have presented a non-exhaustive overview of the behavior of nuclear modification factors for several hard processes. With the exception of some Drell-Yan data, suppression or enhancement patterns are clearly visible in all the energy domains covered.  However, the extent of the observed modification depends on the specific process and observable under consideration.

Here we highlight the key observations and considerations that illustrate the limitations of the nPDF formalism, which models the modification of parton distributions inside nuclei relative to free nucleons, in fully describing the $\hA$ data in Tables~\ref{tab:table_DY}, \ref{tab:table_Quarkonia_FT}, \ref{tab:table_Quarkonia_C}, and \ref{tab:table_OpenHeavy} as a function of $y$ and $p_{T}$. Even in the case of structure functions measured in DIS, it remains unclear how much of the observed behavior arises from leading-twist dynamics as opposed to nuclear-enhanced higher-twist effects~\cite{Frankfurt:2011cs,Qiu:2003pm,Qiu:2004qk}. The nPDF formalism will be discussed in more detail in Section~\ref{sec:nPDF}. The nonexhaustive list below identifies some strong indications that going beyond the nPDF framework is necessary to fully interpret the data.
\begin{itemize}[left=0.1cm]
    \item In the $2 \to 1$ LO kinematics regime ($p_T \lesssim M$), relevant for $gg \to J/\psi$ or $q\bar{q} \to \gamma^{\star}$ (Drell-Yan), the final-state $p_T$ is negligible for the determination of $x_2$ and $Q^2$. As a result, the nuclear modification factor in hadron-nucleus collisions simplifies to
    \begin{equation}
        \label{eq:nonPDF}
        \rha(y, p_T) \simeq R_{i}^{\text{nPDF}}\left(x_2 = \frac{M}{\sqrt{s}} e^{-y}, Q^2 \sim M^2\right),
    \end{equation}
    with $i \in (q, \overline{}{q}, g)$. The Cronin peak observed in $\rha(p_T)$ around $p_T \lesssim 1$--$2$ GeV~\cite{Cronin:1974zm,Vitev:2002pf} is in conflict with nPDF calculations, which predict that $\rha$ has a weak  $p_T$ dependence and thus fails to fully describe the data shown in  Fig.~\ref{fig:E866_DY_Jpsi_pT}.
    \begin{figure}[!ht]  
        \centering  
        \includegraphics[width=\linewidth,clip]{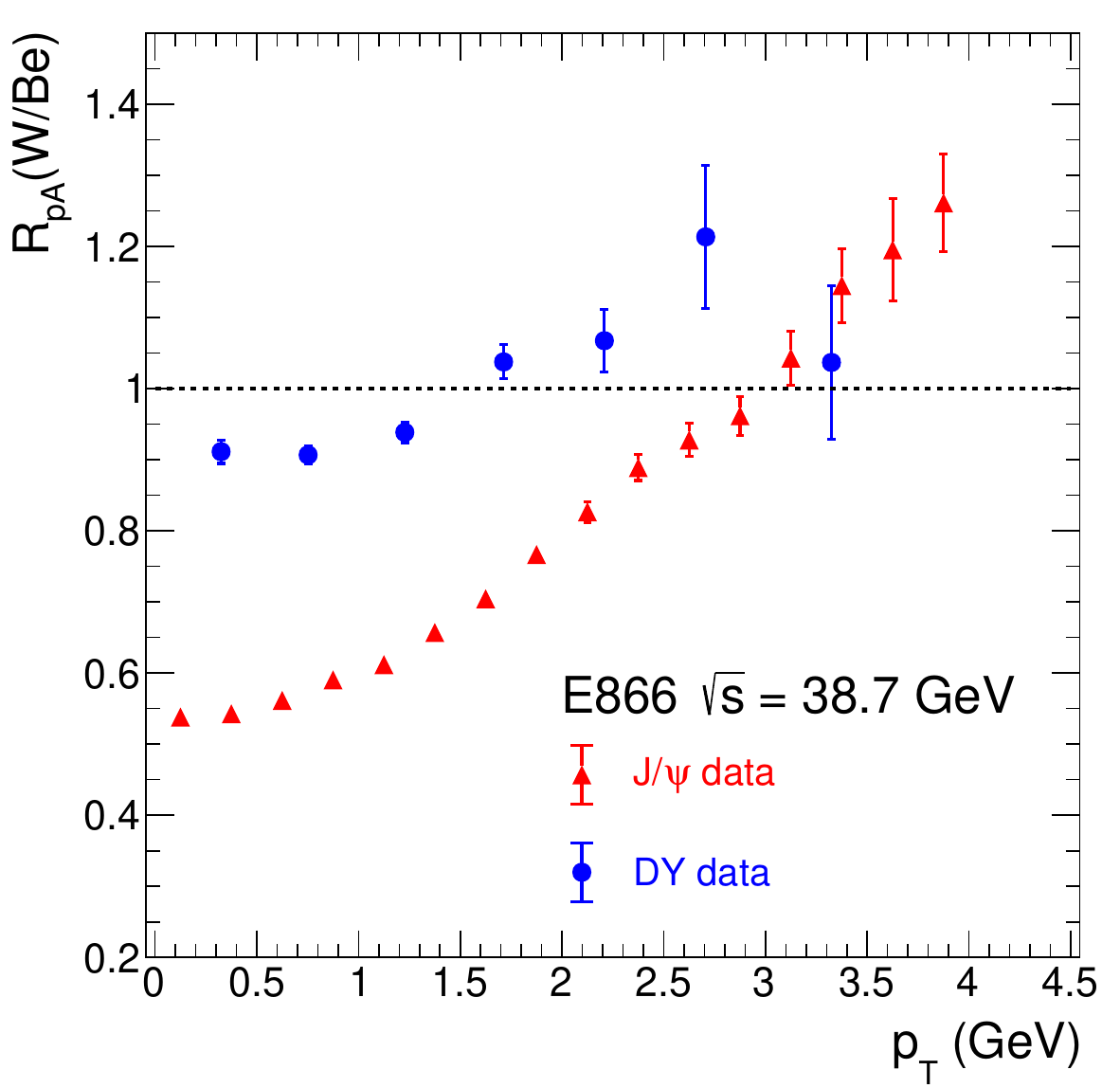}  
        \caption{The $p_T$ dependence of the Drell-Yan  ($4 < M < 8.5$ GeV, blue) and $J/\psi$ ($0.3 \leq x_{\text{F}} \leq 0.93$, red) nuclear modification factor in $p+A$ collisions at $\sqrt{s_{NN}} = 38.7$ GeV, measured by the E866 collaboration \cite{Vasilev:1999fa, Leitch:1999ea}. } 
        \label{fig:E866_DY_Jpsi_pT}  
    \end{figure}
    
     \item The E866 $J/\psi$ data \cite{Leitch:1999ea}  probe the kinematic range $10^{-2} \lesssim x_2 \lesssim 10^{-1}$, where gluon nPDF effects (shadowing and antishadowing) are significant. EPPS16 \cite{Eskola:2016oht} predicts a gluon modification factor of $0.7 \lesssim R_{g}^{\text{nPDF}}(x_2) \lesssim 1.1$ in this $x_2$ range while the measured nuclear suppression, $\rpa^{J/\psi}(\text{W/Be})$ is $ \sim 0.4$ at large $x_F$, exceeding the nPDF expectations (see Fig.~\ref{fig:E866_vs_nPDF}).  Indeed, nPDF effects alone underestimate the observed suppression. Such suppression is, however, consistent with predictions employing Fully Coherent Energy Loss \cite{Arleo:2013xza}.
     
    \begin{figure}[!ht]  
        \centering  
        \includegraphics[width=\linewidth,clip]{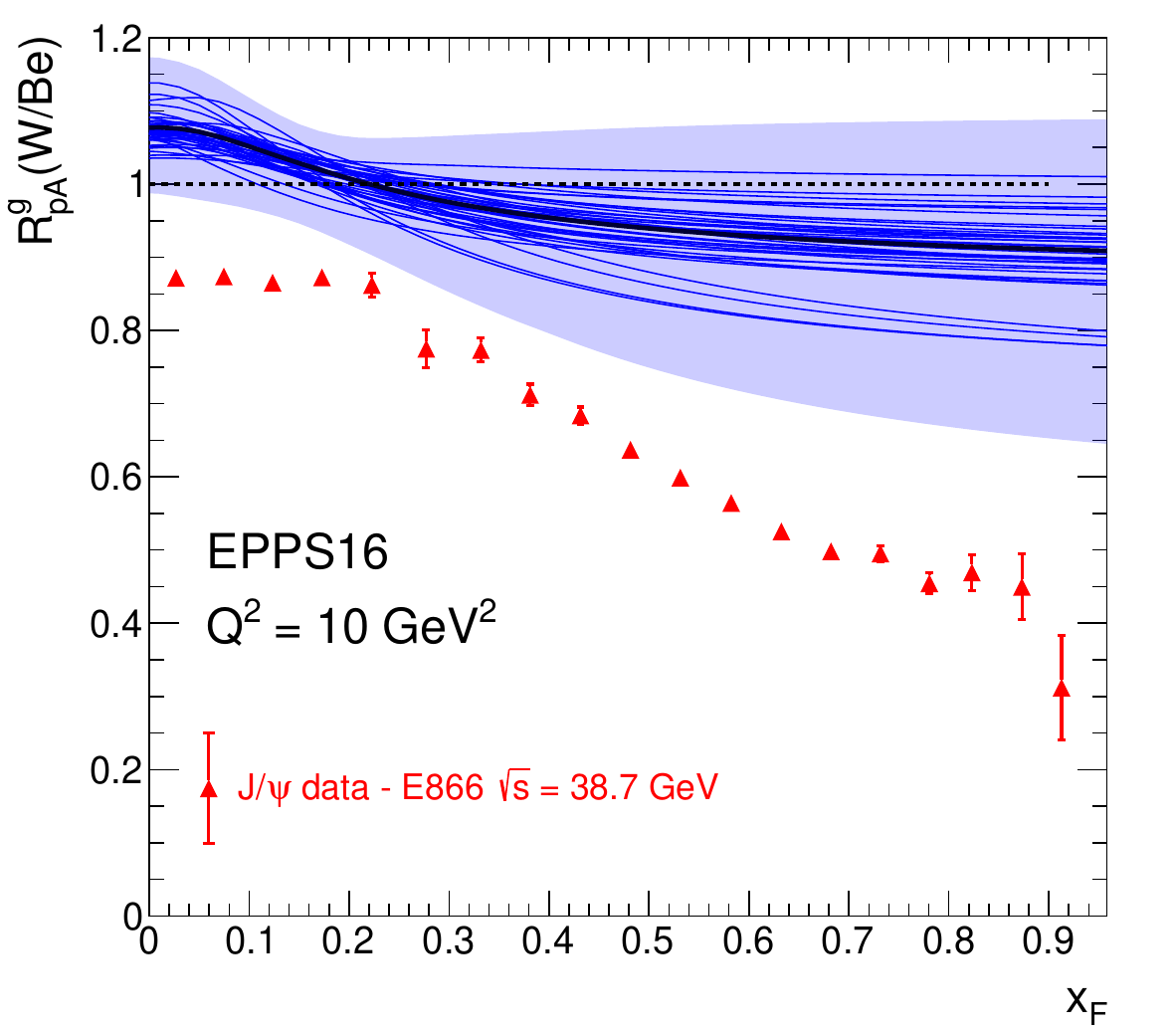}  
        \caption{Comparison between E866 $J/\psi$ data \cite{Leitch:1999ea} (red points) and the EPPS16 NLO gluon nPDF ratio $R_{g}(\text{W}/\text{Be})$ \cite{Eskola:2016oht} (blue band) as a function of $x_F$.}  
        \label{fig:E866_vs_nPDF}  
    \end{figure} 
    
     \item Preliminary E906 Drell-Yan data \cite{Lin:2017eoc} explore the region $0.1 \lesssim x_2 \lesssim 0.3$, where antishadowing effects on sea quarks predict $R^{\text{nPDF}}_{\bar{u}}(x_2) \gtrsim 1$ \cite{Eskola:2016oht, Arleo:2018zjw}. However, the data show suppression at large $x_F$ ($x_2 \sim 0.1$), with $\rpa^{\text{DY}}(\text{W/C}) \sim 0.75$, in tension with nPDF predictions. Such suppression is, however, consistent with predictions from initial-state energy loss~\cite{Neufeld:2010dz, Arleo:2018zjw}.

   \item At LHCb energies, $\rpa^{J/\psi} \sim 0.6$ at forward rapidity ($2.5 < y < 4$) \cite{LHCb:2013gmv, LHCb:2017ygo}, stronger than predicted by nPDF shadowing. This suppression is consistent with observations by the ALICE collaboration \cite{Adam:2015jsa, ALICE:2022zig}, which reported a similar trend in $J/\psi$ suppression at forward rapidity in $p+\mathrm{Pb}$ collisions at the same beam energy. In addition, the observed hierarchy of nuclear modification factors, $\rpa^{\psi{\rm (2S)}} < \rpa^{J/\psi} < \rpa^{\Upsilon(1S)}$ \cite{Arnaldi:2014kta, Moon:2023lgc}, suggests the presence of nuclear effects beyond nPDFs alone.
   
    \item Within the nPDF formalism, $\rha$ should scale with $x_2$ and $Q^2$, i.e. 
    \begin{equation}
        \rha^{\text{nPDF}}(x_2, Q^2, \sqrt{s}) \equiv \rha^{\text{nPDF}}(x_2, Q^2).
    \end{equation}
    However, as shown in Fig.~\ref{fig:scaling_x2_jpsi}, the nuclear dependence of $J/\psi$ production, parameterized as $\sigma_{pA} = \sigma_{pp} A^\alpha$, violates this scaling. The magnitude of the observed suppression clearly depends on the beam energy. A similar trend is observed for Drell-Yan production in fixed-target experiments, indicating the need for nuclear effects beyond those of nPDFs.

\begin{figure}[ht!]
    \centering
    \includegraphics[width=\linewidth,clip]{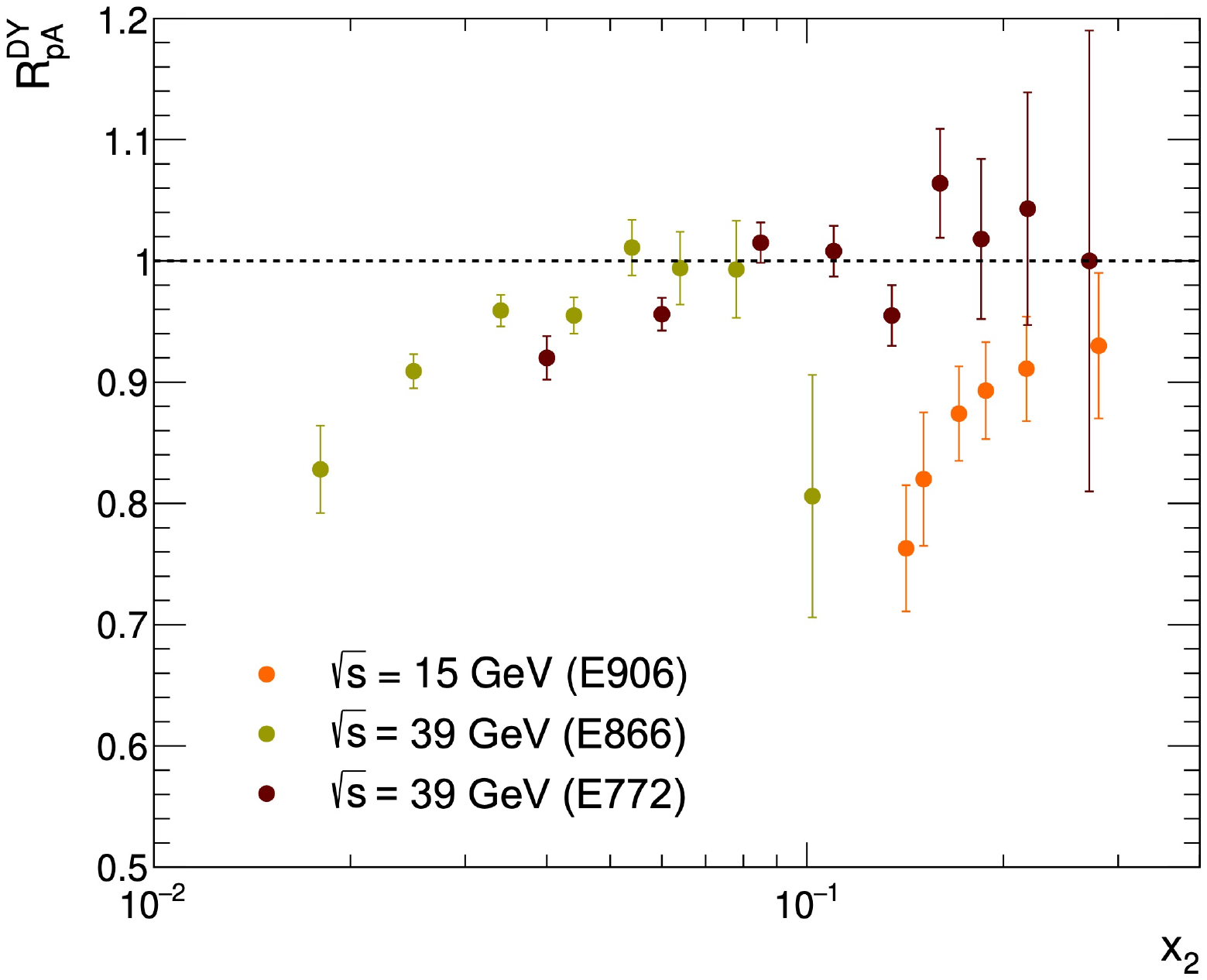}
    \vspace{0.5cm}
    \includegraphics[width=\linewidth,clip]{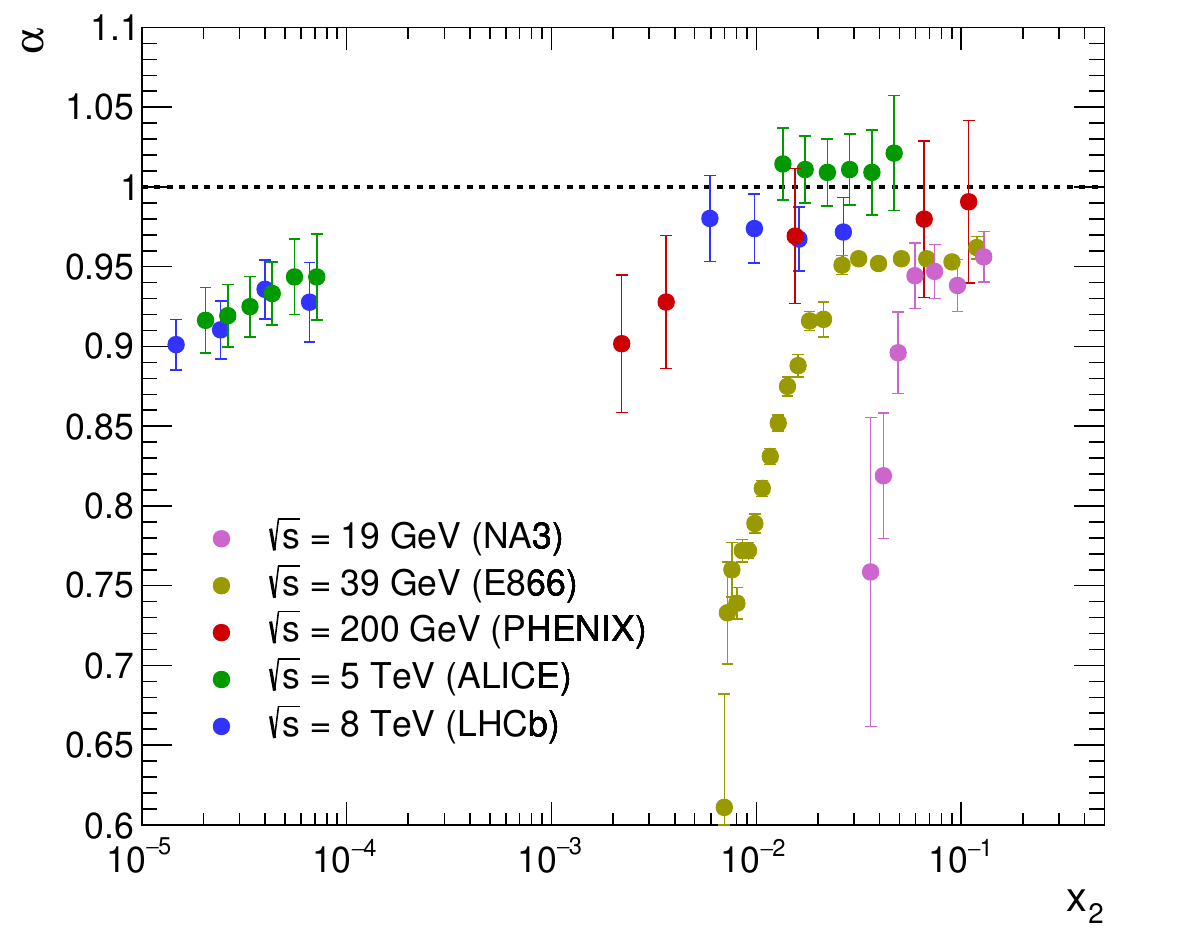}
    \caption{The nuclear dependence of (top) Drell-Yan and (bottom) $J/\psi$ production as a function of $x_2$. The Drell-Yan data, $R_{pA}^{\rm DY}$, are from E772~\cite{Alde:1990im}, E866~\cite{Vasilev:1999fa}, and E906~\cite{Lin:2017eoc}. The $J/\psi$ data, presented as a function of the exponent $\alpha$, are from NA3~\cite{Badier:1983dg}, E866~\cite{Leitch:1999ea}, PHENIX~\cite{Adare:2010fn}, ALICE~\cite{Abelev:2013yxa}, and LHCb~\cite{Aaij:2017cqq}. Adapted from~\cite{Arleo:2018zjw}.}
    \label{fig:scaling_x2_jpsi}
\end{figure}
\end{itemize}

All these observations, across a range of $x_{F}$, $x_{2}$ and $p_{T}$, suggest the need to go beyond the conventional nPDF framework and demonstrate the energy dependence of CNM effects. We attempt to identify the physics mechanisms that could provide a complete and universal description of these data. The following sections introduce the CNM effects that may underlie the observed experimental trends and aim to address the three key questions outlined in Section~\ref{sec:introduction}. We also propose a set of observables designed to disentangle contributions to the observed behavior from individual nuclear effects, thereby facilitating a more precise characterization of the nuclear modifications.   
\section{Disentangling nPDFs from saturation}
\label{sec:nPDF_saturation}

Since the McLerran–Venugopalan effective theory of small-$x$ QCD was introduced in the early 1990s \cite{McLerran:1993ni,McLerran:1993ka,McLerran:1994vd}, the nuclear physics community has long debated whether the observed suppression of the nuclear structure function $F_2$ at small-$x$ reflects ``gluon saturation" or merely conventional ``shadowing." Originally this question arose from comparing $F_2$ in heavy nuclei to that in free nucleons. With the forthcoming EIC, it is timely to recast it more precisely: to what extent is the observed nuclear suppression of $F_2$ driven by perturbative small-$x$ dynamics, and to what extent does it signal genuinely nonperturbative physics? High precision measurements at the EIC should allow us to disentangle and quantitatively assess these two contributions.

If this suppression is genuinely nonperturbative, one must then rely on phenomenological models and possibly lattice methods, which currently face severe limitations. One such model is the Glauber-Gribov model \cite{Armesto:2010kr,Gribov:1968gs,Frankfurt:2002kd, Frankfurt:2003gx, Frankfurt:2003zd} of (virtual) photon-nucleus total cross sections (structure functions) where the exchanged degrees of freedom are nonperturbative color singlet states called ``pomerons" \cite{Capella:2000pe, Capella:2000hq} (or soft pomerons, to be distinguished from hard or BFKL pomerons \cite{Mueller:1994jq}) which carry the quantum numbers of vacuum. In this approach, suppression of the nuclear cross section is due to the destructive interference between the phases of multiple pomeron  exchanges.  The nuclear parton distributions are thus determined at an initial scale $Q_0^2$ and then evolved in $Q^2$ via the DGLAP equations \cite{Lipatov:1974qm,Gribov:1972ri}; this procedure is known as leading-twist shadowing. The evolution of the parton distribution functions is identical for nucleons and nuclei, with all nuclear effects (shadowing) encoded in the initial distributions. These nuclear PDFs, assumed to be universal, are then used within the collinear factorization framework to compute particle production cross sections in nuclear collisions.

On the other hand, if weakly coupled QCD dynamics underlie the observed suppression, it should be calculable directly in terms of quark and gluon degrees of freedom. One such mechanism is gluon saturation, formulated as the Color Glass Condensate (CGC) effective field theory of small $x$ \cite{Iancu:2003xm,Gelis:2010nm,Kovchegov:2012mbw}. Such a saturated state of gluons, characterized by a relatively hard scale, the saturation scale, $Q_s\gg \Lambda_{\rm QCD}$, can be reached in a large nucleus already at moderately small values of $x$ 
($< 4-5 \times 10^{-2}$) without significant evolution in $x$ whereas, in the proton, saturation is expected to be reached at much smaller values of $x$. In this regime of high gluon densities, $Q^2 \lesssim Q_s^2$, nonlinear gluon recombination effects play a crucial role in taming the rapid growth of the gluon distribution at low $x$, leading to  saturation of the distribution, as required by unitarity.  

The saturation scale, which delineates the transition between the dilute and dense regimes of QCD matter, depends on the atomic mass number $A$, the gluon momentum fraction $x$ (which coincides with Bjorken $x$ in structure functions but not necessarily in particle production), and the impact parameter of the collisions with the target nucleus or proton. Importantly, as an effective theory of QCD at small $x$, the CGC framework applies to a broad class of observables beyond structure functions \cite{Morreale:2021pnn,Garcia-Montero:2025hys}.

Note that CGC dynamics can lead to both higher-twist and leading-twist suppression of nuclear cross sections. In the kinematic regime where small-$x$ evolution is significant but the momentum or virtuality remains well above the saturation scale ($Q^2 \gg Q_s^2$), the cross section can be expanded in powers of $1/Q^2$, with the leading contribution corresponding to leading-twist dynamics. Although the system is dilute in this regime, leading-twist shadowing arises due to the presence of a saturation boundary separating the dense and dilute regimes. The extent of the suppression is sensitive to the dynamics near the saturation scale, which can be predicted within the CGC effective theory. Lastly, there has been significant progress toward calculating next-to-leading order (NLO) corrections for various observables using the CGC formalism which is crucial for precise predictions~\cite{Beuf:2016wdz,Beuf:2017bpd,Beuf:2021qqa,Beuf:2021srj,Beuf:2022ndu,Beuf:2022kyp,Beuf:2024msh,Hanninen:2017ddy,Lappi:2016oup,Chirilli:2011km,Chirilli:2012jd,Roy:2019hwr,Roy:2019cux,Mantysaari:2022kdm,Mantysaari:2022bsp,Mantysaari:2021ryb,Iancu:2020mos,Iancu:2022gpw,Kang:2023doo,Caucal:2021ent,Caucal:2023nci,Caucal:2024cdq,Caucal:2024nsb,Bergabo:2022tcu,Bergabo:2023wed,Bergabo:2022zhe,Bergabo:2024ivx,Boussarie:2016ogo,Fucilla:2023mkl,Fucilla:2022wcg, Taels:2022tza, Taels:2023czt, Ayala:2016lhd, Ayala:2017rmh,Wang:2022zdu,Altinoluk:2025dwd}.

\subsection{Nuclear Structure Functions and nPDFs}
\label{sec:nPDF}

In 1981, the EMC collaboration measured the nuclear modification of structure functions in deep inelastic scattering \cite{Aubert:1983xm}. Due to the relation between structure functions and parton distribution functions at leading order and leading twist,
\begin{equation}
   F_2 (x, Q^2) = \sum_i e_i^2 [x f(x,Q^2) + x \bar{f} (x, Q^2)] \, ,
\end{equation}
these data revealed, for the first time, that the PDFs of a proton bound within a nucleus (nPDFs) differ from that of a free proton: $f_{i}^{p/A}(x, Q^2) \ne f_{i}^{p}(x, Q^2)$, where $f_{i}^{p/A}$ represents the PDF of a flavor $i$ parton inside a proton bound in a nucleus. The nuclear modification factor due to this effect is defined as $R_{i}^{A} \equiv f_{i}^{p/A} / f_{i}^p$.\footnote{A slightly different definition is often adopted, $R_{i}^{A} \equiv \left( Z f_{i}^{p/A} + N f_{i}^{n/A} \right) / \left( Z f_{i}^{p} + N f_{i}^{n} \right)$, where $Z(N)$ is the number of protons (neutrons) in nucleus $A$, and $f_{i}^{n/A}(f_{i}^{n})$ are the bound (free) neutron PDFs. This definition takes into account the fact that the proton and neutron distributions in nuclei are always averaged and cannot be accessed separately.} Since then, four main $x$ regions have been identified: at $x \gtrsim 0.7$, Fermi motion dominates with $R^{A}(x) \gtrsim 1$; in the range $0.3 \lesssim x \lesssim 0.7$, the EMC region is characterized by $R^{A}(x) \lesssim 1$; between $0.1 \lesssim x \lesssim 0.3$, antishadowing leads to $R^{A}(x) \gtrsim 1$; and finally, for $x \lesssim 10^{-1}$, shadowing is observed with $R^{A}(x) \lesssim 1$.  Global analyses of nPDFs  have been performed by the DSSZ \cite{deFlorian:2011fp}, EPPS \cite{Eskola:2016oht,Eskola:2021nhw}, KSTSG~\cite{Khanpour:2020zyu}, nCTEQ \cite{Kovarik:2015cma,Duwentaster:2022kpv}, nNNPDF \cite{AbdulKhalek:2022fyi}, and TUJU \cite{Helenius:2021tof} collaborations. The latest analyses also include data from the LHC.  For a recent review, see Ref.~\cite{Klasen:2023uqj}. 

The data used to extract nPDFs come from a number of processes:
\begin{itemize}[left=0.1cm]
    \item \textbf{Deep Inelastic Scattering}: Data such as those from the EMC, SLAC, NMC, and CHORUS experiments  (see Ref.~\cite{Eskola:2016oht} for references to the data), $\ell + A \rightarrow \ell + X$, probe lepton scattering on a nucleus, constraining quark nPDFs in the range $10^{-2} \lesssim x \lesssim 1$ \cite{Eskola:2016oht,Kovarik:2015cma}.
    \item \textbf{Single inclusive hadron production}: Measurements primarily of neutral and charged pions, $\hA \rightarrow \pi + X$, as well as kaons and eta mesons by PHENIX and STAR at RHIC ($\sqrt{s_{NN}} = 200$ GeV)~\cite{Adler:2006wg,Abelev:2009hx,STAR:2006xud,PHENIX:2013kod} and by ALICE at the LHC ($\sqrt{s_{NN}} = 5.02$ and $8.16$~TeV)~\cite{ALICE:2018vhm,ALICE:2016dei,ALICE:2021est} constrain gluon nPDFs in the range $10^{-2} \lesssim x \lesssim 10^{-1}$~\cite{Duwentaster:2021ioo}.
    \item \textbf{Electroweak boson production}: Production of $W$ and $Z$ bosons in \pA collisions, $\pA \to Z, W + X$, measured at the LHC at $\sqrt{s_{NN}} = 5.02$ and $8.16$ TeV probe quark nPDFs and, to some extent, the gluon nPDF in the range $10^{-3} \lesssim x \lesssim 10^{-1}$ in the shadowing and antishadowing regions at scale $Q^2 \sim  M_{W,Z}^2 + p_T^2 \gtrsim 6 \times 10^3 \,{\rm GeV}^2$ \cite{Khachatryan:2015hha,Khachatryan:2015pzs,Aad:2015gta,CMS:2019leu,CMS:2021ynu,ALICE:2016rzo,LHCb:2014jgh}. 
    
    \item \textbf{Dijet production}: CMS measurements at $\sqrt{s_{NN}} = 5.02$~TeV with $p_T \gtrsim 30$ GeV provide constraints on quark and gluon nPDFs in the range $10^{-2} \lesssim x \lesssim 10^{-1}$, depending on the dijet pseudorapidity \cite{Chatrchyan:2014hqa,CMS:2018jpl}.
    \item \textbf{Low mass Drell-Yan pairs}: Data from \pA collisions (E772 \cite{Alde:1990im} , E866 \cite{Vasilev:1999fa}) at $\sqrt{s_{NN}} = 38.7$~GeV in the dilepton mass range $4 < M < 8$ GeV constrain sea quark nPDFs in the shadowing region at $x \sim 0.1$. The Drell-Yan data from $\pi + A$ collisions (NA10 \cite{Bordalo:1987cr,Bordalo:1987cs}, NA3 \cite{Badier:1981ci,Badier:1982zb}, E615 \cite{Heinrich:1989cp}) at $\sqrt{s_{NN}} = 16.2$–$23.1$ GeV constrain the valence quark nPDFs in the EMC and antishadowing regions for $0.1 \lesssim x \lesssim 0.5$. 

    \item \textbf{Heavy flavor production}: LHCb prompt $D^{0}$ data at $0 \lesssim p_T \lesssim 10$~GeV~\cite{LHCb:2017yua} in $p+\mathrm{Pb}$ collisions  constrain the gluon nPDF in the range $10^{-5} \lesssim x \lesssim 10^{-2}$ \cite{Eskola:2019bgf}. These data have been already used in the recent fits by the EPPS \cite{Eskola:2021nhw}, nNNPDF \cite{AbdulKhalek:2022fyi} and nCTEQ \cite{Duwentaster:2022kpv} groups. In addition, the nCTEQ group has studied other heavy flavor processes including $B$ meson and quarkonium production~\cite{Kusina:2017gkz,Kusina:2020dki,Duwentaster:2022kpv}.  
\end{itemize}

The phenomenological extraction of the nuclear modification factor $R^{\text{nPDF}}$ critically depends on the data used in the global analyses. Earlier nPDF determinations relied primarily on DIS processes \cite{Eskola:2003cc,Hirai:2007sx,Schienbein:2009kk}, where the initial lepton interacts purely via quantum electrodynamics (QED) with the nucleus, making them ideal for probing the quark nPDFs. More recently, data in $\hA$ collisions have been incorporated to extend the range of $x$ explored and access the gluon nPDFs more directly.

\begin{figure*}[tbp]
    \centering
    \includegraphics[width=\linewidth,clip]{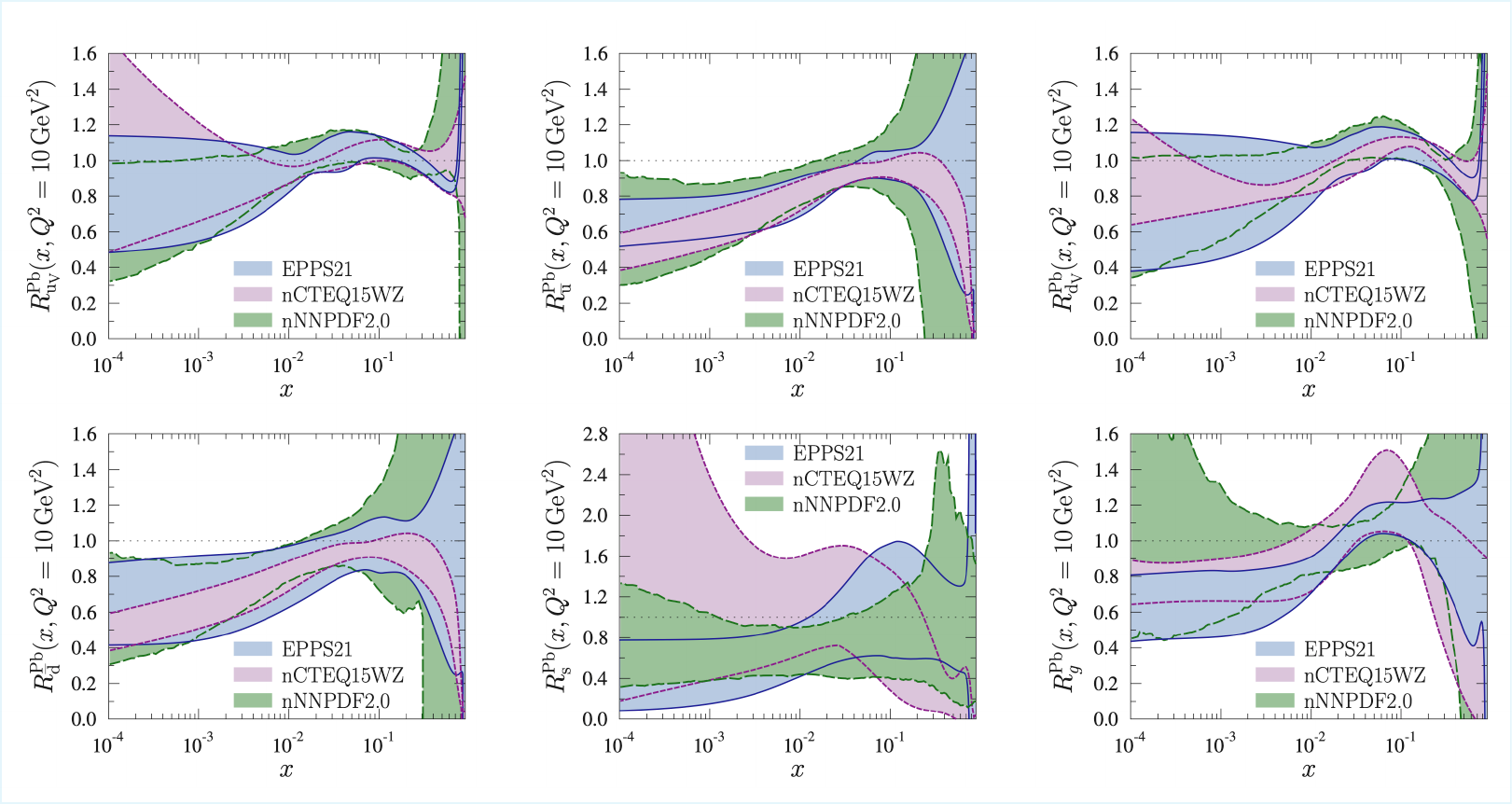}
    \caption{Nuclear modification ratio for quarks and gluons at $Q^2 = 10$ GeV$^{2}$ from EPPS21 \cite{Eskola:2021nhw}, nCTEQ15WZ \cite{Kusina:2020lyz}, and nNNPDF2.0 \cite{AbdulKhalek:2020yuc} taken from \cite{Eskola:2021nhw}.}
    \label{fig:nPDF}
\end{figure*}

Figure~\ref{fig:nPDF} presents a comparison of recent quark and gluon nPDFs at $Q^2 = 10$~GeV$^2$. 
The EPPS21 analysis~\cite{Eskola:2021nhw} incorporates the $D$ production data in their global analysis, providing a strong constraint on $R_{g}^{\text{nPDF}}$ at $x \lesssim 10^{-2}$. Their result agrees with the nCTEQ15WZ analysis~\cite{Kusina:2020lyz} but deviates from NNPDF2.0~\cite{AbdulKhalek:2020yuc}, which does not include these data. These observations suggest that the nPDF evaluations remain highly sensitive to $\hA$ data used in the analyses.

Including the LHC $p+{\rm Pb}$ measurements, the variety and the kinematical reach of the data used in nPDF analyses have grown significantly. It has therefore become increasingly possible to look for deviations from collinear factorization. For example, some of the nuclear effects seen in heavy flavor production in $p+{\rm Pb}$ collisions could be due to dynamical effects such as saturation \cite{Ducloue:2016ywt}, partonic energy loss \cite{Arleo:2021bpv}, or collective effects. It is important to consider to what extent these effects are currently being incorporated in the nPDFs and how to potentially distinguish their "non-factorizable" contribution. However, if the heavy flavor data are left out of the analysis, the small-$x$ gluon nPDF would be largely unconstrained and, as a result, the uncertainties on the baseline upon which these novel effects are supposed to be observed become prohibitively large. In other words, the baseline uncertainties overwhelm the effect one is trying to discover. In principle, this dilemma can be resolved if the remaining data are sufficiently precise and diverse, which is currently not the case. Because the factorizable effects are universal, the presence of effects beyond the nPDF formalism must eventually manifest as the inability to reproduce all the data with a single set of nPDFs in a statistically satisfactory way. It is thus useful to also view the global nPDF analyses as a search of physics beyond collinear factorization and not something that ignores or downplays them. 
  
\subsection{Saturation effects in the small-$x$ regime}
\label{sec:saturation}

At small $x$, the gluon density inside hadrons grows rapidly, leading to nonlinear effects that are not captured by traditional linear QCD evolution equations such as DGLAP \cite{Lipatov:1974qm,Gribov:1972ri} or BFKL \cite{Lipatov:1976zz, Fadin:1975cb, Kuraev:1976ge,Balitsky:1978ic}. This phenomenon, known as gluon saturation, implies that the growth of the gluon distribution eventually slows down and reaches a limiting behavior. To describe this high-density regime, the CGC effective field theory has been developed. It models a high energy hadron as a coherent ensemble of soft gluons with large occupation numbers. By incorporating nonlinear gluon recombination effects, the CGC provides a more accurate and unified description of high-energy scattering processes. These nonlinear dynamics are characterized by the saturation scale $Q_s^2(x)\sim A^{1/3}\, x^{-\lambda}$, which grows as a power law with decreasing $x$. At sufficiently small $x$ (i.e., at high energies), the saturation regime becomes weakly coupled due to the semi-hard scale $Q_s \gg \Lambda_{\text{QCD}}$, making it amenable to perturbative QCD calculations within the CGC framework.

\begin{figure}[tbp]
    \centering
    \includegraphics[width=\linewidth,clip]{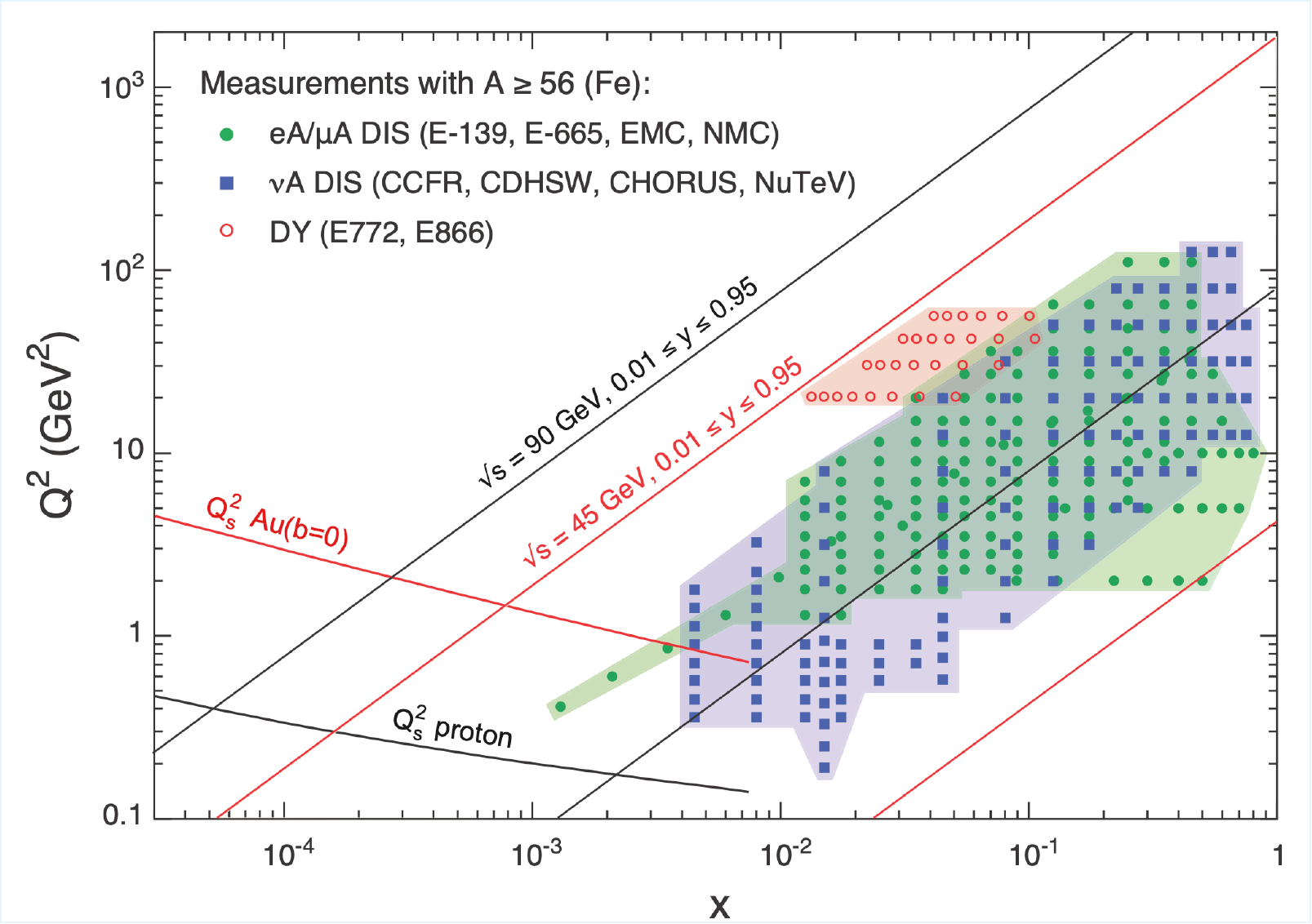}
   \caption{Kinematic phase space of DIS and Drell-Yan data for nuclei heavier than iron as a function of $x$ and $Q^2$, including projections of the saturation scale for  gold nuclei and protons. The kinematic limits of the future EIC are also indicated.  From Ref.~\cite{Accardi:2012qut}.}
    \label{fig:Q2vsX}
\end{figure}

Figure~\ref{fig:Q2vsX} illustrates the current kinematic phase space for existing DIS and Drell-Yan data from nuclei heavier than iron, as a function of $x$ and $Q^2$. The saturation lines $Q^2(x) = Q_{s}^{2}(x)$ demarcate the transition between the linear and nonlinear regimes. The optimal region for studying these nonlinear effects is at $x \lesssim 10^{-2}$ and $Q^2 \sim 1 \text{ GeV}^2$. The kinematic reach of the future EIC, which will significantly extend access to this region, is also shown.

\subsection{Target universality from DIS to $p+A$}
\label{sub:universality-DIS-pA}

In the collinear factorization formalism, valid in the limit $Q^2 \rightarrow \infty$, quark and gluon distribution functions, defined as two-point correlators of quark and gluon fields, are the degrees of freedom that appear in physical cross sections. In this approach, the perturbative evolution of PDFs with respect to $Q^2$ is linear and independent of the nature of the target. However, the initial condition of this evolution is nonperturbative and, as such, encodes information about the structure of the target: in particular, shadowing effects for nuclear targets.

On the other hand, in the kinematic regime where the center-of-mass energy is very large and $Q^2$ is fixed but not too large, typically of the order of the saturation scale, collinear factorization is expected to break down. In this domain, the notion of weakly-interacting partons underlying parton distribution functions becomes inadequate. Instead, one must account for the nonlinear dynamics induced by large gluon densities and resum the large logarithms of the center-of-mass energy $\sqrt{s}$; specifically, $\ln(s/Q^2)$ in DIS and $\ln(\sqrt{s}/p_T)$ in hadronic collisions; which are commonly written as logarithms of $1/x$.  The CGC formalism, where the coupling remains weak, but high-density effects are resummed to all orders, satisfies this requirement.

In the CGC formalism, the universal degrees of freedom are Wilson lines, path-ordered exponentials of the gluon field, which resum the effects of coherent multiple scattering off a dense target. Physical cross sections are expressed in terms of correlators of these Wilson lines. The most prominent are the dipole ($S$) and quadrupole ($Q$) correlators, defined as follows:
\begin{eqnarray}
    S (x_T, y_T) &\equiv& \frac{1}{N_c} 
    \left\langle \Tr\left[V (x_T) V^\dagger (y_T)\right] \right\rangle\,, 
    \nonumber \\
    Q (x_T, y_T, z_T, r_T) &\equiv& 
    \nonumber \\
    &&\hspace{-3em} \frac{1}{N_c}
    \left\langle \Tr\left[V (x_T) V^\dagger (y_T) V (z_T) V^\dagger (r_T)\right] \right\rangle \,,
\end{eqnarray}
where $x_T,y_T,z_T$ and $r_T$ are transverse coordinates.
In multiparticle production in DIS and in proton-nucleus collisions, all other correlators are suppressed by powers of $N_c^2$ \cite{Dominguez:2012ad}. These correlation functions satisfy the JIMWLK evolution equations \cite{Jalilian-Marian:1997qno, Jalilian-Marian:1997ubg} which encode the evolution of the correlators as function of $x$ once an initial condition at $x_0$ is specified.  The large $Q^2$ and small $x$ regimes of QCD are unified in the double limit $Q^2 \to \infty$ and $x \to 0$, where large virtuality and parton density effects interplay.

To leading order in the large $N_c$ approximation, the dipole and quadrupole correlators are the only objects that appear in physical cross sections within the CGC formalism. As such, they are universal quantities that can be probed by different processes, such as DIS and forward particle production in $\pA$ collisions~\cite{Gelis:2002nn,Kang:2014lha}. The universality of the cross sections is because the dipole and quadrupole correlators contain the small $x$ dynamics of the target independent of the projectile. It is then possible to extract the characteristics of the  correlators from DIS and apply them to other processes, conferring the CGC formalism with predictive power. This is analogous to the collinear factorization framework, which is valid at high $Q^2$ but is expected to break down at very small $x$. Moreover, the universality of the Wilson line correlators mirrors that of the gluon distribution function governed by DGLAP evolution in the collinear factorization approach.

There have been sustained efforts to investigate higher-twist corrections to collinear factorization, particularly those enhanced by the size of the nucleus, scaling as $A^{1/3}$~\cite{Qiu:2003vd,Qiu:2004qk,Qiu:2004da,Fu:2023jqv,Fu:2024sba}.\footnote{Target mass corrections are important for understanding higher-twist effects at large $x$.  These corrections can also be systematically included within the collinear factorization framework~\cite{Ruiz:2023ozv}.} More recently, analogous studies have been initiated in the high-energy (small $x$) regime, focusing on subeikonal, energy suppressed contributions which are important for both unpolarized \cite{Altinoluk:2014oxa,Altinoluk:2015gia,Altinoluk:2015xuy,Agostini:2019avp,Agostini:2019hkj,Altinoluk:2020oyd,Altinoluk:2021lvu, Chirilli:2018kkw,Chirilli:2021lif,Altinoluk:2022jkk,Altinoluk:2023qfr,Altinoluk:2024zom} and polarized observables \cite{Kovchegov:2015pbl,Kovchegov:2016weo,Kovchegov:2017lsr,Kovchegov:2018znm,Cougoulic:2022gbk,Borden:2023ugd,Adamiak:2023okq,Li:2023tlw,Kovchegov:2022kyy,Santiago:2023rfl,Li:2024xra}, as well as for connecting the collinear and high-energy (small $x$) frameworks~\cite{Boussarie:2021wkn,Boussarie:2020fpb,Mukherjee:2023snp,Duan:2024qev,Caucal:2024bae,Jalilian-Marian:2017ttv,Jalilian-Marian:2018iui,Jalilian-Marian:2019kaf}.

These efforts motivate the search for observables that can be commonly accessed in both DIS and proton–nucleus collisions, enabling a consistent characterization of the small $x$ structure of nuclei for a variety of probes. 

\subsection{Observables}
\label{sec:saturation-observables}

In this section, we explore several key observables that may distinguish between nuclear parton distribution function and gluon saturation effects.  So far, the extraction of nPDFs at small $x$ has not taken other physical effects into account. Determining the relative contribution of each effect is crucial, as there are limited small $x$ data available, {\it e.g.} at $x \sim 10^{-5}$ in $\hA$ collisions, as discussed in Sec.~\ref{sec:nPDF}.

\subsubsection{Structure functions in DIS}

CGC-based observables have a characteristic $x$ dependence. While the theoretical uncertainties associated with the perturbative expansion are, in principle, controllable, additional uncertainties arise from modeling the nonperturbative inputs required to initialize the small $x$ evolution. The initial conditions are typically parametrized and fitted to experimental data at an intermediate $x$ value, $x_0 \sim 10^{-2}$~\cite{Albacete:2009fh,Albacete:2010sy,Iancu:2015joa,Beuf:2020dxl}.
Once this initial condition is fixed, the CGC framework provides first principles predictions of the $x$ dependence of the structure functions $F_2$ and $F_L$, generally measured in DIS.

The EIC will cover a large kinematic range in $x$, about three orders of magnitude in the region where $x$ can be considered small, which should make it possible to distinguish between shadowing generated by the perturbative small $x$ evolution of the target via the CGC and DGLAP evolution of the $x$-dependent parameterization of shadowing models.  The $x$ dependence of these two approaches is quite different.  For example, DGLAP evolution does not generate any additional shadowing while, in the CGC, it is dynamically generated. Unfortunately, the $F_2$ structure function necessarily receives sizable contributions from nonperturbatively large dipoles~\cite{Mantysaari:2018nng} at small $x$, even at intermediate $Q^2$, making it difficult to distinguish perturbative contributions from nuclear shadowing.  The longitudinal structure function $F_L$ is under much better theoretical control.  However, determinations of $F_L$ are much more difficult because it requires measurements at different energies.

\begin{figure}
    \centering
    \includegraphics[width=\linewidth,clip]{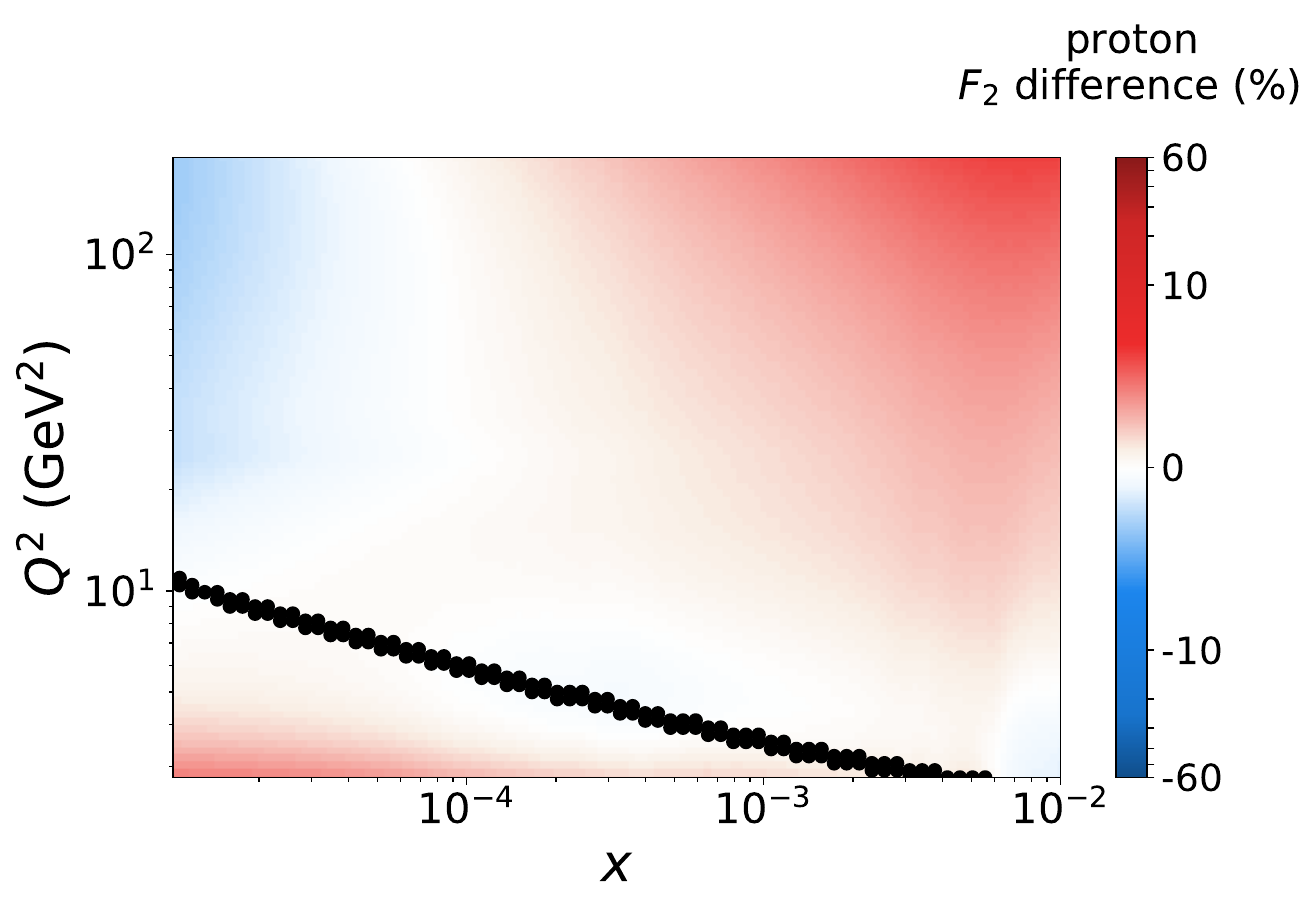} \\
    \includegraphics[width=\linewidth,clip]{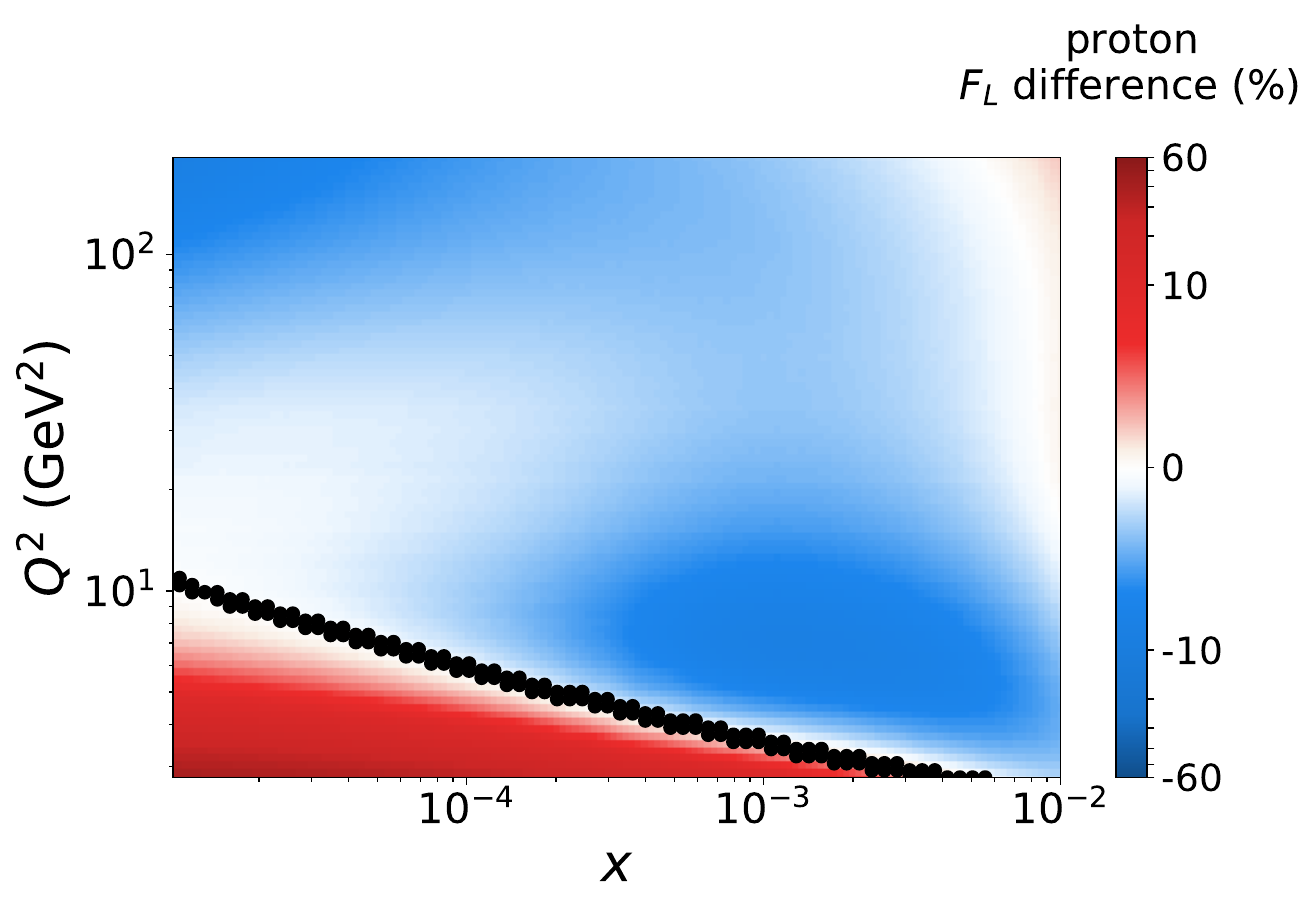}

    \caption{The percent difference between the CGC structure functions and the matched $F_2$ (a) and $F_L$ (b) from DGLAP evolution for a proton as a function of $x$ and $Q^2$. The black dots indicate the region where the two frameworks are matched by construction. From Ref.~\cite{Armesto:2022mxy}.}
    \label{fig:BK-nPDF-proton}
\end{figure}

The CGC framework and that of leading-twist shadowing follow distinct trajectories in the $(x, Q^2)$ plane. While both allow freedom in the parametrization of the initial condition, at scale $Q_0$ for DGLAP evolution and at an initial momentum fraction $x_0$ for the CGC, the underlying QCD dynamics are fundamentally different. Consequently, their predictions diverge as one evolves away from the initial condition, offering a potential means of disentangling the sources of nuclear shadowing~\cite{Armesto:2022mxy}. This difference is illustrated in Fig.~\ref{fig:BK-nPDF-proton}, where the contrast between CGC and DGLAP dynamics becomes particularly evident for the longitudinal structure function $F_L$.

Moreover, a key aspect of QCD factorization and high-energy factorization, upon which leading-twist shadowing and gluon saturation, respectively, are based, is the universality of the gluon distributions probed in different processes. This universality imposes additional constraints on both frameworks, providing a critical test of their internal consistency and predictive power.

\begin{figure}
    \centering
    \includegraphics[width=\linewidth,clip]{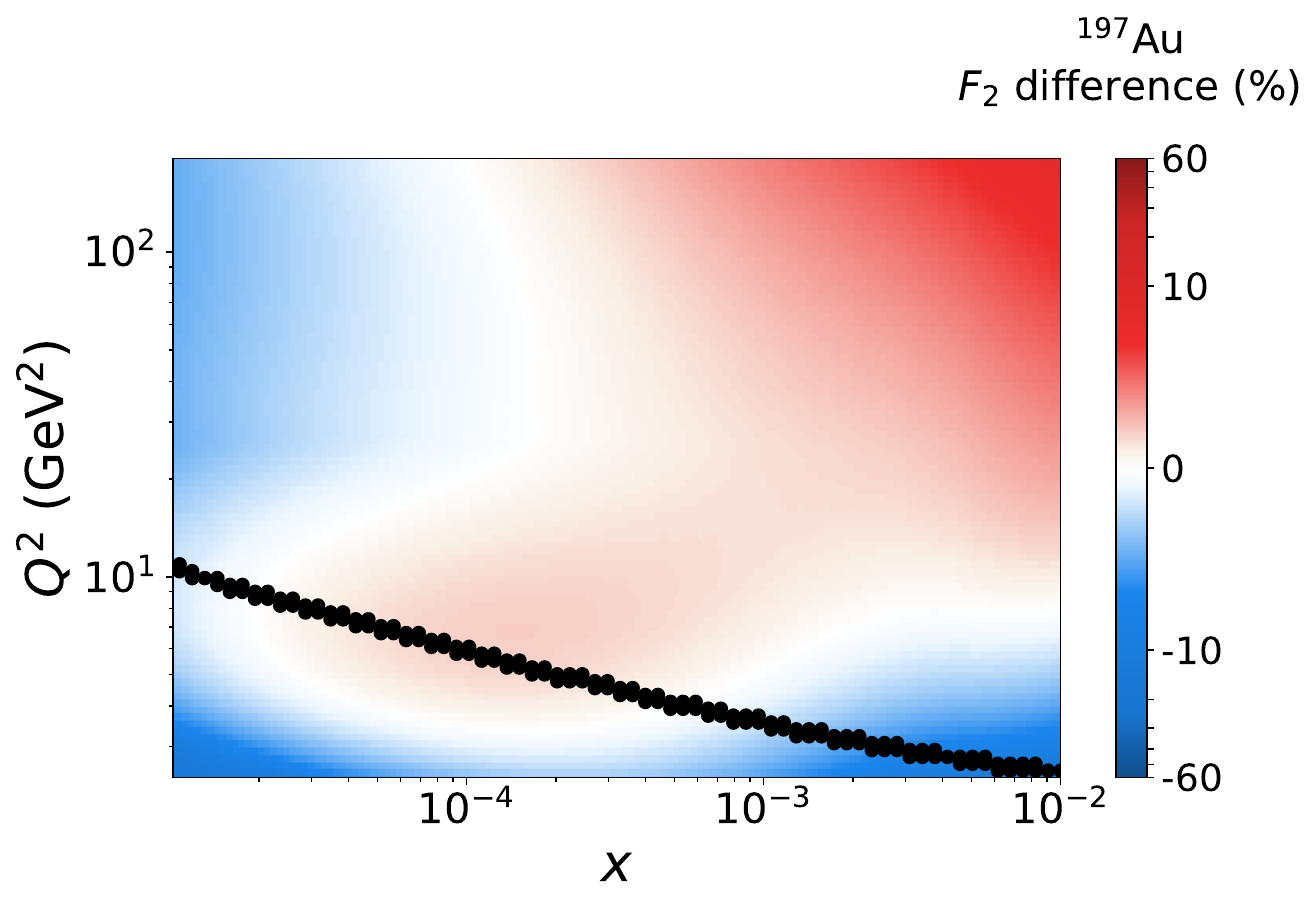} \\
    \includegraphics[width=\linewidth,clip]{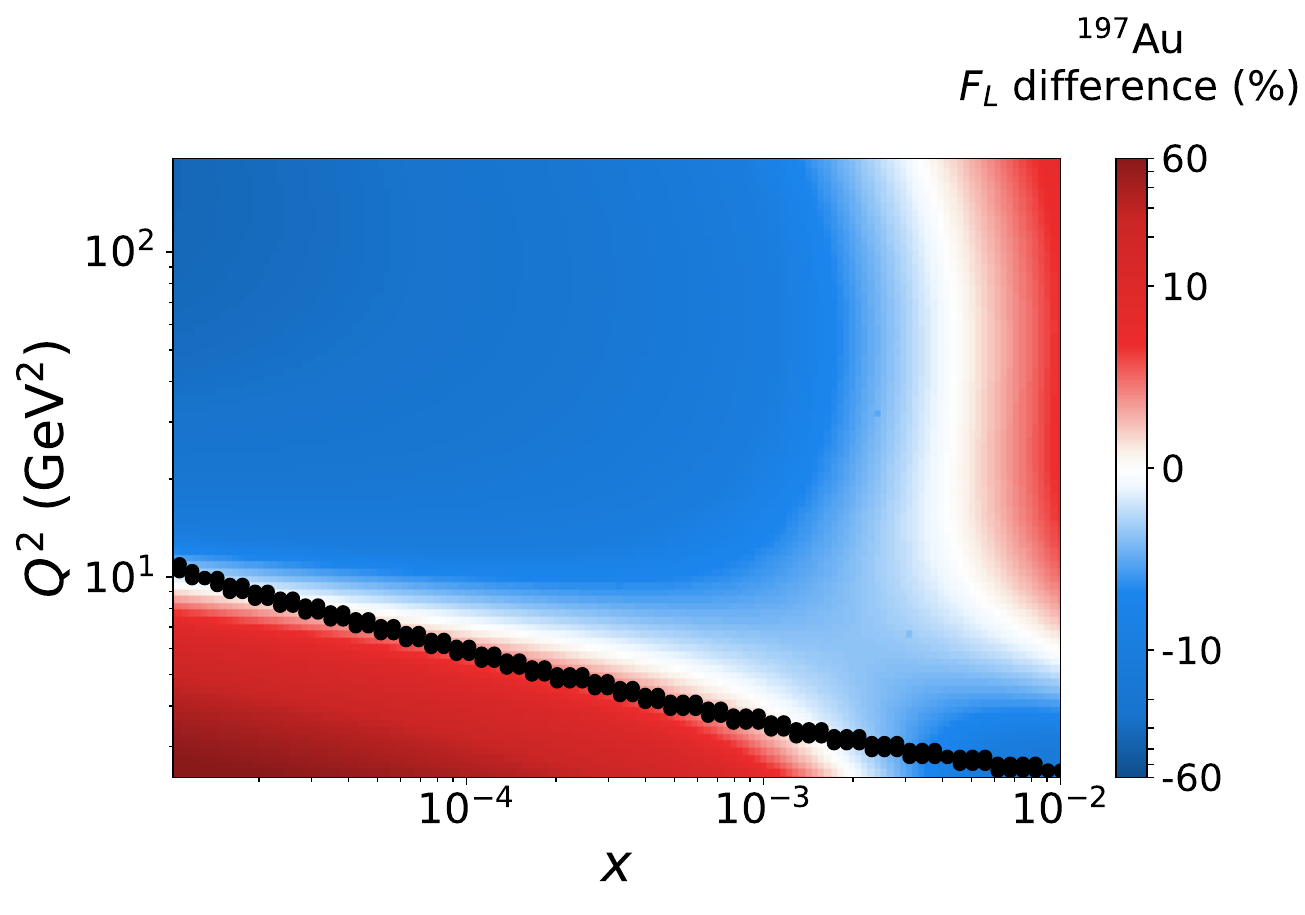}
    \caption{Same as Fig.~\ref{fig:BK-nPDF-proton} for a gold nucleus. From Ref.~\cite{Armesto:2022mxy}.}
    \label{fig:BK-nPDF-gold}
\end{figure}

\subsubsection{Single and two-particle production in $\eA$ and $\pA$}

Alongside the fully inclusive processes discussed in the previous section, saturation and shadowing also leave their imprints on more differential observables, such as single and double-inclusive particle production in DIS and forward production in proton-nucleus collisions.

\paragraph{Two-particle azimuthal correlations.} A promising production channel for observing saturation is the disappearance of the away-side peak in two-particle azimuthal correlations in $e+A$ or $p+A$ collisions at forward rapidity~\cite{Kharzeev:2004bw,Jalilian-Marian:2004vhw,Marquet:2007vb}. In the strict back-to-back limit, CGC calculations, expanded to leading power in the ratio between the transverse momentum imbalance and the relative dijet momentum, smoothly match to TMD factorization~\cite{Dominguez:2011wm}. In particular, back-to-back dihadron production at LO probes the Weizs\"{a}cker-Williams gluon distribution while direct photon-jet correlations in forward $\pA$ collisions are sensitive to the dipole gluon TMD~\cite{Dominguez:2011wm}. Beyond LO in the CGC, two-particle correlations also probe the sea quark TMD distribution~\cite{Caucal:2025xxh}. This correspondence between CGC predictions and TMD factorization provides a powerful framework for addressing the question posed at the beginning of Sec.~\ref{sec:nPDF_saturation}, whether the suppression of the away-side peak is purely nonperturbative in origin (as encoded in the nonpertubative part at small transverse momenta of the TMDs or in the nPDF initial condition of Collins-Soper-Sterman evolution~\cite{Collins:1981uk,Collins:1981uw,Collins:1984kg,Collins:2011zzd}) or whether gluon saturation itself demonstrates the perturbative mechanism at work.

There is, however, a complication to the previous general picture: when the two particles are back-to-back in the transverse plane, higher order QCD corrections, enhanced by large logarithms in the ratio of the relative transverse momentum to the transverse momentum imbalance, further decorrelate the away side peak by a vacuum-like Sudakov suppression factor (or Collins-Soper-Sterman evolution~\cite{Collins:1981uk,Collins:1981uw,Collins:1984kg,Collins:2011zzd} in the TMD language).

While this complication makes interpretation of the cause of suppression more ambiguous, it is systematically included both in CGC calculations and in standard TMD factorization with increasing accuracy, enabling the search for saturation effects ``on top of" the modifications caused by soft gluon radiation, present even in the absence of multiple soft scattering or nonlinear effects. The interplay between Sudakov resummation and gluon saturation or small $x$ evolution has been considered in Refs.~\cite{Mueller:2013wwa,Taels:2022tza,Caucal:2023fsf,Altinoluk:2024vgg}.

Thus, in discussions of two-particle azimuthal correlations, it is useful to distinguish between broadening and suppression: saturation is expected to lead to both broadening (due to multiple scattering) and suppression (due to small $x$ radiation and nonlinear evolution) whereas radiative corrections associated with large-angle soft gluon emissions (the Sudakov effect) primarily result in broadening alone.  The CGC approach provides a physical mechanism, gluon saturation, to generate the dependence of the nuclear suppression on the $p_T$ of the associated hadron. In the nPDF paradigm, in turn, the $p_T$ dependence enters through the average $x$ value probed (the smaller the $p_{T}$, the smaller the $x$), as well as through the value of the factorization scale, typically proportional the the hadron $p_{T}$.  In addition, DGLAP evolution tends to reduce the nuclear effects at larger factorization scales.

Numerical predictions based on the TMD framework supplemented by CGC models of the small $x$ TMD gluon distribution have been moderately successful at describing the two-particle correlation data in $p+A$ collisions~\cite{Albacete:2018ruq,Albacete:2010pg,Lappi:2012nh,Stasto:2011ru,Stasto:2018rci}, albeit with large theoretical uncertainties due to the naive LO approximation used. At LO, the CGC formalism predicts broadening of the away-side dihadron correlation.  However, parton shower effects and the intrinsic $k_T$ of the fragmentation control the width of the correlation~\cite{Cassar:2025vdp} and these effects are expected to reduce the broadening due to gluon saturation. However, recent studies \cite{Perepelitsa:2025qpz} based on collinear factorization, nPDFs and parton showers in the final state are also able to describe the majority of the dihadron and jet suppression effects in $\pA$ data at RHIC and the LHC. In such studies, the final state parton shower, accounting for the Sudakov resummation, is essential to reproduce the broadening of the away-side peak seen in the data.

There are additional caveats to keep in mind. First, the effect of partially coherent energy loss in cold matter may be significant. This effect is not yet included in the saturation formalism although fully coherent energy loss is accounted for in higher order $\alpha_s$ corrections within the CGC~\cite{Bergabo:2021woe}.  Shadowing models can mimic this effect by shifting the parton energy by hand. This will be further discussed in Sec.~\ref{sub:elosssmallx}.

\begin{figure}
\centering
\includegraphics[width=\linewidth,clip]{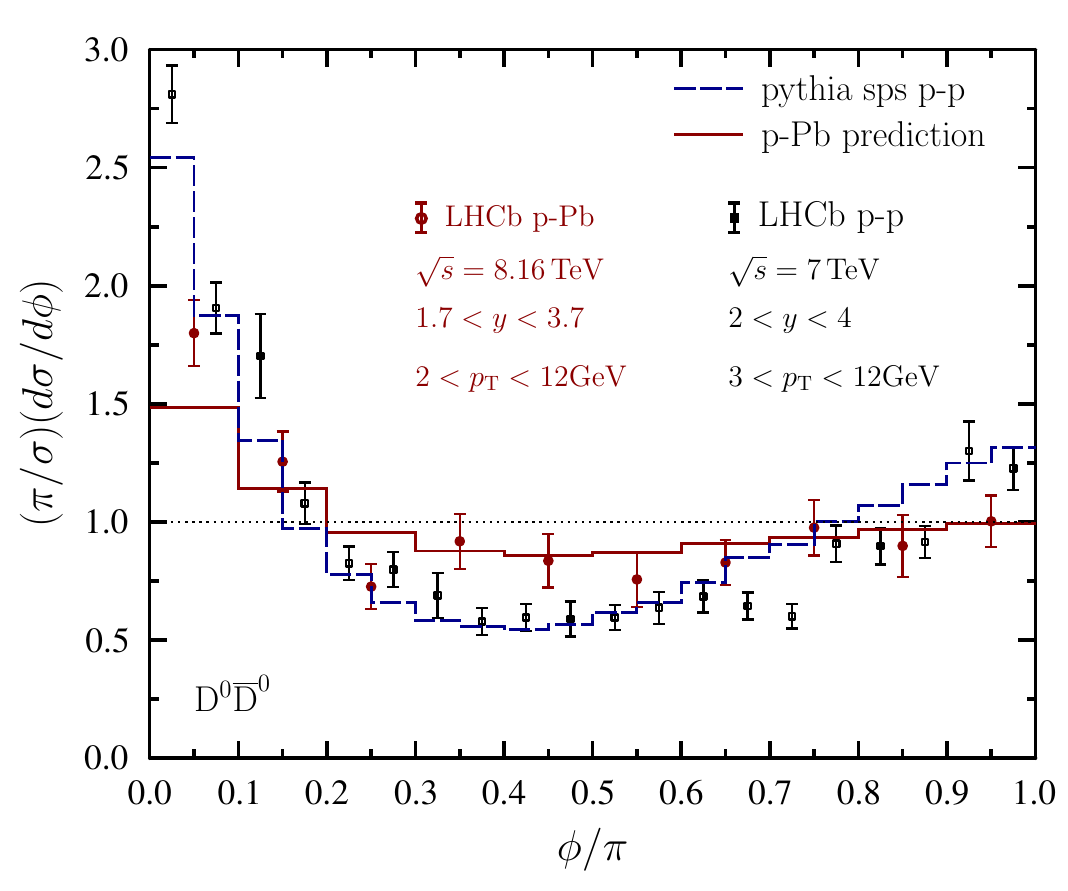}
\caption{Dependence of $D^0 \overline{D}^0$ production in $p+p$ and $p+\mathrm{Pb}$ collisions on the azimuthal separation $\phi$ between the $D^0$ and $\overline{D}^0$ measured by LHCb \cite{LHCb:2020jse}. The calculations are based on collinear factorization \cite{Helenius:2019uge}. The blue curve is a PYTHIA result with multiparton interactions turned off, thus only accounting for single-parton scattering. The red curve is a linear combination of PYTHIA single particle prediction and a flat pedestal due to double-parton scattering. From Ref.~\cite{HPcernInitialStages}.}  
\label{fig:doublecorr}
\end{figure}

Second, double-parton scattering (DPS) can also at least partially obscure the interpretation of saturation effects, particularly in $\pA$ and even more substantially in ${\rm d} +A$ collisions in which double-parton scattering is enhanced \cite{Strikman:2001gz}. (Note that this DPS enhancement was not considered in Ref.~\cite{Perepelitsa:2025qpz}.) One possibility would be to focus on $D^0 \overline{D^0}$ correlations, as recently measured by LHCb \cite{LHCb:2020jse}. The away-side peak also seems to disappear going from $p+p$ to $p+{\rm Pb}$ collisions at $p_{T} > 2$~GeV.  However, this transition can also be explained by nPDF + DPS \cite{Helenius:2019uge}, see Fig.~\ref{fig:doublecorr}. It would be interesting to investigate this observable from the CGC perspective. The LHCb measurement also differs from other correlation measurements, such as by STAR \cite{STAR:2021fgw} by not normalizing the double yield ${d N}/{d^2 p_T d^2 q_T d y_1 d y_2 }$ by the single inclusive (trigger) yield.  This is notable because shadowing or saturation can affect the single and double yields differently \cite{Perepelitsa:2025qpz}.  Direct photon-hadron/jet~\cite{Jalilian-Marian:2012wwi,Jalilian-Marian:2005qbq,Jalilian-Marian:2005tod,Kolbe:2020tlq,Ganguli:2023joy} or DIS lepton-hadron/jet~\cite{Liu:2018trl,Tong:2022zwp} correlations could prove valuable because they involve only one strongly interacting system in the final state.

Finally, fragmentation of the partons to hadrons typically results in low $p_{T}$ hadrons produced from higher $p_T$ partons.  The final state hadron momentum depends on $z$, the ratio of the longitudinal momentum of the parton carried by the hadron to the initial parton momentum.   At midrapidity at LHC energies, $z \lesssim 0.5$ \cite{dEnterria:2013sgr}, while, away from midrapidty the average $z$ is higher because of the more restricted phase space for parton radiation at higher rapidities. Thus, depending on the kinematics, the underlying partonic process may be in a kinematic region outside the saturation regime or even outside the TMD domain of validity.
In this respect, $D^0 \overline{D}^0$ correlations should be advantageous because $\langle z \rangle \sim 0.7$ in $D$ meson production (see e.g. Fig.~4 of Ref.~\cite{Helenius:2018uul}) because charm fragmentation function has a different $z$ dependence than those of light quarks. Thus hadronization does not blur the underlying parton kinematics for $D$ mesons as much as it does for pions. A promising alternative, albeit challenging from the experimental point of view and not immune to DPS, is low $p_T$ back-to-back dijet production. See the following discussion as well as the $\pA$ studies in Refs.~\cite{vanHameren:2019ysa,Al-Mashad:2022zbq,vanHameren:2023oiq}.

\begin{figure}[!t]
\centering
\includegraphics[width=\linewidth,clip]{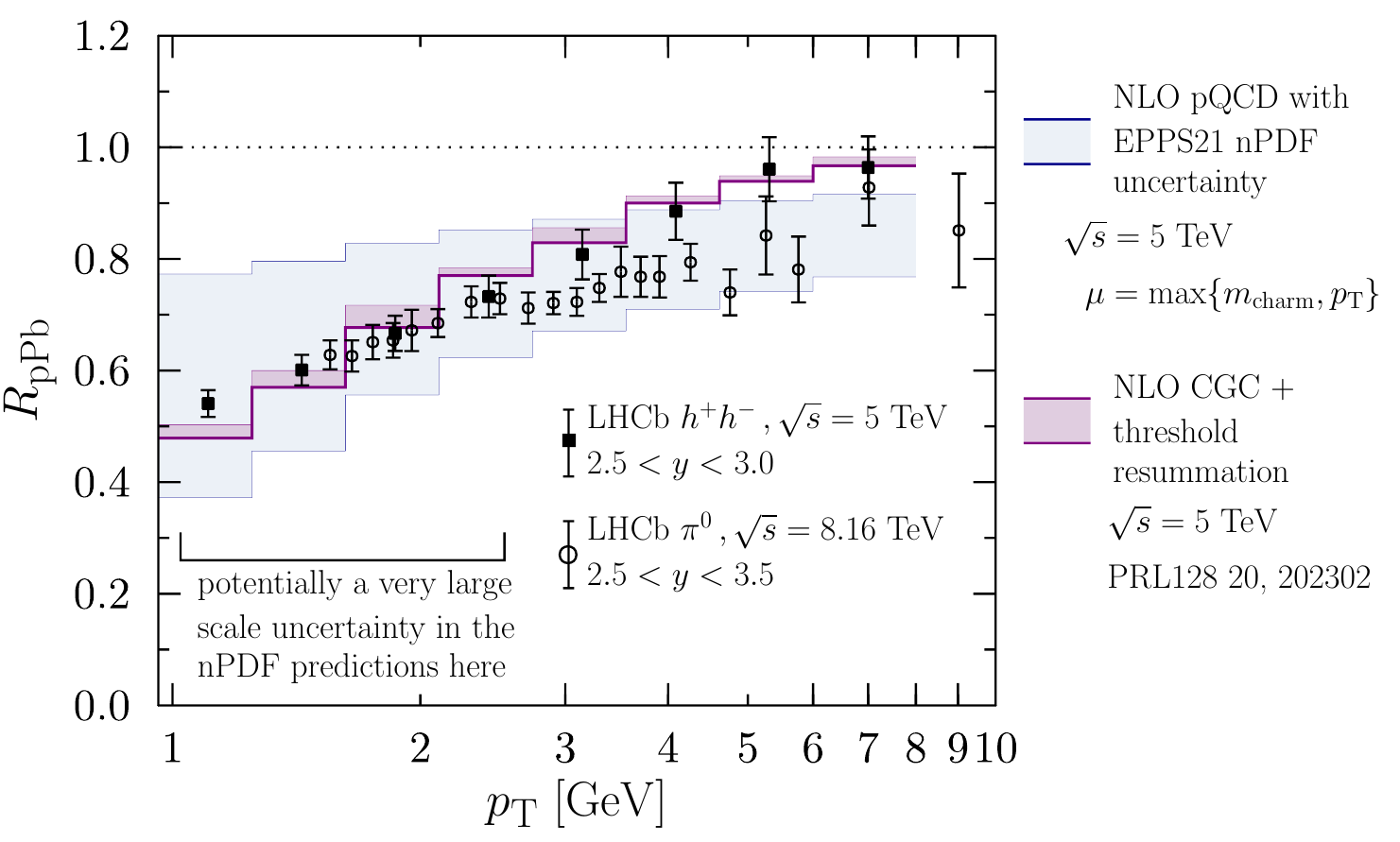}
\includegraphics[width=\linewidth,clip]{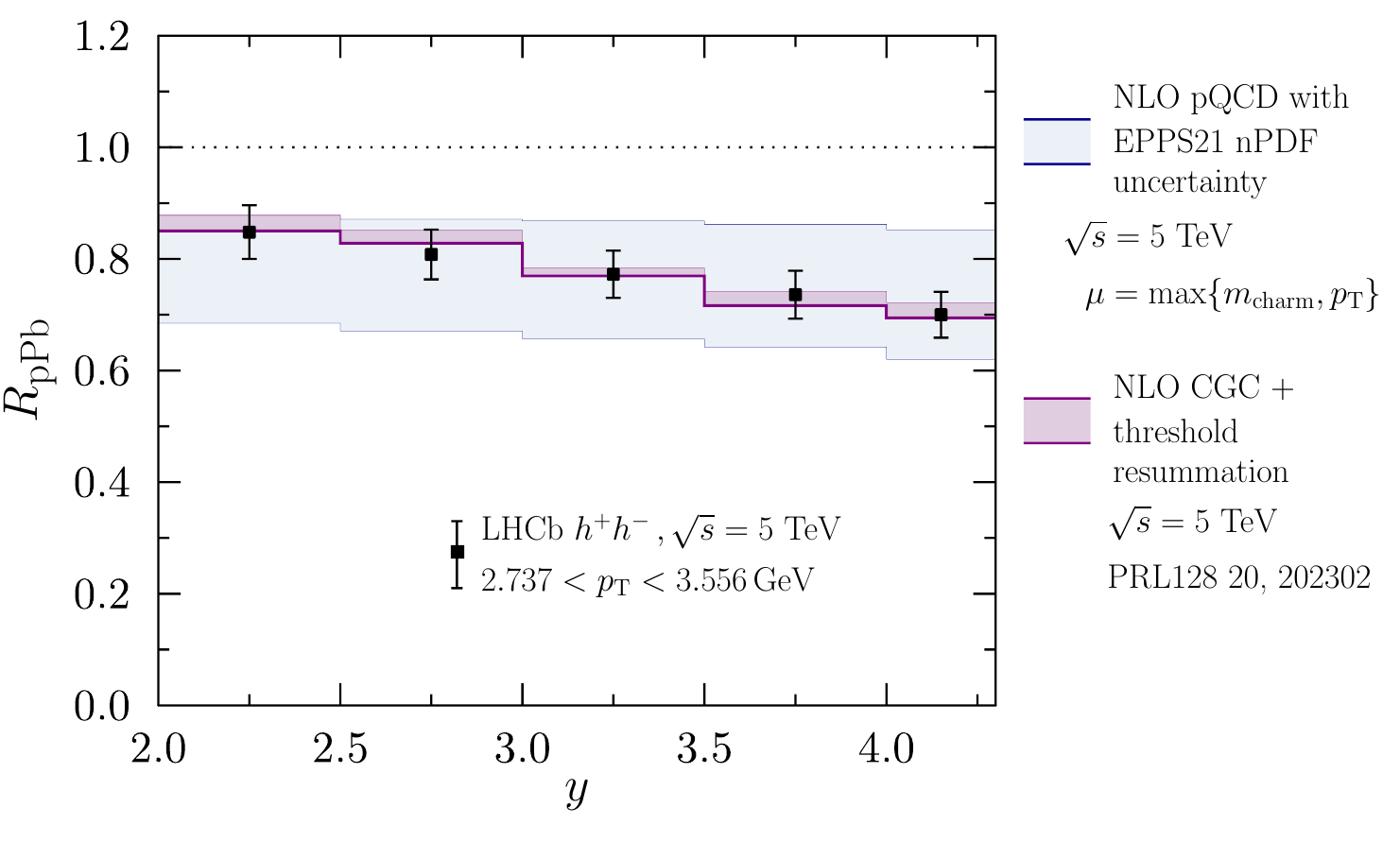}
\caption{
The nuclear modification factor of forward charged hadron production in $p +{\rm Pb}$ collisions as a function of $p_{T}$ (top) and $y$ (bottom) at $\sqrt{s_{NN}}= 5$~TeV from the EPPS21 nPDFs at factorization scale $\mu={\rm max}\{m_c, p_{T} \}$ compared to NLO CGC calculations including high energy, Sudakov and threshold resummation \cite{Shi:2021hwx}. The blue band shows the nPDF uncertainty while the red band shows the CGC uncertainties arising from varying the factorization scale, assuming the $p+\mathrm{Pb}$ and $p+p$ cross sections are fully correlated. The calculations are compared to LHCb charged hadron data \cite{LHCb:2021vww}. The corresponding $\pi^0$ data at $\sqrt{s_{NN}}= 8.16 \,{\rm TeV}$ \cite{LHCb:2022tjh} are shown for comparison as a function of $p_T$.  Note the large nPDF uncertainty at low $p_T$.
}  
\label{fig:RpA_combo}
\end{figure}

\paragraph{Single-inclusive hadron production in $\pA$}

The saturation and nPDF approaches also lead to distinct predictions of single-inclusive hadron production. Here, the features of inclusive hadron production in $p+{\rm Pb}$ collisions employing the CGC and nPDF approaches are compared.

Figure \ref{fig:RpA_combo} compares the nuclear modification factors in the CGC \cite{Shi:2021hwx} and collinear factorization approaches as functions of $p_{\rm T}$ and $y$. 
The CGC uncertainty bands are dominated by variations of the factorization scale, assuming the  $p+\mathrm{Pb}$ and $p+p$ cross sections are correlated.  
The uncertainty bands in the nPDF calculations reflect the variation of the nPDF parameters added in quadrature.  This variation is a reflection of the precision of the data used to extract the nPDFs and is the leading uncertainty above $p_{T} \gtrsim 3$~GeV. At lower $p_{T}$, the scale uncertainties become very large but are not well defined because the factorization scale becomes smaller than the nPDF parametrization scale.

In general, the CGC predictions for single-inclusive hadrons tend to have somewhat steeper slopes as a function of $p_{T}$ and $y$ compared to the nPDF predictions.  The trends associated with charged hadron production in the CGC approach are in somewhat better agreement with the LHCb data \cite{LHCb:2021vww}. However, including the baryon contribution to charged hadron production results in a poorer description below $p_T \sim 10$~GeV using standard independent fragmentation functions \cite{dEnterria:2013sgr}. The discrepancy with the data is probably most notable for the $p/\pi$ ratio which exhibits a bump at $p_{T} \sim 3$~GeV \cite{ALICE:2019hno} that cannot be described by modifying the fragmentation functions. Indeed, the pion $p_{\rm T}$ slope is flatter \cite{LHCb:2022tjh} and better in line with the nPDF calculations.

Because the CGC and nPDF calculations both depend on the fragmentation functions, forward pion measurements differential in $p_T$ and $y$ would be more beneficial than inclusive charged hadrons alone. Based on Fig.~\ref{fig:RpA_combo}, the nPDF and CGC calculations could be distinguished by systematic studies of the $p_{T}$ and $y$ slopes of the suppression factor at $p_{T} \gtrsim 3$~GeV. In $D$ meson studies this $p_{T}$ restriction could be relaxed because the finite charm quark mass allows the perturbative calculation to be continued to $p_{T} = 0$.  Existing LHCb measurements in the forward region~\cite{LHCb:2017yua,LHCb:2022dmh} could be studied from this perspective.

\paragraph{Single inclusive hadron or jet production in $e+A$}

In single inclusive hadron production (SIDIS), a final-state hadron with $p_T$ and $z$ 
could also distinguish between saturation and shadowing effects.

Future SIDIS measurements at the EIC are known~\cite{Aschenauer:2019kzf} to significantly reduce the uncertainties in the PDFs in $e+p$ collisions, particularly for radiative production of light sea quarks. Analogous measurements in $\eA$ collisions should therefore provide valuable information about sea quark relative to gluon shadowing of the nPDFs.

At small $x$, the dominant parton-level production channel for SIDIS is $\gamma^*g\to q\bar q$ where the kinematics of one of the final state quarks is integrated over. SIDIS has the advantage of probing smaller values of $x$ for $t$-channel gluons compared to dihadron production, discussed in the previous section, because the gluon $x$ is comparable to Bjorken $x$ for $p_T\ll Q$.

In fact, when $p_T \ll Q$, SIDIS also exhibits TMD factorization~\cite{Collins:1992kk,Ji:2004wu,Collins:2011zzd,Boussarie:2023izj} for the sea quarks: the CGC cross section can be expressed~\cite{Marquet:2009ca} as product of a hard factor for the $\gamma^*q\to q$ subprocess at scale $Q$ and a TMD sea quark distribution~\cite{McLerran:1998nk,Venugopalan:1999wu,Mueller:1999wm} resulting from $g \to q \overline q$ splitting in the nuclear target with the $q$ or $\overline q$ measured with final-state $p_T$.  SIDIS can thus probe saturated sea quarks in the target wavefunction with $p_T \sim Q_s\ll Q$~\cite{Marquet:2009ca}. An additional interesting regime of gluon saturation is the very forward limit~\cite{Iancu:2020jch}, where $1 - z \ll 1$, such that $(1 - z) Q^2 \sim Q_s^2$ with $Q \gg Q_s$. In this regime, the CGC predicts strong suppression in $R_{eA}(z)$ as $z\to 1$~\cite{Iancu:2020jch}.

Due to the hierarchy of scales imposed in these situations, radiative corrections are dominated by Sudakov logarithms, similar to two-particle correlations, as previously discussed. The all-order exponentiation of these logarithms will also suppress the cross section, competing with the suppression caused by saturation.  As for two-particle correlations, these corrections make the physical interpretation of the suppression more difficult. Theoretical frameworks including both suppression mechanisms, either within the TMD factorization formalism~\cite{Altinoluk:2024vgg} or by coupling the CGC to a final-state parton shower~\cite{Zheng:2014vka,Cassar:2025vdp}, are the only way to conclusively determine the role of nonlinear effects in future experimental data.

In the absence of a specific hierarchy of scales imposed on the final state, the Sudakov effects disappear. However, the interesting regime for saturation is typically semi-hard scales where $p_T \sim Q \sim Q_s$ and $z \sim 0.5$. In this regime, no approximation is applicable and the all-twist CGC cross section must be employed. While LO estimates are simple to obtain because the SIDIS cross-section involves only dipoles, NLO predictions are challenging to compute numerically due to the complexity of the formulas~\cite{Bergabo:2022zhe,Caucal:2024cdq,Bergabo:2024ivx}. A direct comparison to the leading-twist nPDF approach necessitates at least NLO precision.  More developments are needed to evaluate CGC cross sections at full NLO.

Single-inclusive \textit{jet} production~\cite{Gutierrez-Reyes:2018qez,Gutierrez-Reyes:2019msa,Gutierrez-Reyes:2019vbx,Arratia:2019vju}, which has a reduced dependence on nonperturbative effects in the final state because no fragmentation functions are required, is related to SIDIS. At low $p_T$, $p_T\ll Q$, jets can probe the sea quark distribution, much like single inclusive hadron production~\cite{Caucal:2024vbv}. Recently, jet distance measures have been developed to ensure TMD factorization beyond LO using jets~\cite{Arratia:2020ssx,Caucal:2024vbv}. However, jet measurements at low $p_T$ remain challenging~\cite{Page:2019gbf}.  (At HERA, jets have been measured down to $4-5$~GeV in the Breit frame.) In addition, the phase space for low $p_T$ jets at high $Q^2$ and small $x$ shrinks drastically in the EIC kinematics. At moderate $x$, single-inclusive jet production at low $p_T$ is also sensitive to jet energy loss in CNM~\cite{Li:2020rqj}, see the discussion in the following section. Lastly, it is also possible to study saturation effects in jet SIDIS in the target fragmentation region of DIS in the Breit frame \cite{Caucal:2025qjg}.  Furthermore, nucleon energy correlators have been proposed as a new way to study saturation dynamics in this kinematic regime \cite{Liu:2023aqb}.

\subsubsection{Diffractive vector meson production in $\gamma + A$}

\begin{figure}
\centering
\includegraphics[width=\linewidth,clip]{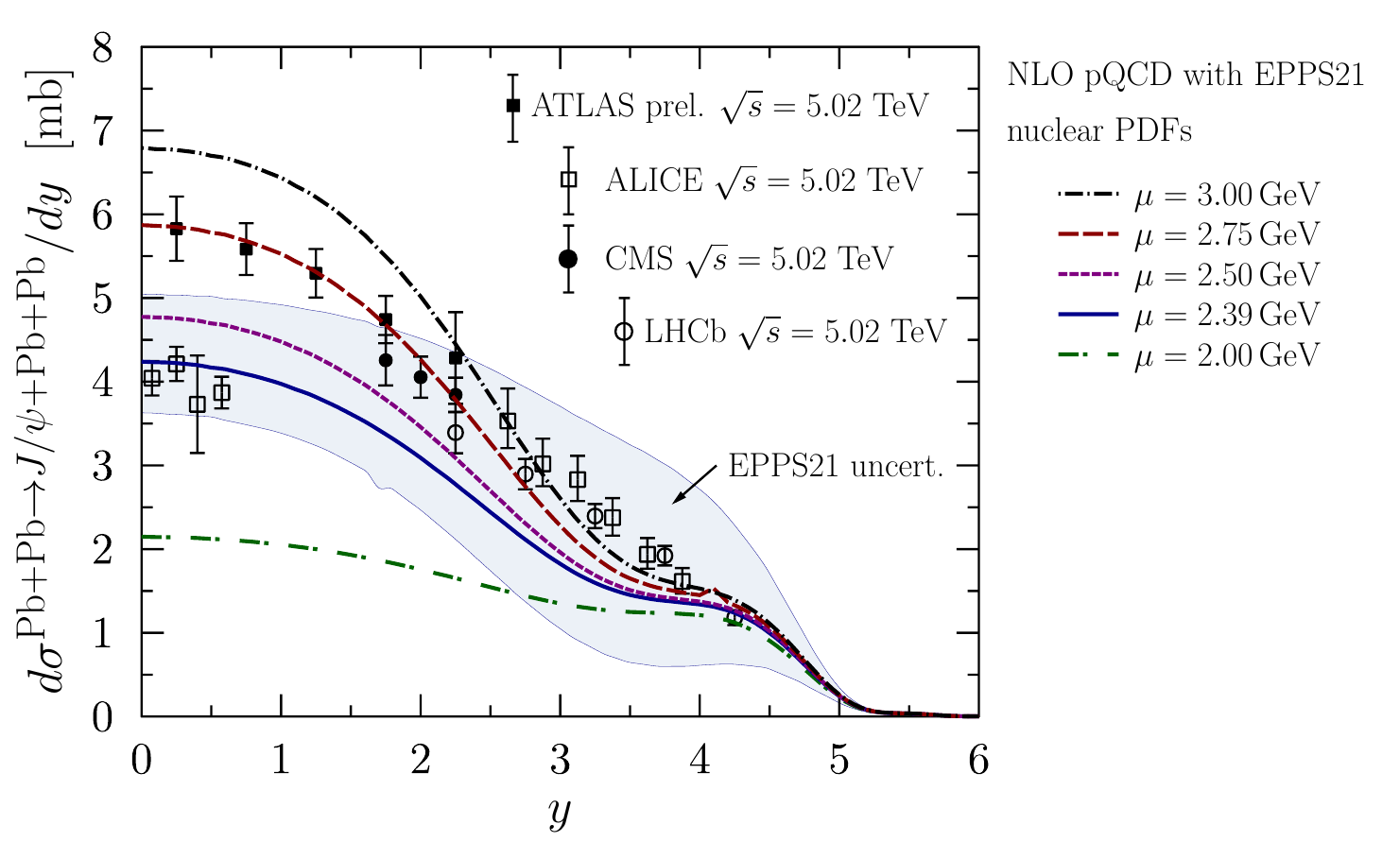}
\includegraphics[width=\linewidth,clip]{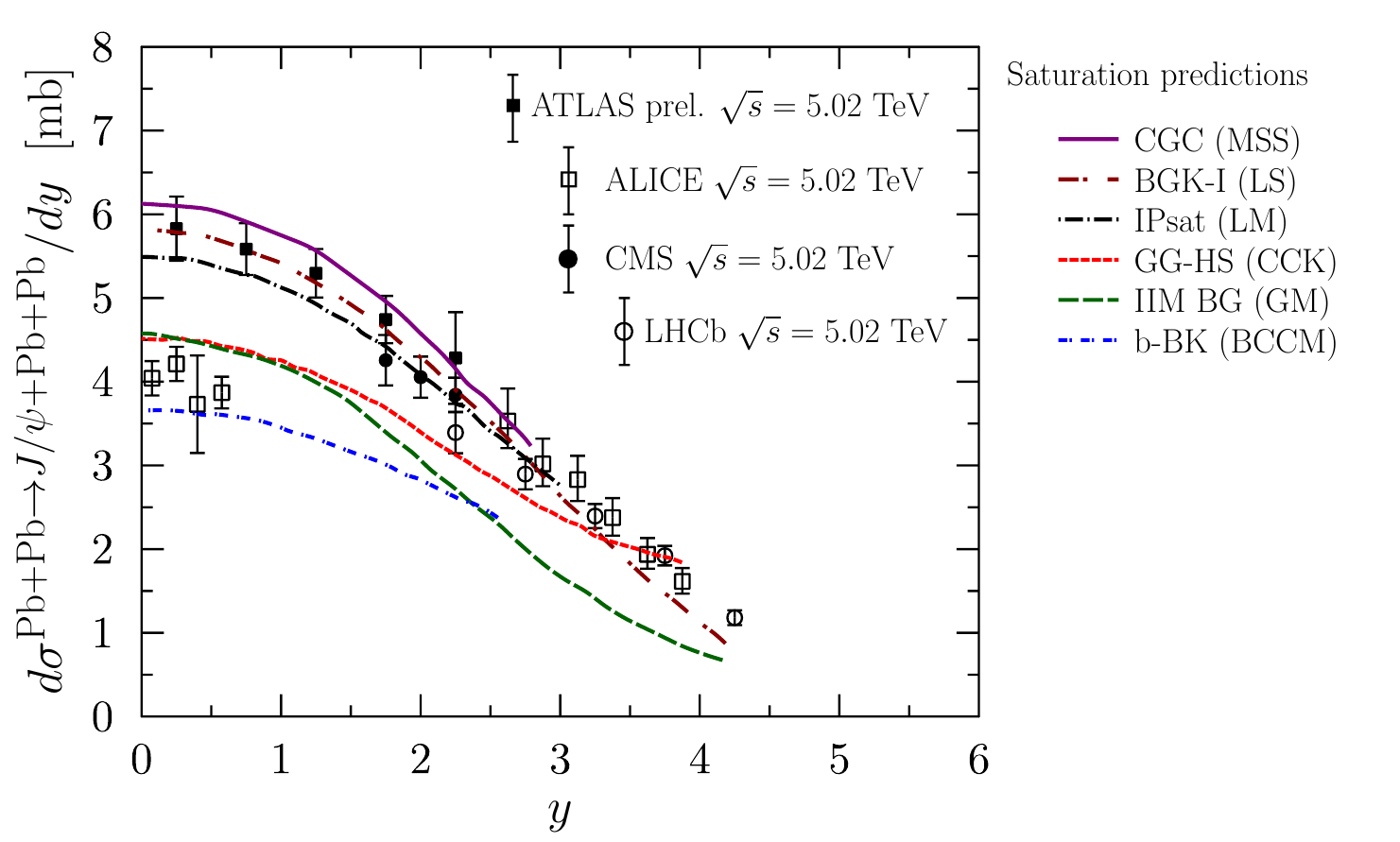}
\caption{A compilation of exclusive $J/\psi$ production measurements in Pb+Pb UPCs at $\sqrt{s_{NN}}=5.02 \, {\rm GeV}$ \cite{ALICE:2021gpt,ALICE:2019tqa,LHCb:2021bfl,CMS:2023snh,ATLAS:2025uxr}. (Top) The curves represent NLO calculations with EPPS21 nPDFs using different factorization and renormalization scales $\mu$. The light blue band is the EPPS21 uncertainty at $\mu=2.39$~GeV. The NLO calculations are from Refs.~\cite{Eskola:2022vpi,Eskola:2022vaf}. (Bottom) The calculations, taken from Ref.~\cite{Schenke:2024gnj}, represent different realizations of saturation-based calculations, depending on their description of the dipole scattering with the nucleus, are based on Refs.~\cite{Mantysaari:2022sux,Bendova:2020hbb,Luszczak:2019vdc,Lappi:2013am,Cepila:2017nef,SampaiodosSantos:2014puz}.}
\label{fig:Jpsiupc}
\end{figure}

Diffractive vector meson production in $\gamma + A$ interactions serves as a complementary and particularly sensitive probe of gluon structure at small $x$ \cite{Mantysaari:2020axf} relative to the observables discussed earlier. Unlike semi-inclusive observables, diffraction demands a color neutral exchange, which at LO is achieved by at least two-gluon exchange, making it more sensitive to saturation and shadowing effects. There are two kinds of reactions.  In the first, coherent interactions, the nucleus remains intact and the average gluon distribution is probed. In the second, incoherent interactions are characterized by nuclear breakup and are sensitive to event-by-event fluctuations of the gluon densities  \cite{Good:1960ba,Caldwell:2009ke,Mantysaari:2016jaz, Mantysaari:2020axf}.  Diffractive $J/\psi$ production is particularly compelling because the charm mass provides a hard scale, allowing perturbative treatment, but the scale is also comparable to the saturation scale in $\gamma + A$ collisions.

In the saturation formalism, this process is studied within the dipole picture, where, at LO, the photon splits into a $q\overline q$ pair \cite{Kowalski:2006hc} (and $q\overline{q}g$ at NLO \cite{Mantysaari:2022kdm}) which interacts with the nucleus through multiple coherent scatterings and then recombines into the vector meson. The saturation physics is encoded in the $q\overline{q}$ interaction with the nucleus, characterized by the dipole amplitude.  The energy dependence of the dipole is governed by the BFKL equation in the linear regime and the BK/JIMWLK equation in the nonlinear regime.

The cross section for diffractive vector meson production in $\gamma + A$ interactions has been studied as a function of the center of mass energy and rapidity of the produced vector meson in ultraperipheral collisions of heavy-ions at RHIC and the LHC. In the linear regime, governed by BFKL evolution, the cross section is predicted to grow as a power of the energy while saturation effects are expected to suppress this growth. The suppression is expected to be strong when the typical transverse size of the vector meson is larger than or comparable to the inverse of the saturation scale \cite{Mantysaari:2017slo}.  At the energies and rapidities reached in these colliders, the growth of the $J/\psi$ cross section in $\gamma + p$ collisions in UPCs is consistent with a power law because the typical transverse size of $J/\psi$ is smaller than the inverse of the saturation scale in the proton \cite{Penttala:2024hvp,Mantysaari:2024zxq}. On the other hand, strong nuclear suppression is seen in $J/\psi$ production in $\gamma + {\rm Pb}$ collisions at the LHC.  The suppression is stronger than predicted by the current saturation-based approaches, see the lower panel of Fig.~\ref{fig:Jpsiupc}. These saturation-based predictions differ in their description of the dipole amplitude but are usually constrained by fits to $\gamma + p$ data. (The acronyms here refer to the models in the legend of the bottom panel of Fig.~\ref{fig:Jpsiupc}.)  The CGC (MSS) calculation starts with initial conditions taken from an impact-parameter dependent MV model with the rapidity and energy evolution following the JIMWLK equation \cite{Mantysaari:2022sux}.  The b-BK (BCCM) model employs an impact-parameter dependent dipole amplitude that evolves according to impact-parameter dependent BK evolution \cite{Bendova:2020hbb}. The BGK-I (LM) \cite{Luszczak:2019vdc}, IPsat (LM)  \cite{Lappi:2013am}, GG-HS (CCK) \cite{Cepila:2017nef}, and IIM BG (GM) \cite{SampaiodosSantos:2014puz} calculations implement different phenomenological models of the impact parameter, rapidity and dipole size dependence of the dipole amplitude. For more details about the model implementation, see Ref.~\cite{Schenke:2024gnj}.

Exclusive $J/\psi$ production in Pb+Pb collisions probes the nPDFs predominantly at $10^{-5} \lesssim x \lesssim 10^{-2}$ and scales similar to those in low $p_{T}$ open charm production \cite{Eskola:2022vpi,Eskola:2022vaf}. The coefficient functions are known up to NLO in strong coupling \cite{Ivanov:2004vd} when the $J/\psi$ is treated nonrelativistically. Extraction of the nPDFs, including testing dynamical modeling as in leading-twist shadowing \cite{Frankfurt:2011cs,Guzey:2024gff}, based on UPC data is complicated by several factors. Convergence of the perturbative expansion has proven to be quite difficult due to coincidental cancellations between the LO and NLO gluon contributions. As a result, the calculated cross sections are rather sensitive to the factorization and renormalization scales, as shown in the upper panel of Fig.~\ref{fig:Jpsiupc}. Better stability seems to be achieved by including resummation of BFKL-type terms \cite{Flett:2024htj}. Other complications arise because generalized nuclear parton distributions enter the calculations. These GPDs can be approximated by nPDFs but the associated theoretical uncertainty is difficult to reliably assess and at least induces some degree of model dependence. Other theoretical uncertainties include the role of nonrelativistic corrections to the treatment of the $J/\psi$ wave function as well as theoretical modeling of the exclusivity condition. All in all, an accurate description of exclusive $J/\psi$ production in Pb+Pb UPCs is still a distant goal which hinders the determination of the dynamical effects of saturation. Currently there also appears to be tensions between the ALICE \cite{ALICE:2021gpt} and preliminary ATLAS \cite{ATLAS:2025uxr} data, see the difference in the data at midrapidity in Fig.~\ref{fig:Jpsiupc}.

Looking forward, studying the energy dependence in different colliding systems, such as $\gamma + {\rm O}$, as well as the production of lighter and heavier vector mesons such as $\phi$ and $\Upsilon$ could help to further discriminate among different approaches to the observed nuclear suppression and thus distinguish between saturation and nonperturbative shadowing. Moreover, the production ratios of excited vector meson states to the ground states, particularly $\psi$(2S) to $J/\psi$, offers additional discriminatory power. In Ref.~\cite{Peredo:2023oym} it was argued that the distinctive node in the $\psi$(2S) wavefunction enhances the sensitivity to nonlinear effects, providing a promising ameans to discriminate between linear shadowing and nonlinear saturation dynamics. Lastly, further measurements of incoherent reactions, where the nucleus breaks up, and which are sensitive to fluctuations, could also shed some light on small-$x$ dynamics \cite{ALICE:2025cuw}.   
\section{Radiative parton energy loss}
\label{sec:eloss}

Reactions with nuclei are essential for the study of QCD. Cold nuclear matter effects can be investigated in $\eA$~\cite{AbdulKhalek:2021gbh} and $\pA$ collisions~\cite{Albacete:2013ei,Albacete:2017qng}.  The modification of hadron production cross sections at low and intermediate $p_T$, known as the Cronin effect~\cite{Cronin:1974zm}, was established as early as the mid 1970s. In the 1990s, fixed-target measurements of Drell-Yan and charmonia at Fermilab (see Sec.~\ref{sec:data}) found a substantial suppression of the cross sections in nuclei at large $x_F$. While initially attributed only to nPDF effects, it was noted early on that attenuation of the production rates of such states is also compatible with CNM energy loss~\cite{Gavin:1991qk}.      

In the late 1990s and early 2000s important developments in many-body perturbative QCD produced theoretical frameworks to evaluate non-Abelian parton energy loss due to soft gluon emission~\cite{Gyulassy:1993hr,Zakharov:1996cm,Baier:1996sk,Gyulassy:2000fs,Wang:2001ifa,Arnold:2002ja,Djordjevic:2003zk,Vitev:2007ve}. Subsequently, the energy loss mechanism has been extensively employed to study Drell-Yan \cite{Arleo:2002ph, Neufeld:2010dz, Xing:2011fb, Arleo:2018zjw} and $J/\psi$  \cite{Gavin:1991qk, Arleo:2012hn,Arleo:2012rs,Arleo:2013zua}  suppression at large $x_F$ in minimum bias $\pA$ collisions as well as in jet studies \cite{Kang:2015mta} in central $\pA$ collisions at very high energies. 

In the remainder of this section, we describe the various energy loss regimes and observables that are sensitive to these regimes. We further touch upon recent progress on understanding in-medium radiation beyond the soft gluon emission limit and its connection to other perturbative approaches such as saturation.

\subsection{Energy loss regimes}

The dynamics of high-energy processes in dense QCD are governed by multiple scatterings of energetic partons as they traverse QCD matter. In addition to transferring $p_T$ via successive gluon exchanges with the target, these interactions can also induce soft gluon radiation, leading to parton energy loss. During the quantum-mechanical formation time of the radiated gluon, such multiple scatterings can act coherently as a single effective scattering center. This coherence gives rise to destructive interference effects, resulting in suppression of the radiative spectrum, a phenomenon known as the Landau–Pomeranchuk–Migdal (LPM) effect~\cite{Gyulassy:2000fs,Zakharov:2000iz}.
 
These processes are widely studied as one of the dominant mechanisms for high $p_T$ hadron and jet suppression in heavy-ion collisions. However, they may also play a significant role in forward particle production of quarkonium~\cite{Arleo:2012hn,Arleo:2012rs,Arleo:2013zua}, light hadrons~\cite{Arleo:2020hat,Arleo:2020eia}, and heavy flavor mesons~\cite{Arleo:2021bpv} in $\pA$ collisions.

The medium-induced gluon radiation spectrum, the difference between gluon radiation in protons and nuclei,  exhibits different scaling properties depending on the typical gluon formation time $t_f \sim \omega /k_T^2$, where $\omega$ and $k_T$ are the energy and the transverse momentum of the radiated gluon, respectively. 
This gluon formation time needs to be compared to the two other length scales appearing in the process, the parton mean free path in the medium, $\lambda$, and the length of the medium traversed, $L$.  In practice, qualitative analytic behavior can be derived in the limits of very large parton energy and large numbers of scatterings.  Numerical corrections can be large for lower energies and a smaller number of scatterings~\cite{Gyulassy:2000fs,Gyulassy:2000er}. 

It is useful to recall the different regimes identified in the medium-induced gluon spectrum depending on the gluon formation time $t_f$~\cite{Baier:1996kr}:
\begin{description}[left=0.1cm]
\item[Incoherent or Bethe-Heitler regime] 
In this regime, $t_f \lesssim \lambda$, each scattering center acts as an independent radiation source. This regime is important for very soft gluons, $\omega \ll \mu^2 \lambda = \hat{q} \lambda^2$, where $\hat{q}= 2 \mu^2/\lambda$ is the medium transport coefficient and $\mu$ is the inverse of the interaction range \cite{Baier:1996kr, Zakharov:1997uu};

\item[Partially coherent or LPM regime]  Here $\lambda \ll t_f \lesssim L$, and a group of scattering centers with $t_f / \lambda \gg 1$ acts as a single radiator, leading to relative suppression of the gluon radiation intensity with respect to the intensity in the incoherent regime. This  condition leads to $\mu^2 \lambda \ll \omega \lesssim \hat{q} L^2$. The average LPM energy loss scales parametrically as $\Delta E_{_{\textnormal{LPM}}} \propto  \hat{q} L^2 \ln(E/\mu^2 L)$, with a logarithmic dependence on the parton energy;

\item[Fully coherent regime]  Here $t_f \gg L$ and all scattering centers in the medium act as a coherent radiation source. In this large gluon energy regime, $\omega \gg \hat{q}L^2$, the average energy loss $\Delta E_{_{\textnormal{FCEL}}} \propto (\hat{q}\,L)^{1/2}\,(E/M_T)$ becomes proportional to the parton energy $E$~\cite{Arleo:2010rb,Arleo:2012rs}, overwhelming the LPM energy loss, $\Delta E_{_{\textnormal{FCEL}}} \gg \Delta E_{_{\textnormal{LPM}}}$. The phenomenology of FCEL has been investigated in quarkonium~\cite{Arleo:2012hn,Arleo:2012rs,Arleo:2013zua}, light hadron~\cite{Arleo:2020hat,Arleo:2020eia}, and heavy flavor meson~\cite{Arleo:2021bpv} production in $\pA$ collisions. The effects of FCEL, which explicitly break QCD factorization, are sizable and therefore cannot be neglected.  Given this, including measurements of hadron production in $\pA$ collisions in global analyses of nPDFs~\cite{Kusina:2017gkz,Eskola:2019bgf,Kusina:2020dki} without proper implementation of FCEL might lead to biased nPDF extractions.

\item[Bertsch-Gunion (BG)-like regime]  Initial-state interactions can partially cancel radiation effects, especially at low frequencies. In this regime, This medium-induced radiation \cite{Vitev:2007ve} exhibits Bertsch-Gunion~\cite{Gyulassy:1999ig, Wiedemann:1999fq} scaling properties, $\omega (dI/d\omega)  \sim L$, without additional power suppressions. The type of energy loss is manifested in Drell-Yan, hadron, and jet suppression at large $x_F$~\cite{Neufeld:2010dz,Kang:2017frl,Li:2018xuv}.  
\end{description}

\begin{figure}[htbp]
\centering
\includegraphics[width=\linewidth,clip]{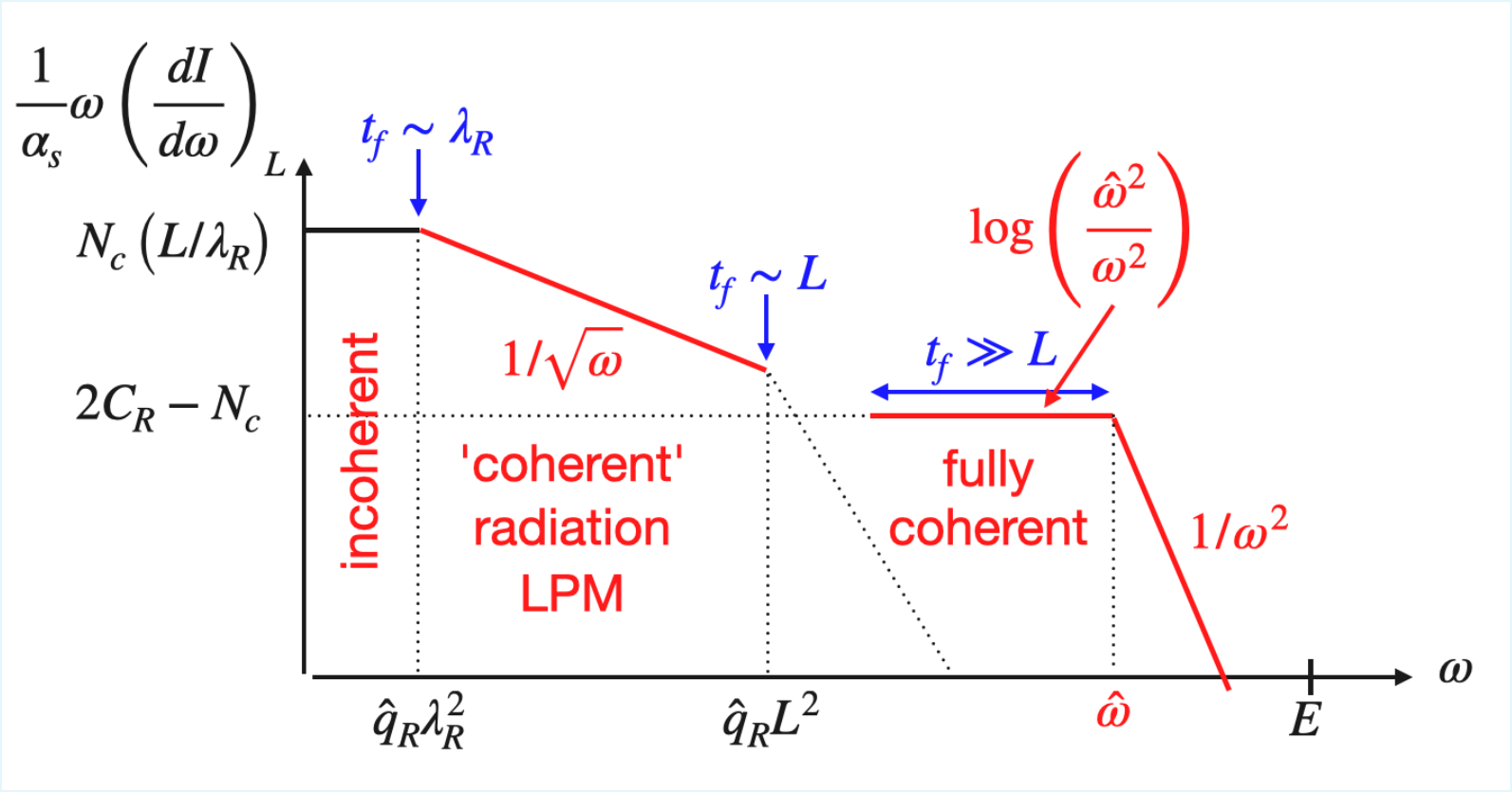}
\caption{Schematic log-log plot showing the parametric dependence of the $\alpha_s$-normalized induced radiation spectrum $(\omega/\alpha_s) (dI/d\omega)$ of parton production in the color representation $R$ at small angles in the regimes of gluons formation times described here. Taken from Ref.~\cite{Arleo:2016cdm}.}
   \label{fig:radiationspectrum}
\end{figure}

\begin{figure}[th]
\centering
\includegraphics[width=\linewidth,clip]{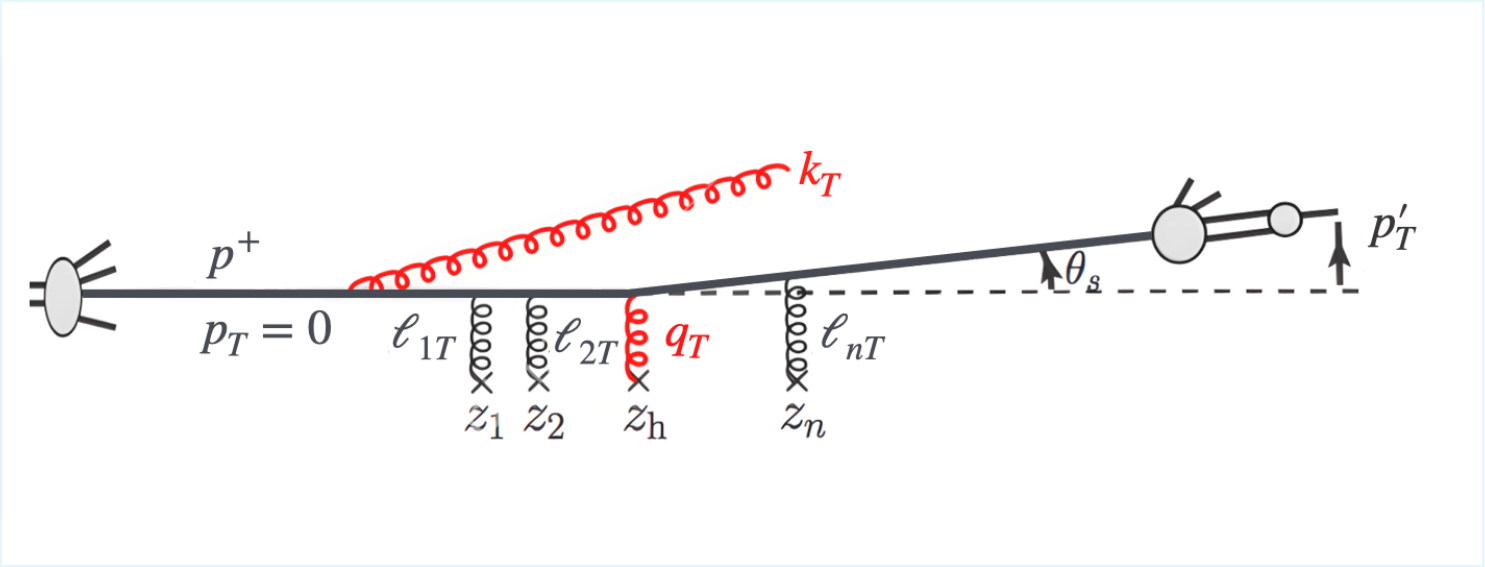}
\caption{Induced gluon radiation from a fast parton experiencing multiple scatterings in a dense nucleus. Taken from~\cite{Arleo:2016cdm}.} 
\label{fig:eloss_cartoon}
\end{figure}

Figure~\ref{fig:radiationspectrum} schematically summarizes the parametric dependence of the medium-induced radiation spectrum. 
It can be shown that if a parton suddenly decelerated (accelerated), leading to initial-state, IS, or (final-state, FS) energy loss, the medium-induced gluon spectrum decreases as $\omega (dI/d\omega) \sim (1/\omega)$ at $\omega \gg \hat{q}\,L^2$, leading to $\Delta E_{_{\textnormal{IS}}} = \Delta E_{_{\textnormal{FS}}} = \Delta E_{_{\textnormal{LPM}}}$. On the contrary, whenever the hard process appears to be the result of small-angle deflection of a high energy parton, as is typically the case for hadron production in $\pA$ collisions when both the initial-state and the final-state partons carry a color charge (see Fig.~\ref{fig:eloss_cartoon}), the associated medium-induced gluon radiation arises from the interference of gluon radiation before and after the hard scattering. In that case, the induced spectrum is dominated by the fully coherent radiation regime, $\tf \gg L$, leading to $\Delta E = \Delta E_{_{\textnormal{FCEL}}}$.

Disentangling the different regimes of radiative parton energy loss is key to understanding its effects. 
At high collision energies, typically at the LHC, the effect of LPM energy loss in nuclear matter should be negligible because the incoming parton energy $E$ is extremely large in the nuclear rest frame, leading to negligible fractional energy loss, $\Delta E_{_{\rm LPM}} / E \sim \ln E/E \ll 1$. Thus the LPM regime would best be probed in \textit{low energy} nuclear collisions.

Conversely, FCEL affects all hard processes in $\pA$ collisions, even at the LHC, since $\Delta E_{_{\textnormal{FCEL}}} / E$ is finite in the high-energy limit. Hard processes with color neutral final states (Drell-Yan or weak vector boson production) or collisions with a color neutral projectile, as in $\eA$ scattering or exclusive processes in ultraperipheral heavy-ion collisions  are exceptions. 
Table~\ref{tab:energy-losses} summarizes the different energy loss processes, along with their respective parametric dependencies, and gives examples of observables and colliding systems where these processes can ideally be studied.

\begin{table*}
    \centering
    \resizebox{\linewidth}{!}{%
    \begin{tabular}{p{2.5cm}ccccc}
\hline
\hline
         &  $\Delta E$ & Observables & Systems & Facilities & References\\
\hline
Initial state LPM & $\hat{q}\,L^2\,\ln E$ & Drell-Yan & $\hA$ & E906, COMPASS & \cite{Arleo:2018zjw} \\
Final state LPM    & $\hat{q}\,L^2\,\ln E$ & $h$, jets & $\eA$ & CLAS, HERMES, EIC & \cite{Arleo:2003jz} \\
FCEL    & $(\hat{q}\,L)^{1/2}\,(E/M_T)$ & $h$, jets & $\hA$ & SPS, FNAL, RHIC, LHC & \cite{Arleo:2012hn,Arleo:2012rs,Arleo:2013zua} \\
BG-like  &  $f(\hat{q}\,L) L\,  E$  & high $x_F$  & $\hA$ &   FNAL, RHIC, LHC  &  \cite{Gavin:1991qk,Neufeld:2010dz,Kang:2015mta}\\
        &  & Drell-Yan, $h$, jets &  &    &  \\
\hline
\hline
    \end{tabular}
    } 
    \caption{Energy loss processes in collisions on nuclear targets, their parametric dependencies, and the systems and facilities in which they might be probed.  (No example of Bethe-Heitler observables is given.)}
    \label{tab:energy-losses}
\end{table*}

At the end  of this section, we outline several observables and their potential to more precisely probe and constrain medium-induced gluon radiation and energy loss in the LPM, FCEL or BG-like regimes, and separate their effects from nPDFs.

\subsection{Cold nuclear matter transport coefficient $\hat{q}$}\label{sec:qhat}

The strength of induced gluon radiation in cold nuclear matter is governed by the transport coefficient $\hat{q}$ within the Baier–Dokshitzer–Mueller–Peigné–Schiff and Zakharov (BDMPS–Z) formalism\footnote{Note that in Eq.~\eqref{eq:transportcoefficient}, $\Tilde{\mu}^2$ already accounts for the  typical transverse momentum squared.} 
~\cite{Baier:1996kr,Baier:1996sk,Zakharov:1996fv,Zakharov:1997uu}, which quantifies the scattering power of the nuclear medium. It is related to the gluon distribution $xg(x, Q^{2})$ inside each nucleon,
\begin{equation}\label{eq:transportcoefficient}
\hat{q} \equiv \frac{\Tilde{\mu}^{2}}{\lambda}
= \frac{4\pi^{2}\,\alpha_{s}\,C_R}{N_{c}^{2}-1}\;\rho\,xg\bigl(x,Q^{2}\bigr)\,,
\end{equation}
where $Q^{2} \sim \Delta p_{T}^{2}$ denotes the characteristic transverse momentum scale, $C_R$ is the parton color factor ($C_F = (N_{c}^{2} - 1)/2N_{c}$ for a quark and $C_A = N_{c}$ for a gluon), and $\rho$ is the nuclear density. 

\begin{figure}[!h]
    \centering
    \includegraphics[width=\linewidth,clip]{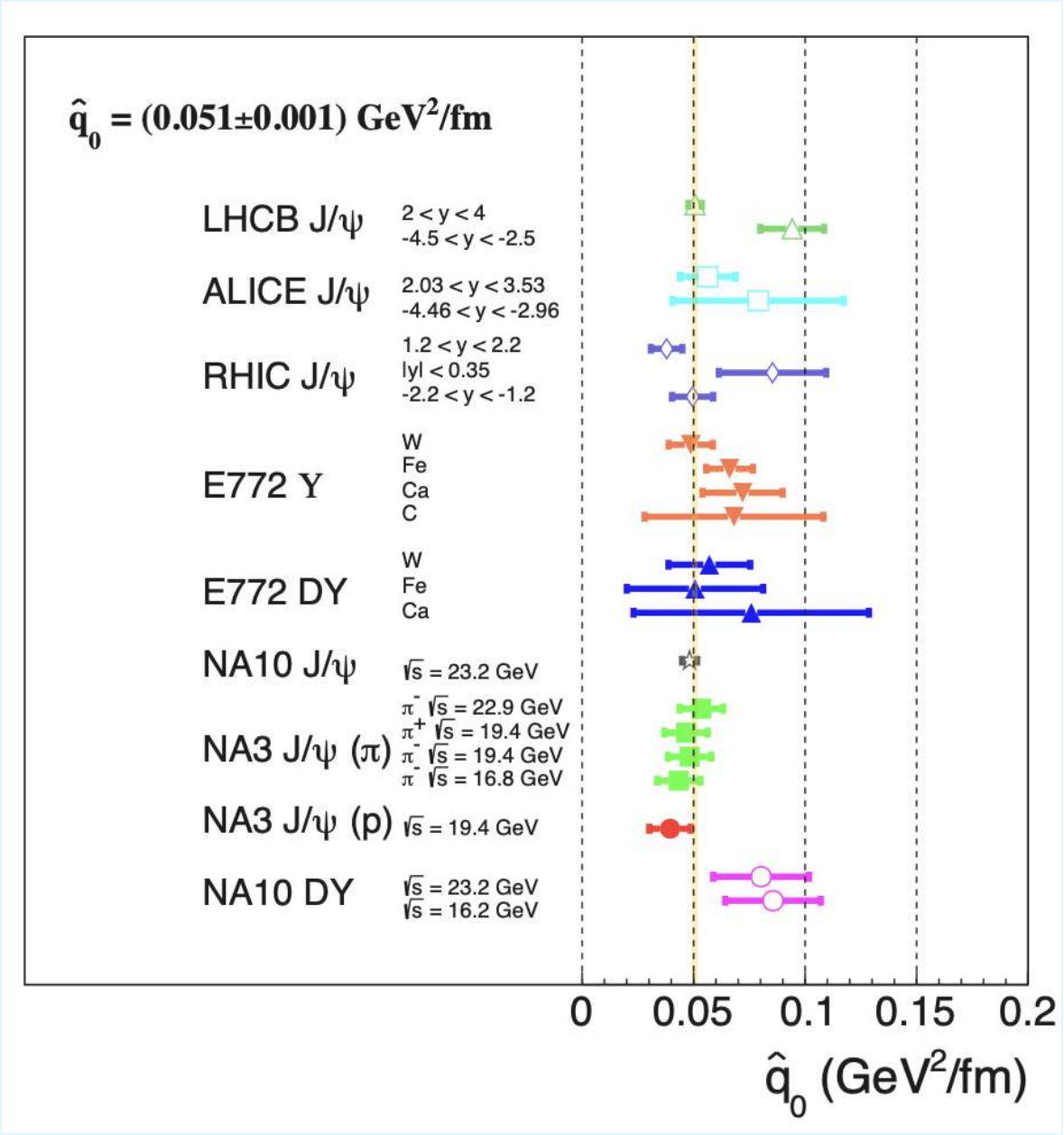}
    \caption{Extraction of the transport coefficient using $p_T$ broadening data, assuming the $c\overline c$ pair is produced in a color-octet state, as discussed in Ref.~\cite{Arleo:2020rbm}.}
    \label{fig:transport_broadening}
\end{figure}

Other scales, such as the inverse range of the interaction $\mu$ or the mean free path $\lambda$  in different formalisms such as Gyulassy-Levai-Vitev (GLV)~\cite{Gyulassy:1999zd,Gyulassy:2000fs,Gyulassy:2000er}, can be related to $\hat{q}$, but may also appear individually in logarithmic or sub-leading terms. In this case the leading power dependence of the transport coefficient is  $\hat{q} = 2{\mu^{2}}/{\lambda}$, where the dimensionality of the random walk is included explicitly. Regardless of the definition, a fit to HERMES hadron attenuation data using generalizations of the GLV formalism yields $\hat{q} = 0.05$~GeV$^2$/fm directly comparable to the extractions shown in Fig.~\ref{fig:transport_broadening}.  

The value of  $\hat{q}$ can be determined from different processes, such as fully coherent energy loss, initial-state and final-state energy loss, and transverse momentum broadening. Uncertainties in its determination arise from the type of energy loss formalism used and how it is implemented phenomenologically as well as the treatment of the nuclear geometry. Quite remarkably, a number studies find roughly the same value of $\hat{q}$ within uncertainties. FCEL studies of $J/\psi$ suppression in $\pA$ collisions~\cite{Arleo:2012rs} find $0.05 \lesssim \hat{q}_0 \lesssim 0.09\,\text{GeV}^2/\text{fm}$ where$\hat{q}_0 \equiv \hat{q}(x=10^{-2})$, depending on the nPDF effects included. Using these values in the calculation of initial-state quark energy loss leads to a satisfactory description of the preliminary E906 Drell-Yan $\pA$ data~\cite{Arleo:2018zjw}.  The value of $\hat{q}_0$ can also be determined from studies of the transverse momentum broadening in particle production in nuclear collisions, $\Delta p_T^2$, see e.g.\ Ref.~\cite{Kopeliovich:2010aa}, 
\begin{equation}
         \Delta p_T^2 = \left\langle p_T^2 \right\rangle_{hA} - \left\langle p_T^2 \right\rangle_{hp}\, .
\end{equation}
This value is indirectly related to multiple parton scattering in the nuclear medium, $\Delta p_T^2 \propto \hat{q}_{A} L_{A} - \hat{q}_p L_p$. As shown in Fig.~\ref{fig:transport_broadening}, using world data on $J/\psi$ and Drell-Yan transverse momentum broadening in $\hA$ collisions from SPS to the LHC, the transport coefficient is found to be $\hat{q}_0=0.051\,\text{GeV}^2/\text{fm}$ or $\hat{q}_0=0.075\,\text{GeV}^2/\text{fm}$ depending on the assumed color state of the produced $c\overline c$ pair that fragments into a $J/\psi$~\cite{Arleo:2020rbm}. The lower value of $\hat q^2_0$ assumes color octet production while the larger value assumes that the pair is produced as a color singlet. 

\subsection{Renormalization group approach to  energy loss}

In the past decades, successful energy loss phenomenology was developed for both hot and cold QCD matter. Such theoretical models, however, have clear  limitations. They cannot be systematically improved to higher perturbative orders and do not give a correct description of in-medium parton showers. For example, in the case of open heavy flavor production, the $g\rightarrow Q\bar{Q}$ process, which is especially important at high energies, is not included. To address these deficiencies,     
full medium-induced splitting kernels~\cite{Sievert:2018imd,Sievert:2019cwq} were derived using effective field theory and lightcone wavefunction approaches. These analogs of DGLAP splitting functions in matter further require implementation in cross section evaluations that go beyond energy loss phenomenology and bridge the gap between particle and nuclear theory. For instance, numerical in-medium DGLAP evolution was validated on the example of HERMES measurements~\cite{Li:2020sru} and used to make the first prediction of the centrality dependence of light and heavy meson production in $e+A$ collisions at the EIC~\cite{Li:2023dhb}.

\begin{figure}[!t]
\centering
\includegraphics[width=\linewidth,clip]{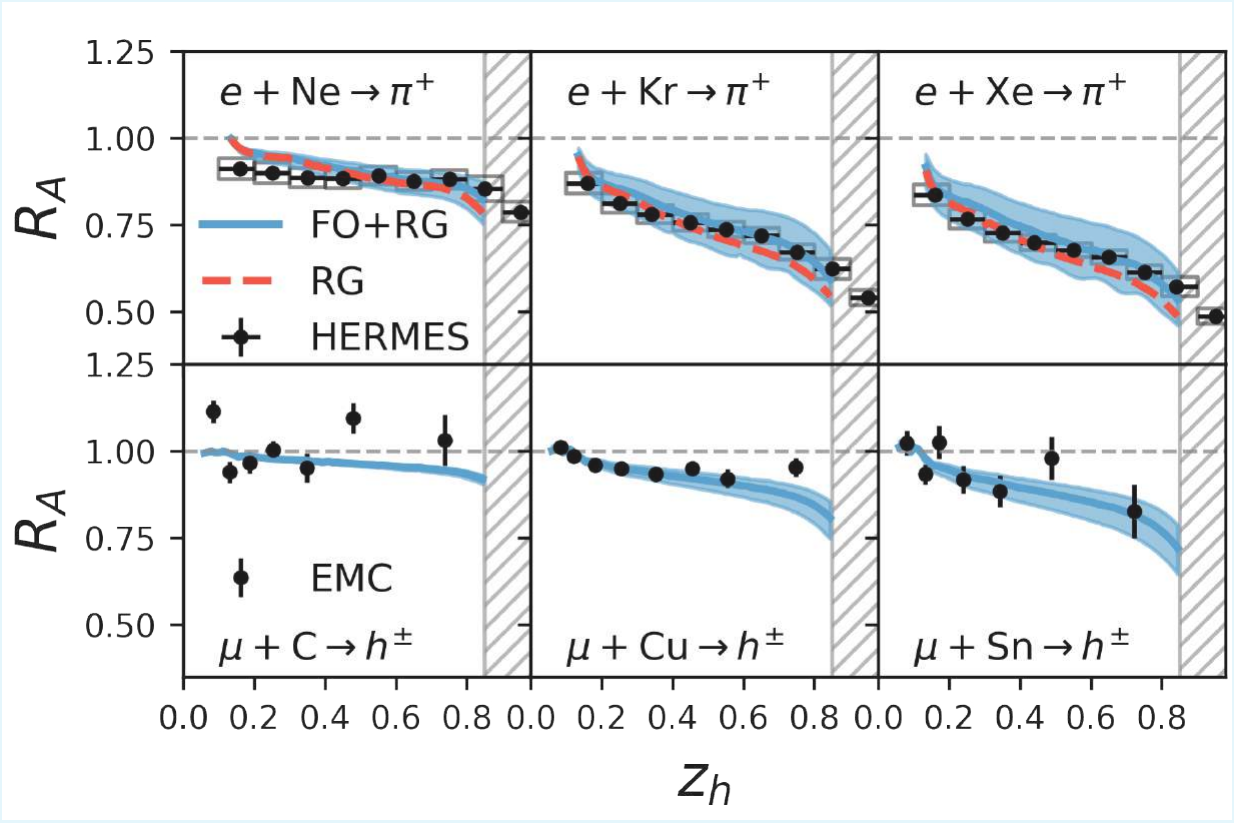} \ \  
\includegraphics[width=\linewidth,clip]{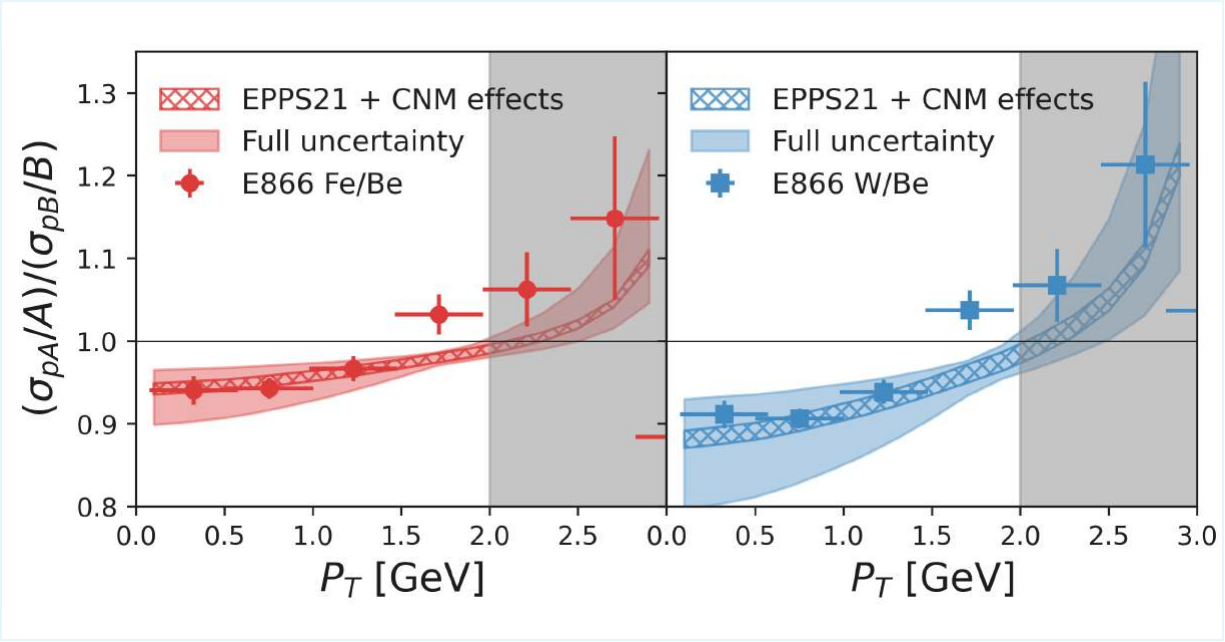} 
\caption{Top: Modification of the differential hadron cross section in SIDIS for several nuclei with different center-of-mass energies and vitualities~\cite{Ke:2023ixa}. Bottom: Ratio of $p_T$-differential Drell-Yan production cross sections on heavy to light nuclei compared to TMD models including in-medium broadening and radiative corrections~\cite{Ke:2024ytw}.}   
\label{fig-Evolution} 
\end{figure}

While the fully numerical approaches in the literature give a good description of jet quenching data, they do not provide insight into the types of logarithms that are being resummed or the fixed-order contributions to the modification of the measured cross sections. 
Most recently, a renormalization group (RG) analysis of collinear hadron production in nuclear DIS  was performed in the limit where the parent parton energy $E$ is large while the medium opacity $L/\lambda$ remains small. This resulted in the derivation of a novel set of  evolution equations~\cite{Ke:2023ixa} that resum leading $\ln(E/\mu^2 L)$ enhanced medium contributions arising from multiple emissions near the endpoints of the splitting functions at first order in opacity. The RG evolution framework  can be applied to study fragmentation in $\eA$ collisions as a function of $z_h=E_h/\nu$, as shown in the top panel of Fig.~\ref{fig-Evolution}.  The nuclear modification factors, normalized by the inclusive cross sections to eliminate initial-state effects,
\begin{align}
R_A = 
\frac{
  \left( \dfrac{d\sigma_{eA \rightarrow h}}{d\nu \, dQ^2 \, dz_h} \middle/ \dfrac{d\sigma_{eA}}{d\nu \, dQ^2} \right)
}{
  \left( \dfrac{d\sigma_{ed \rightarrow h}}{d\nu \, dQ^2 \, dz_h} \middle/ \dfrac{d\sigma_{ed}}{d\nu \, dQ^2} \right)
}
\, ,
\end{align}
measured by HERMES and EMC at different center-of-mass energies and hadron production virtualities, can both be described  well. The calculation demonstrates the importance of fixed-order contributions at intermediate and large values of the fragmentation fraction $z$, and the advantages of RG evolution in handling large separation of scales. The renormalization group approach predicts significant suppression at large values of $x$, corresponding to forward rapidities at the EIC.

A major hadronic physics goal of the past decade has been to understand the three-dimensional structure of nucleons. Such structure is encoded in transverse momentum dependent parton distribution functions (TMDs) that can be extracted through global analyses ~\cite{Anselmino:2013lza,Echevarria:2014xaa,Echevarria:2020hpy,Bacchetta:2017gcc, Bacchetta:2022awv}.  Despite the progress in our understanding of TMD observables in elementary reactions with free nucleons,  investigation of the three-dimensional structure of nuclei has just begun~\cite{Alrashed:2021csd,Alrashed:2023xsv}. 

Importantly, the first TMD investigation of the perturbative dynamics arising from  forward scattering between the partonic constituents of the proton and the nuclear medium has recently appeared. The new contributions to low $p_T$ Drell-Yan production were obtained up to NLO in $\alpha_s$ and to first order in the medium opacity. The derivation includes both collisional broadening and initial-state energy loss.   The collinear and rapidity divergences related to parton showers in the nucleus and the coherent nature of radiation due to the LPM  effect lead to an in-medium RG equation that encodes the $p_T$ dependence of parton energy loss. In addition, the rapidity evolution leads to a BFKL 
evolution equation with restricted phase space for the forward scattering cross section.   

A phenomenological application  of this formalism~\cite{Ke:2024ytw} is shown in the bottom panel of Fig.~\ref{fig-Evolution} using the E866 measurements with Be, Fe, and W  targets as an example.
The theoretical calculation implements the experimental kinematics. A comparison of the full theory to the nuclear modification factors W/Be and Fe/Be  finds overall agreement with the data. The calculation thus captures the correct $A$  dependence of the modifications as well as the interplay between $p_T$ broadening and energy loss.  

\subsection{Impact of the radiative energy loss on nPDF evaluation}
\label{sec:FCEL_nPDF}

As discussed in Section~\ref{sec:nPDF}, nPDFs are extracted from a variety of QCD processes in lepton-nucleus and proton-nucleus collisions. Over the last few years, it has been shown that the production of open-charm mesons and charmonia in $\pA$ collisions at LHC energies provides significant constraints on nPDF extractions, especially in the still poorly constrained gluon sector at small $x$ \cite{Kusina:2017gkz,Eskola:2021nhw,Kusina:2020dki,AbdulKhalek:2022fyi}. These processes, however, are also sensitive to medium-induced gluon radiation and, hence, to fully coherent energy loss \cite{Arleo:2012rs,Arleo:2013zua,Arleo:2021bpv}. Thus, some fraction  of the $A$  dependence of $D$ meson and $J/\psi$ production can be attributed to radiative energy loss and should therefore not be absorbed into the nPDFs. Despite the good description of world data in current global fits, analyses based on QCD factorization and neglecting factorization-breaking effects such as FCEL may lead to a biased nPDF extraction.

\begin{figure}
    \centering
    \includegraphics[width=1\linewidth]{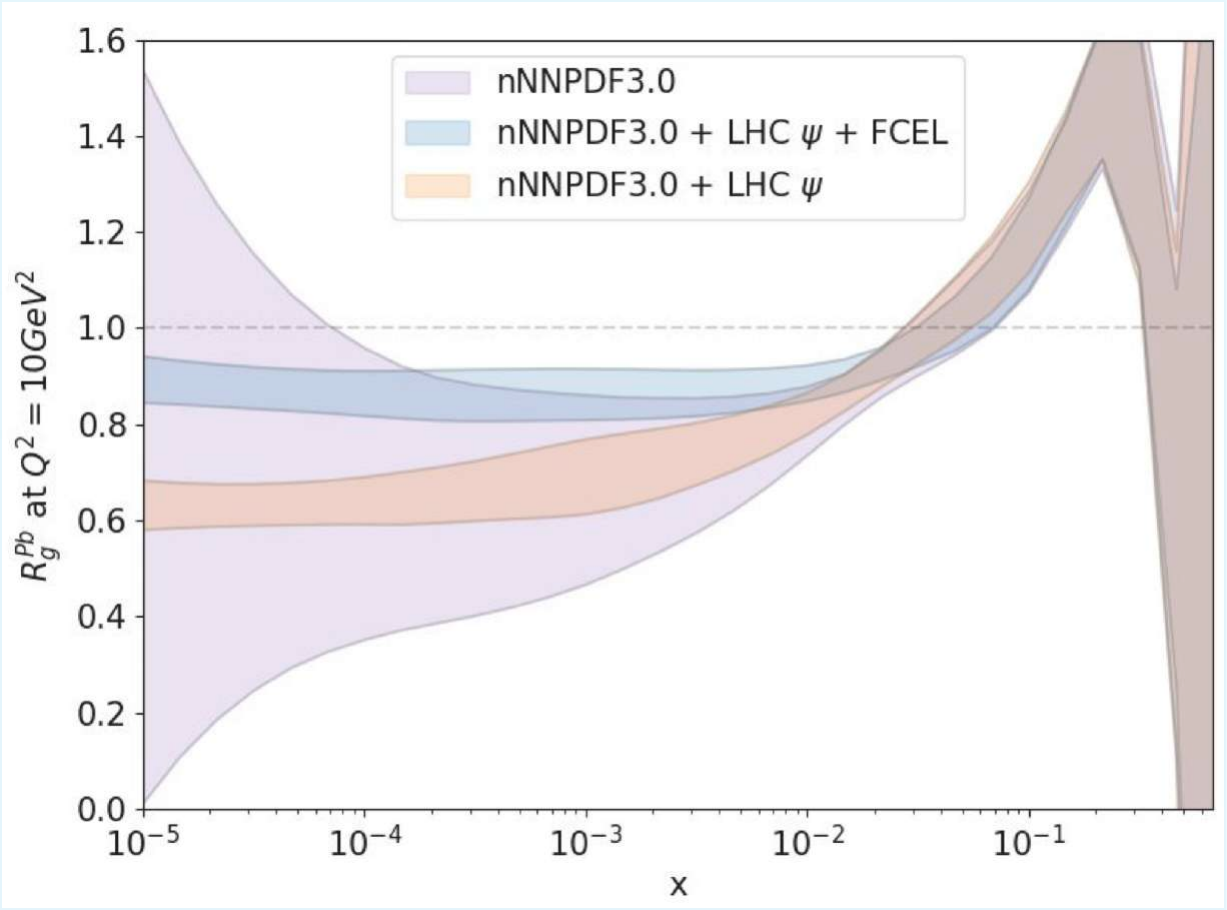}
    \caption{A comparison of the default nNNPDF3.0 set (without including the LHCb $D$ meson data)~\cite{AbdulKhalek:2022fyi} (purple band) with sets obtained after Bayesian reweighting with the ALICE $J/\psi$ data \cite{ALICE:2022zig} with (blue band) and without (orange band) including FCEL. Taken from~\cite{ArleoAvez:2023}.}
\label{fig:npdf_fcel}
\end{figure}

This challenges the current strategy for extracting nPDFs from global fits. To circumvent this issue, one possibility is to restrict the fits to observables insensitive to FCEL, such as electroweak bosons, Drell-Yan, and, to a lesser extent, jet production in $\pA$ collisions, in addition to DIS measurements. This, however, weakens the nPDF constraints, especially for small $x$ gluons, as illustrated by the purple band in Fig.~\ref{fig:npdf_fcel}, showing the nNNPDF3.0\_noLHCbD set, which excludes the LHCb $D$ meson data.

An alternative strategy would be to include the heavy flavor data in global nPDF fits, but only after incorporating FCEL effects into the analyses. This approach was recently explored in a case study~\cite{ArleoAvez:2023} where $J/\psi$ production was computed at leading order in the color evaporation model with  several nPDF sets included in the $p+A$ calculations: EPPS21~\cite{Eskola:2021nhw}, nCTEQ15WZSIH~\cite{Kusina:2020dki}, and nNNPDF3.0\_noLHCbD~\cite{AbdulKhalek:2022fyi}. Preliminary results are shown in Fig.~\ref{fig:npdf_fcel} for a Bayesian reweighting of the nNNPDF3.0\_noLHCbD set, using ALICE $J/\psi$ measurements~\cite{ALICE:2022zig} and assuming either the presence (blue band) or absence (orange band) of FCEL. In both cases, the $J/\psi$ data strongly constrain the gluon nPDF at small $x$, confirming earlier findings~\cite{Kusina:2017gkz,Kusina:2020dki}. The figure shows that the gluon nPDF ratio using the $J/\psi$ data with FCEL included is clearly separated from the ratio obtained with the $J/\psi$ data and no FCEL.  Without FCEL, the gluon nPDF ratio is approximately $R_g \gtrsim 0.6$ at $x \sim 10^{-5}$, similar to the nNNPDF3.0 set which includes LHCb $D$ meson data (not shown). In contrast, $R_g \simeq 0.9$ when FCEL is taken into account.\footnote{In this example, the transport coefficient at $x = 10^{-2}$ is taken to be $\hat{q}_0 = 0.07$~GeV$^2$/fm.} A more systematic and in-depth investigation is in progress.

In conclusion, nPDF sets reweighted with heavy flavor data are shown to be very sensitive to the inclusion of FCEL effects, demonstrating that taking FCEL into account can significantly impact the extraction of the gluon nPDF in particular.  This highlights the need to consistently implement such effects in global fits incorporating heavy-flavor probes of $\hA$ collisions in order to determine the correct level of small $x$ leading-twist gluon shadowing.

\subsection{Energy loss at small $x$}
\label{sub:elosssmallx}

The medium transport parameters (interaction range $\mu$, mean free path $\lambda$, or transport coefficient $\hat{q}$) must be sufficiently large for CNM energy loss to be significant. In a microscopic picture, this, in turn, implies that the nuclear target must be a dense system of partons. In the saturation picture, such a dense state can be probed either by probing large $A$ at intermediate $x$ (the MV regime) or at small $x$ through BK-JIMWLK evolution. The relevant quantity is the coherence length (longitudinal wavelength of the radiated gluon),  inversely proportional to $x$. If the coherence length is longer than the nuclear size, this FCEL is contained in NLO corrections to the CGC~\cite{Munier:2016oih,Bergabo:2021woe}. On the other hand, if the coherence length of the radiation is larger than the nucleon size but smaller than the nuclear size, we have “partially coherent” energy loss. This could be a serious issue for the CGC because the treatment of this regime from first principles is unknown in the saturation framework. Much of our knowledge of gluon saturation in nuclei comes from forward rapidity RHIC data for $p \, ({\rm d} )+ {\rm Au}$ collisions where, due to the steeply falling $p_T$ spectra, even a small energy loss contribution could result in major modification of the spectra. While this is not relevant for total cross sections (structure functions) it could play a significant role for forward rapidity observables. How to separate energy loss effects from suppression generated by genuine gluon saturation dynamics in regions of phase space where both can contribute is an open question. 

\subsection{Observables}\label{sec:energy-loss-observables}

Exploiting the distinct parametric dependencies of LPM energy loss, fully coherent energy loss, the Bertsch-Gunion regime and nuclear parton densities is useful for disentangling the various physical effects. We illustrate this with different observables.

\begin{itemize}[left=0.1cm]
\item Colorless processes  either in the initial-state or final-state could be particularly relevant because they are  insensitive to the fully coherent regime and thus solely be sensitive to LPM energy loss and nPDF effects. This is the case for hadron production in low energy semi-inclusive DIS (SIDIS) on nuclei~\cite{Arleo:2003jz} where propagating quarks lose energy in the final state before fragmenting into hadrons. In addition, normalizing the hadron yields in SIDIS by the number of DIS events drastically reduce the impact of nPDF effects.
The Drell-Yan process in $\pi +A$ and $\pA$ collisions is also particularly relevant. At high energies, LPM energy loss effects on Drell-Yan should be negligible, making it a clean nPDF probe~\cite{Arleo:2015qiv}. Conversely, nPDF effects in low energy $\pi +A$ and $\pA$ collisions are either small or tightly constrained by DIS measurements, making the Drell-Yan process well suited for studies of LPM energy loss in the large $x_F$ regime~\cite{Arleo:2002ph,Neufeld:2010dz, Xing:2011fb,Arleo:2018zjw,Ke:2024ytw}. 

\item FCEL is not universal effect because it depends on the color structure of the partonic process. Thus FCEL cannot be absorbed into the wavefunction of the target nucleus. This non-universality could best be seen by comparing measurements in $\eA$ and $\pA$ collisions: the $A$ dependence of hadron production should be affected by FCEL only in $\pA$.  However, nPDF effects should be similar in both systems, provided the typical momentum fractions probed in the target nucleon are equal: $x\simeq (M^2+Q^2)/(W^2+Q^2)$ in $e+A$ and $x_2 \simeq M/\sqrt{s}\,e^{-y}$ in $p+A$ for production of a  particle with mass $M$. Specifically, one would expect the nuclear production ratio of, {\it e.g.} $J/\psi$ production to be\footnote{Assuming LPM energy loss to be small for high energy $J/\psi$ production.}
 \begin{eqnarray}
   R_{eA}^{J/\psi}(x) &\simeq& R_{eA}^{\text{nPDF}}(x) \approx R_g^A(x) ; \\[0.3cm]
R_{pA}^{J/\psi}(y, x_2) &\simeq& R_{pA}^{\text{nPDF+FCEL}}(y, x_2) \\ & \approx &  R_g^{\text{A}}(x_{2}) 
    \times R_{pA}^{\text{FCEL}}(y, x_2) \,.
    \end{eqnarray}
\item The hadronic structure of the photon can be resolved in photoproduction. In this case, a colored parton coming from the photon participates in the hard scattering, making resolved photoproduction similar to hadroproduction, at least in terms of color flow. Such contributions would therefore be sensitive to FCEL effects. It would thus be valuable to investigate the nuclear dependence of forward jet and hadron production in direct and resolved photoproduction from nuclear targets~\cite{Arleo:2010rb}.
\item There is a strong correlation between the small $x$ gluon and sea quark distributions.  Reference~\cite{Arleo:2015qiv} showed that the double ratio $R_{pA}^{J/\psi} / R_{pA}^{\text{DY}}$ at forward rapidity was close to or somewhat above unity for the available nPDF sets~\cite{Arleo:2015qiv} but significantly smaller than unity for FCEL. This observable could therefore help separate FCEL from nPDF effects in this region.

\item The ratio of the $J/\psi$ and $\Upsilon$ nuclear modification factors, $R^{J/\psi}_{pA}(x)/ R^{\Upsilon}_{pA}(x)$, is useful for studying the final-state mass dependence of the FCEL regime where the average energy loss scales like $1/M_T$ where $M_T$ is the quarkonium transverse mass. The impact of nPDFs effects is somwhat reduced because the differences in nPDF evolution are small for finite values of $M_T$. 

\item The observed $p_T$ broadening, $\Delta p_T^2$ in Drell-Yan and quarkonium production or SIDIS can be used to extract the transport coefficient $\hat q$ more precisely and thus provide constraints on energy loss.
This observable is mainly independent of nPDF and energy loss effects at low $p_T$. Indeed, in this kinematic regime, the dependence of $d\sigma/dp_T$ on nPDFs and energy loss is small; $R_{hA}^{\text{nPDF+E.loss}}(p_T \lesssim M)$ remains approximately constant. As a result, these effects tend to cancel out in the calculation of $\Delta p_T^2$ (see Ref.~\cite{Arleo:2020rbm} for further discussion).

\item Finally, Bertsch-Gunion type energy loss, even with a partial reduction due to coherence effects~\cite{Vitev:2007ve}, may suppress energetic final states at high $x_F$, including light hadrons, heavy flavors and jets.  
The observed scaling behavior of the nuclear modification factor as a function of the total jet energy, $p_T \cosh y$, in several $\pA$ rapidity intervals can naturally be explained  by CNM energy loss~\cite{Kang:2015mta}.
This type of medium-induced bremsstrahlung can also lead to azimuthal asymmetries between produced particles in $\pA$ collisions~\cite{Gyulassy:2014cfa}. 
\end{itemize}
   
\section{Final-state interactions beyond energy loss}
\label{sec:finalstate}

This section discusses final-state interactions that can be active in addition to the energy loss discussed in the previous section.  The processes described in this section mainly pertain to open heavy flavor and quarkonia, with absorption and comover suppression germane only to quarkonium.

\subsection{Collisional dissociation of open heavy flavor and quarkonia}

Collisional and radiative parton-level processes can be complemented by hadron-level interactions. Heuristically, the physics picture that underlies such models is easy to understand.  The time to form a hadron in its rest frame, $\tau_{\rm form}$ is on the order of its size, $\sim 1$~fm, which must be boosted to the lab frame by the Lorentz $\gamma$ factor. At low energies, even light hadrons can have short formation times. This fact was exploited in a model of hadron absorption~\cite{Kopeliovich:2003py,Accardi:2009qv} applied to HERMES fixed target $e+A$ measurements.  For example, if a
finite mass parton, $M_Q$, fragments into a hadron of mass $M_h$ and a light parton,  the formation length, $\Delta y^+ \sim\tau_{\rm form}$, is~\cite{Adil:2006ra}
\begin{eqnarray} 
\Delta y^+ &\simeq& 
\frac{2z(1-z)p^+}{ {\bf k}^2 + (1-z)M_h^2 - z(1-z)M_{Q}^2   } \;. \quad
\label{tfrag}
\end{eqnarray} 
Here $p^+$ is the large light-cone momentum fraction of the parton, ${\bf k}$ is the deviation of splitting from collinearity, and $z = k^+/p^+$ is the fragmentation  fraction.
The $z$ dependence can be exploited to control the formation length in back-to back dihadron~\cite{Markert:2008jc}, photon-hadron or jet-hadron correlations~\cite{STAR:2023ksv}.  
Equation~(\ref{tfrag}) can provide qualitative information about whether energy loss or dissociation/absorption might be more important in a particular kinematic regime.   

The hadron and parton mass dependence in Eq.~(\ref{tfrag}) suggests that, for open heavy flavor~\cite{Adil:2006ra,Sharma:2009hn} and quarkonia~\cite{Sharma:2012dy,Aronson:2017ymv}, the formation length can be considerably shorter than the nuclear size, $\Delta y^+ < 2R_A$, even for large boosts. This further implies that in-medium hadron-level dynamics may be dominant at small and moderate $p_T$. A microscopic model that evaluates the dissociation rate from the distortion of the meson wave function due to collisional interactions in matter has been formulated and phenomenologically applied to successfully describe the suppression patterns of $D$ and $B$ mesons; their semileptonic decays; and quarkonium suppression in $A+A$ collisions. 

\begin{figure}[!t]
\centering
\includegraphics[width=0.8\linewidth]{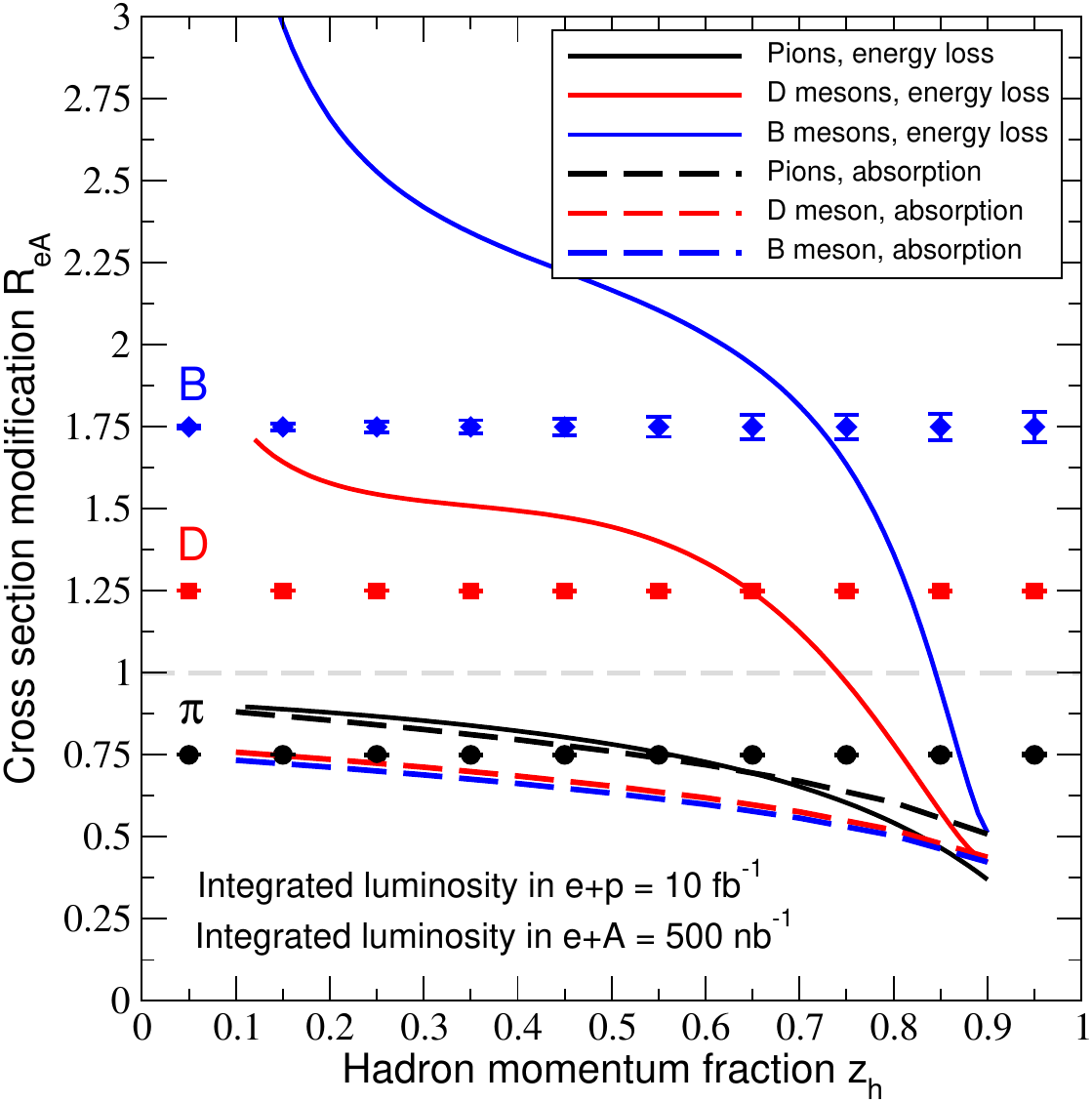} \ \  
\includegraphics[width=0.8\linewidth]{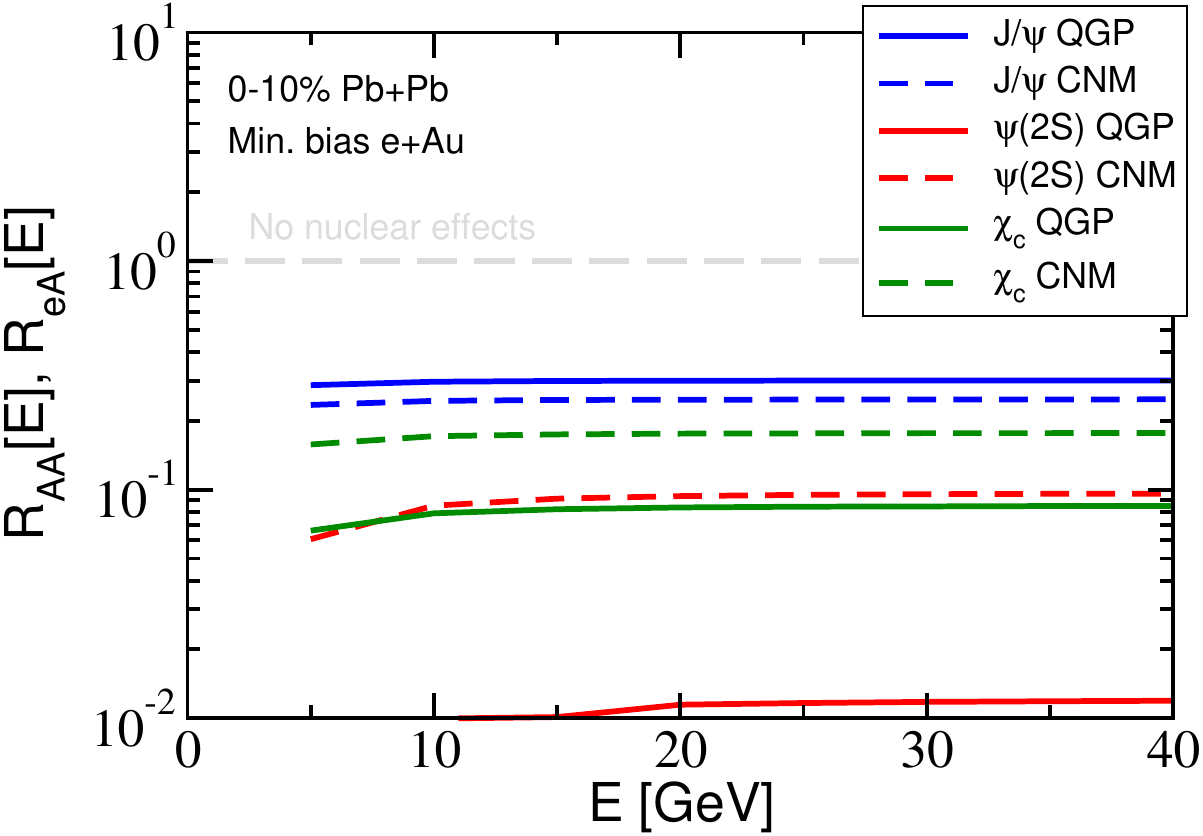} 
\caption{Top: Simplified calculations of parton energy loss (solid lines) and absorption within nuclei (dashed lines) on the momentum fraction $z_h$ distribution of hadrons in the acceptance of the forward silicon tracker at the EIC. The projected statistical precision (points) of forward rapidity $B$ and $D$  measurements with  integrated luminosities in $e+p$ ($e+{\rm Au}$) collisions is 10~fb$^{-1}$ (500~nb$^{-1}$) illustrates the power of the EIC to discriminate between energy loss and absorption~\cite{Li:2020sru}. Bottom: Comparison between the suppression of several charmonium states in CNM (dashed lines) and the QGP (solid lines) due to dissociation. Adapted from Ref.~\cite{Boer:2024ylx}.}   
\label{fig-AbsorbDissoc} 
\end{figure}

Similar formation considerations apply to CNM and nuclear DIS. On general grounds, one would expect that the larger center-of-mass energies at the EIC relative to HERMES will lead to larger boosts so that the nuclear modification factor for light hadrons, such as pions and kaons, will be predominantly driven by energy loss. Nevertheless, this remains an open question that cannot currently be unambiguously answered. The characteristic shape of the charm and bottom quark fragmentation functions may be key to identifying the origin of CNM meson suppression~\cite{Li:2020sru}. As shown on the top panel of Fig.~\ref{fig-AbsorbDissoc}, absorption always leads to a reduction of the differential cross section as a function of $z$.  In contrast, in-medium parton shower modifications or energy loss lead to enhanced $D$ and $B$ production at intermediate and small $z$. In addition, the narrower $B$ meson fragmentation function is shifted to larger $z$, signifying that less $b$ quark energy is lost to fragmentation.  This shift is responsible for the larger $B$ modification relative to $D$ mesons. It is expected, as indicated by the simulated error bars for projected yearly luminosities, that this physics can be studied at the EIC. 

In contrast to open heavy flavor, at the parton level, quarkonia originate from a heavy $Q\overline Q$ pair produced  in a specific color and angular momentum state at very early times, inversely proportional to the virtuality of the hard process. Consequently, interactions of the heavy quarks with the medium start early and the energy dependence of the suppression is very weak~\cite{Boer:2024ylx}.  A preliminary comparison of the suppression of $J/\psi$, $\psi$(2S), and $\chi_c$ in CNM, $R_{eA}$, and in the QGP, $R_{AA}$, is shown on the bottom panel of Fig.~\ref{fig-AbsorbDissoc}. While weakly bound states like the $\psi$(2S) and $\chi_c$ are more suppressed in the QGP due to thermal wavefunction effects, the CNM and QGP effects on the $J/\psi$ are similar. This is because the thermal screening of the potential for the most tightly bound $Q\overline Q$ states is very small, while collisional dissociation depends on the interplay between the quarkonium binding and the space-time dynamics of the nuclear matter transport coefficients.   In all cases, the nuclear modification is significant, a factor of two or more, and should be easily observable experimentally.   The future EIC will provide the first opportunity to study quarkonium modification in $\eA$ collisions.

\subsection{Absorption and Comovers}

This section covers two final-state effects on the $Q \overline Q$ or formed quarkonium state.  The argument for the effects is generic even though they have most often been applied to the charmonium states.  In the early days of $J/\psi$ studies with hadron beams on nuclear targets, the assumption often was that the $J/\psi$ cross section on a nuclear target would simply scale by the mass number of that target $\sigma_{pA} = A \sigma_{pp}$.  However, the first such studies, including with photon beams \cite{Anderson:1976hi}, showed
that this dependence was not, in fact, linear but less than linear, quantified
as $\sigma_{pA} = A^\alpha \sigma_{pp}$ with $\alpha < 1$, as already discussed in Sec.~\ref{sec:data}.
All the unknown nuclear effects were assumed to be subsumed into the exponent $\alpha$, with no effects on the production cross section itself.
As has been learned over the years, this is not the case and several important cold nuclear matter effects may be incorporated into the perturbative calculation of the cross sections, such as

\begin{widetext}
\begin{equation}
\begin{split}
\sigma_{pA} = \sum_{i,j} \int \frac{d\tau}{\tau} \int d^2b \, dz \, d\epsilon \, dx_1 \, dx_2 \, 
& \delta(x_1 x_2 - \tau) \delta(x_F' - x_F - \delta x_F) \delta(x_F' - x_1 + x_2) \\
& \times P(\epsilon) S_A^{\rm abs} S_{\rm co} 
F_i^p(x_1) F_j^A (x_1', \vec b, z) \widehat{\sigma}_{ij} \, \, .
\end{split}
\label{eq:sigma_pA}
\end{equation}
\end{widetext}

Here $P(\epsilon)$ denotes the energy loss (see Sec.~\ref{sec:eloss}); $F_i^p$ and $F_j^A$ are the
parton densities in the proton and the nucleus respectively (see Sec.~\ref{sec:nPDF}), assumed to be functions of $M_Q$, factorization scale $Q$, internal transverse momentum $k_T$, and, for the nucleus, impact parameter $b$; and $\hat \sigma$ is the partonic cross section as a function of $M_Q$, $Q$ and renormalization scale $\mu_R$.  The terms $S_A^{\rm abs}$ and $S_{\rm co}$ are the survival probabilities for nucleon
absorption and comover dissociation respectively. These survival
probabilities are assumed to depend on the location of the hard interaction in the nucleus.

In the remainder of this section, absorption by nucleons and dissociation by comovers are discussed in turn.  The potential impact parameter dependence of nuclear modifications of the parton densities is discussed in Sec.~\ref{sec:shad_bdep}.

\subsubsection{Absorption by Nucleons}

In $p+A$ collisions, the proto-$J/\psi$ may interact with nucleons and be dissociated before it can escape the target, referred to as nuclear absorption.
The survival probability for absorption by nucleons may be expressed as \cite{rvrev}
\begin{eqnarray}
S^{\rm abs}(b) =  \exp \left\{
-\int_z^{\infty} dz^{\prime} \rho_A (b,z^{\prime}) \sigma_{\rm abs}(z^\prime
-z)\right\} \, \, , 
\label{sigfull}
\end{eqnarray}
where $b$ is the impact parameter, $z$ is the $c \overline c$
production point, $S^{\rm abs}(b)$ is the nuclear absorption survival probability, and $\sigma_{\rm abs}(z^\prime -z)$ 
is the nucleon absorption cross section along its path through matter. 
The exponent $\alpha$ can be related to 
$\sigma_{\rm abs}$ by integrating Eq.~(\ref{sigfull}) over the nuclear volume, giving $\alpha = 1 - (9\sigma_{\rm abs}/(16\pi r_0^2))$. This is an oversimplification
since other effects are ignored and final-state charmonium dissociation by hadron interactions can be accomplished both by nucleons and comoving secondary hadrons with a similar $A$ dependence \cite{Vogt:1990us}.
The value and dependence of $\sigma_{\rm abs}$ has been modeled several different ways, depending on the production dynamics of the charmonium state.

If it is produced as a small color singlet, it grows until reaching its formation time when it becomes a fully formed bound state. In this case, absorption is assumed to only play a role if the state can be formed inside the target.  Thus, at most energies, the state is only produced inside the target, where absorption is effective, at negative $x_F$.
The larger the state, the longer its formation time and the larger its asymptotic absorption cross section, assuming $\sigma_{\rm abs}$ grows as the square of the quarkonium radius.  

If, on the other hand, the proto-charmonium state,
$|(c \overline c)_8 g \rangle$, is assumed to be in color singlet through its characteristic octet lifetime,
all the charmonium states pass through matter with a constant absorption cross section, particularly at forward $x_F$.  At negative $x_F$, the octet path length is shorter and it can neutralize its color and be absorbed as a color singlet with a different cross section.  Here, the $(c \overline c)_8$ configuration is not yet on shell.

Finally, in nonrelativistic QCD, the charmonium states are an admixture of singlet and octet production. In this case, the color singlet dominated $\chi_c$ states would have a near linear $A$ dependence at forward $x_F$ while the $J/\psi$ and $\psi$(2S), dominated by octet production, would have a less than linear $A$ dependence at forward $x_F$.  In this case, the quarkonium states would come on shell after the characteristic formation time for color singlet states and after the octet lifetime for color octet states.  Before these characteristic times, the $c \overline c$ pair interacts with the nuclear matter.

All of these scenarios were examined in detail in
Ref.~\cite{RV_HeraB}.  Unfortunately the $\chi_c$ $A$ dependence has not yet been measured with precision, a measurement that could distinguish between types of production and thus also of absorption.

Since no other cold nuclear matter effects were initially considered in the extraction of the absorption cross section, their subsequent inclusion revealed that the effective absorption cross section could be significantly underestimated for low energy fixed-target experiments, particularly at values of $\sqrt{s_{NN}}$ where $x_2$ is in the EMC or antishadowing region where the nuclear gluon distribution is enhanced relative to that of the proton \cite{Lourenco:2008sk}. At forward rapidity, the $J/\psi$ formation time, $t_{\rm form}$, exceeds the nuclear size, so that nuclear absorption does not affect $\rpa$. In this region, only nPDF effects, fully coherent energy loss, and potentially production by intrinsic charm, see Sec.~\ref{sec:other_effects}, can contribute to the modification of $J/\psi$ production in $h+A$ collisions. However, at midrapidity or negative rapidity, where the formation time is less than $L$, the distance traversed by the $J/\psi$  through nuclear matter, $t_{\text{form}} \lesssim L$, nuclear absorption becomes relevant. This condition is expressed as
\begin{equation}
t_{\text{form}} = \gamma \tau_{\text{{form}}}= \frac{E}{m_Q} \tau_{\text{form}} \lesssim L,
\label{eq:production}
\end{equation}
where $\gamma$ is the Lorentz factor and $E$ is the total energy of the $J/\psi$.   Thus criteria ensures that the $J/\psi$ is fully formed within the nuclear target, making  absorption relevant.  A similar relation would hold for other quarkonium states.

\begin{figure}[tbp]
    \centering
    \includegraphics[width=\linewidth,clip]{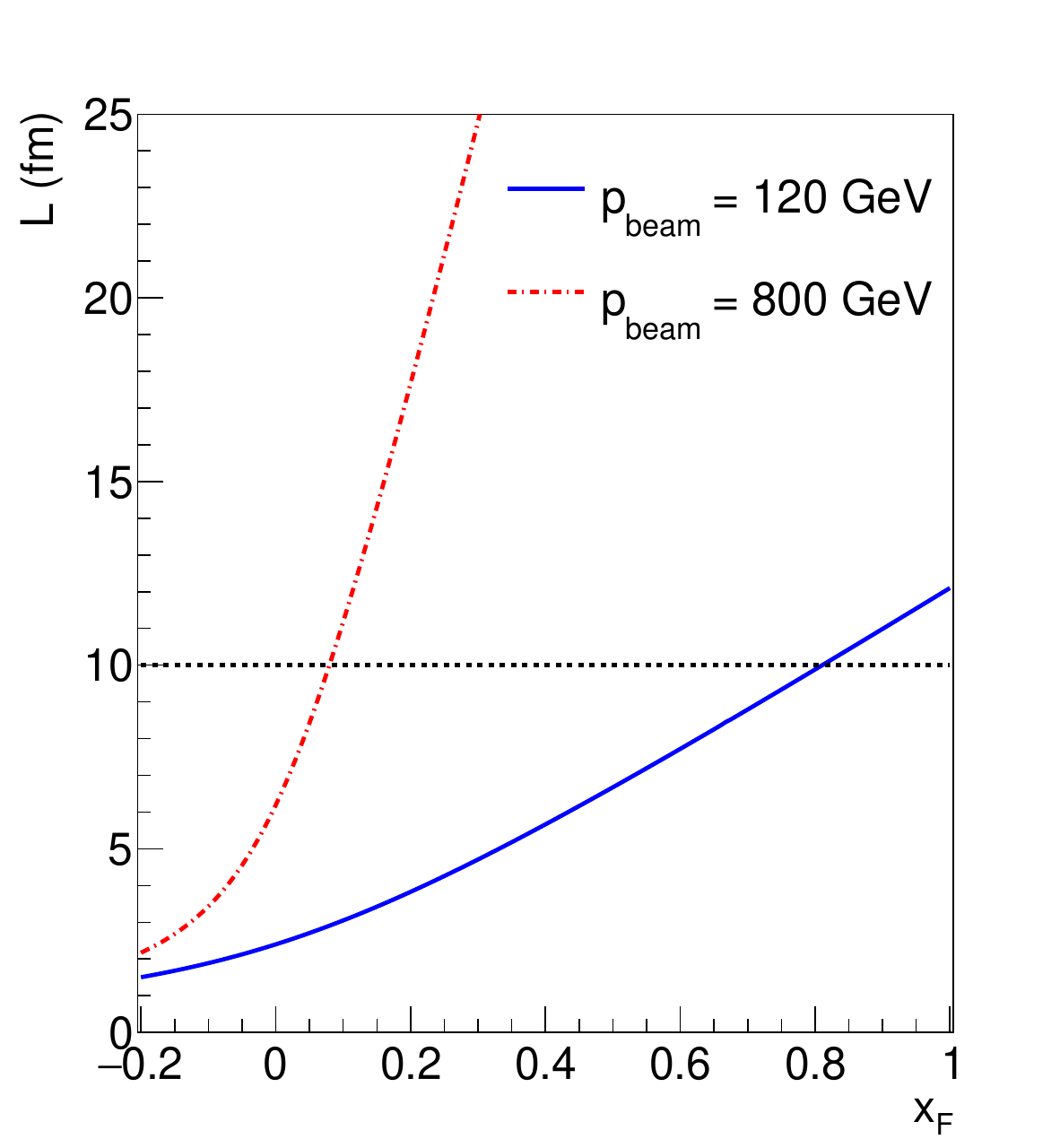}
    \caption{The distance through nuclear matter traversed by the $J/\psi$ before its formation, $L$, as a function of $x_F$, assuming $\tau_{\rm form}$ = 0.3 fm using Eq.~(\ref{eq:production}) for two fixed-target energies. The horizontal line at $L = 10$~fm shows the maximum path length for a nucleus of $A \sim 200$.  For further details, see~Eq.~(3.3) in \cite{Arleo:2012rs}.}
    \label{fig:jpsi_production}
\end{figure}

Figure~\ref{fig:jpsi_production} shows the production length of the $J/\psi$ as a function of $x_F$ for fixed-target beam energies, assuming $\tau_{\rm form}  = 0.3$~fm. At $p_{\text{beam}} = 800$ GeV, the $J/\psi$ is predominantly formed outside the nucleus at large $x_F$ since $\langle  L \rangle \sim (3/2)R_A \sim 10$~fm. As the beam energy decreases, $J/\psi$ formation is increasingly likely within the nuclear target. These simple considerations provide insight into the potential impact of nuclear absorption effects on quarkonium production in $\hA$ collisions.

\begin{figure}[tbp]
  \centering
  \includegraphics[width=0.8\linewidth]{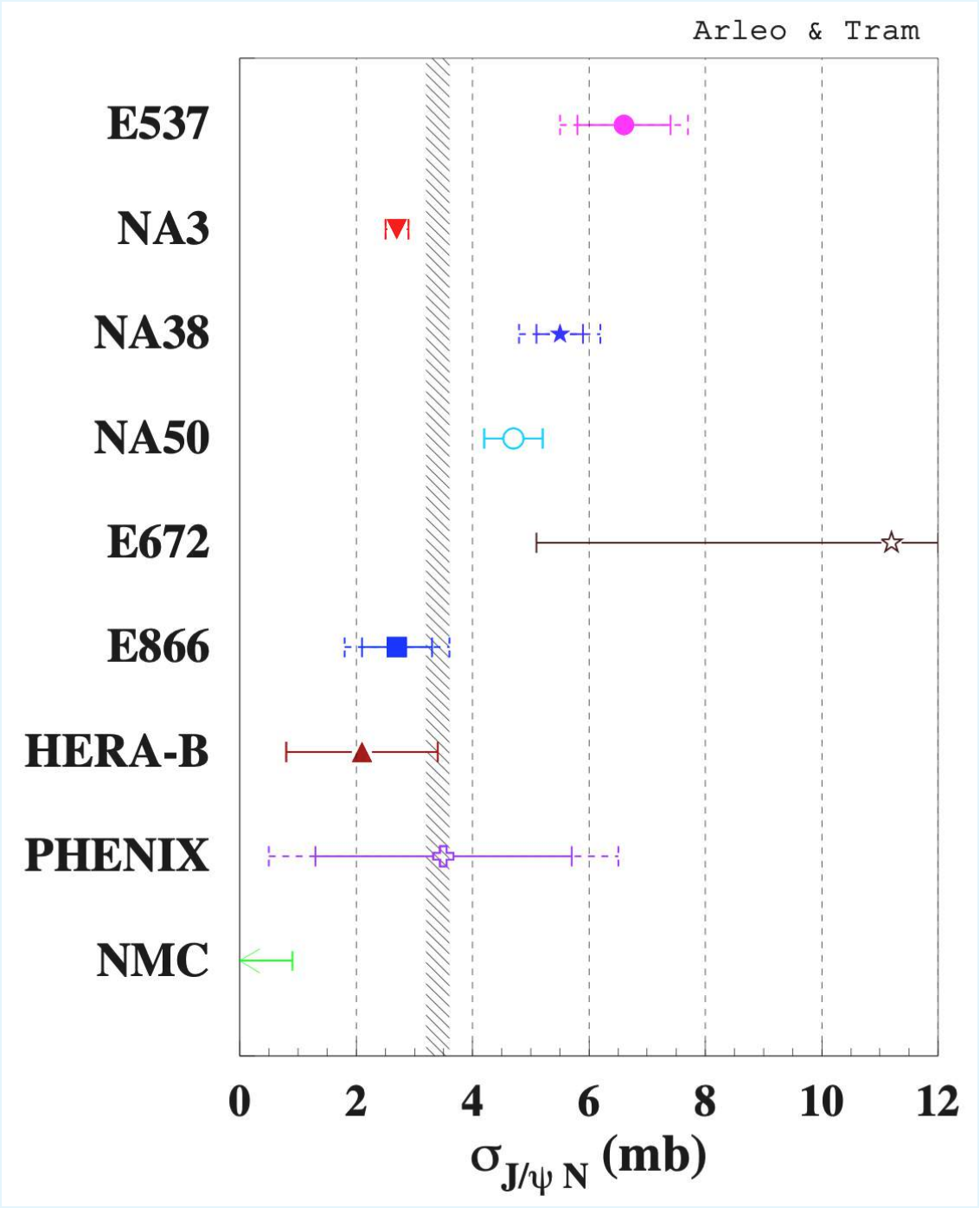} 
  \vspace{0.5cm}
  \includegraphics[width=0.8\linewidth]{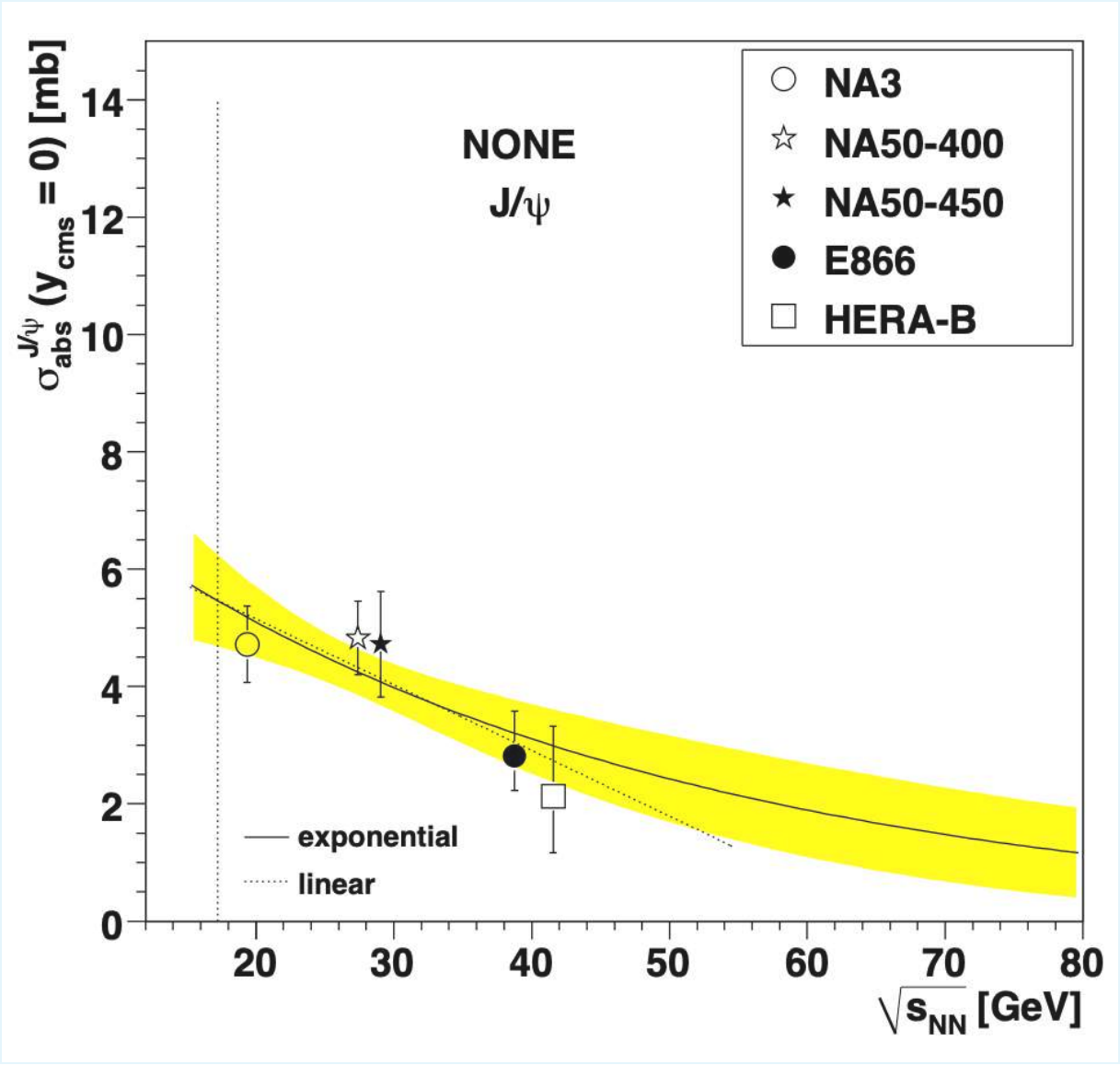}
  \caption{Extraction of $\sigma_{\text{abs}}$ without including nPDF or FCEL effects using $J/\psi$ data from $\hA$ collisions.  (Top) Results from Ref.~\cite{Arleo:2006qk}.  (Bottom) Calculations from Ref.~\cite{Lourenco:2008sk}.}
  \label{fig:nuc_abs}
\end{figure}

Figure~\ref{fig:nuc_abs} presents two global extractions of $\sigma_{\text{abs}}$ using midrapidity data \cite{Arleo:2006qk, Lourenco:2008sk}, both of which converge to a value of $\sigma_{\text{abs}}$ of a few millibarns, without considering other nuclear effects.

The extraction of $\sigma_{\text{abs}}$ is affected by both nPDF and FCEL contributions. When $\rha^{\text{nPDF}} \lesssim 1$ (in the EMC or shadowing) regions, $\sigma_{\text{abs}}$ may be underestimated, if $\rha^{\text{nPDF}} \gtrsim 1$ (in the antishadowing region), absorption could be overestimated. Moreover, if FCEL is not properly distinguished from absorption, $\sigma_{\text{abs}}$ may also be overestimated because $\rha^{\text{FCEL}} < 1$.

\subsubsection{Breakup by Comovers}

The absorption models just discussed do not generally distinguish between $J/\psi$ and $\psi$(2S) suppression at forward $x_F$ because, at this point, the final quarkonium state is fully formed well outside the nucleus. However, absorption by comovers, whether these comovers are characterized as hadrons or partons \cite{Elena}, does do so because it assumes that the comovers interact with formed charmonium states, consistent with larger interaction cross sections for the $\psi$(2S), especially at energies where the nucleon absorption cross section is assumed to be negligible \cite{Lourenco:2008sk} but comover suppression is active \cite{Ferreiro:2014bia}.

In the comover model \cite{Capella97,Armesto98,Armesto99,Capella00,Capella05,Capella:2006mb}, the suppression arises from scattering of the
nascent charmonium with produced particles
that travel along the path of the $c\bar{c}$
\cite{Capella97,Gavin:1996yd}.
The comover survival probability also depends on the time during which the charmonium state can interact with the comovers.  The result depends on the
ratio $\tau_f/ \tau_0$, the final over the initial time, effectively the lifetime of the system. Because the proper time is inversely proportional to the density, $\tau_f/ \tau_0 = \rho^{\rm co}(\vec b,z)/\rho_{pp}$, where the final time at which the interactions cease is when the comover density has decreased to the particle density in $p+p$ collisions at the same energy.  Thus, 
\begin{eqnarray}
\label{eq:survivalco}
S^{co}_{\psi}(\vec b,z )  \;=\; \exp \left\{-\sigma_{\rm co}
  \, \rho_{\rm co}(\vec b,z)\, \ln
\left[\frac{\rho_{\rm co}(\vec b,z)}{\rho_{pp}}\right] \right\} \; .
\end{eqnarray}
Dissociation by comovers is stronger where the comover density is higher, in
more central collisions and higher center-of-mass energies, as well as, in collider mode, in the nucleus-going direction, defined as negative rapidity or
$x_F$.

The values of $\sigma_{\rm co}$ were fixed \cite{Armesto98} from fits to fixed-target data to be $\sigma_{\rm co}^{J/\psi}=0.65$ mb and $\sigma_{\rm co}^{\psi{\rm (2S)}}=6$ mb.  These same values have also been used to describe data from  RHIC \cite{Capella:2007jv} and the LHC \cite{Ferreiro:2012rq}.  Comover suppression has also been applied to excited states of bottomonium at the LHC \cite{Ferreiro:2018wbd}.  Comover suppression can explain differences in suppression patterns between excited quarkonium states in $p+A$ collisions, particularly where absorption effects should be small.  Since the comover effects depend on the density of the produced particles in the collision, this mechanism results in different dependencies on charged particle multiplicity or impact parameter.

\subsection{Observables}
\label{sec:final-state-observations}
Several observables can help isolate collisional dissociation, absorption, and comover effects, if they exist. These are also particularly useful for distinguishing  these effects from other competing suppression mechanisms such as energy loss or nPDF modifications which are independent of whether the quarkonia are produced in an excited state or the ground state.

\begin{itemize}[left=0.1cm]
    \item Initial-state,  final-state, and fully coherent energy loss and nPDF effects are expected to be similar for $J/\psi$ and $\psi$(2S) in $\pA$ collisions. Consequently, the double ratio,
    \begin{equation}
        \mathcal{R} = R^{Q\overline Q_i} / R^{Q\overline Q_j}  \, \, ,\label{eq:double_ratio}
        \end{equation} 
        where $Q \overline Q_i$ and $Q\overline Q_j$ are two different quarkonium states,  would help isolate the effects of in-medium dissociation or nuclear absorption when $t_{\text{form}} \lesssim L$.

\item In higher energy collisions, where $t_{\rm form} \gg L$ and absorption is likely negligible, measurements of $\mathcal{R}$ as a function of the density of produced particles could help isolate comover effects.  In addition, since comover suppression is expected to apply only to quarkonium production, a comparison between $\mathcal{R}$ in Eq.~(\ref{eq:double_ratio}) and double ratios of open heavy flavor mesons could be helpful.
         
   \item In calculations of collisional dissociation, once the binding energy of the heavy quarks in matter is fixed~\cite{Makris:2019ttx}, $\mathcal{R}$ can be predicted for individual quarkonium states based on the width of the quarkonium wavefunction. More weakly bound quarkonium states experience more suppression than tightly bound ones~\cite{Sharma:2012dy,Aronson:2017ymv}, following a hierarchy that can be quantitatively tested, see the bottom panel of Fig.~\ref{fig-AbsorbDissoc}.  

    \item The $\chi_c$ is primarily produced as a color singlet, while the $J/\psi$ and $\psi$(2S) predominantly originate from color octets. As a result, their final-state interactions in nuclei are expected to differ significantly. FCEL and nPDF effects are effectively eliminated in $\mathcal{R}$ in Eq.~(\ref{eq:double_ratio}).  
   
    \item Finally, the hadron production ratio between $e+A$ and $e+p$ in semi-inclusive DIS can  isolate absorption effects if final-state energy loss is small for high energy hadrons. 
\end{itemize}

By leveraging these observables, nuclear absorption effects can be isolated more effectively, particularly in kinematic regions where the competing effects of energy loss and nPDF modifications are minimized.
   
\section{Other nonperturbative effects}
\label{sec:other_effects}

In this section, we focus on two nonperturbative initial-state effects. The first concerns the role of the impact parameter in leading-twist nuclear shadowing in the nPDF framework in $\hA$ and $\eA$ collisions. So far, the spatial dependence, the location within the nucleus where the probe interacts, has largely been neglected in standard calculations. The second nonperturbative effect is intrinsic charm, the existence of which has recently received renewed support from LHCb measurements~\cite{Ball:2022qks}.

\subsection{Impact Parameter Dependent Shadowing}
\label{sec:shad_bdep}

The nuclear modifications of the parton densities are derived from nuclear DIS data and minimum-bias $h+A$ collisions, as in EPPS21 \cite{Eskola:2021nhw}.
However, given that the parton distributions are
modified in the presence of the nuclear medium, it is also reasonable to assume that the effect may further depend on where in the nucleus the parton originated, from a nucleon closer to the nuclear surface or embedded deeply in its volume. No detailed studies of parton distributions in the nucleus as a function of collision centralities exist so far. The first experimental study of spatial dependence of parton distributions relied on dark tracks in emulsion to tag more central collisions \cite{E745}. Studies of the impact parameter dependence of shadowing were first proposed in Ref.~\cite{Emelyanov:1997guf}.

In Eq.~(\ref{eq:sigma_pA}), the parton densities in the nucleus can be factored into the nuclear density distribution, a shadowing function, and the
unmodified proton parton densities:
\begin{eqnarray}
F_i^A(x,Q^2,k_T,\vec{b},z) & = & \nonumber \\
&& \hspace{-3em} \rho_A(s) \, S_i(x,Q^2,\vec{b},z) \, f_i^p(x,Q^2,k_T)
\label{eq:FiA}
\end{eqnarray}
where $s = \sqrt{b^2 + z^2}$, $b$ is the transverse impact parameter, $z$ the longitudinal position, and $k_T$ the transverse momentum. If the parton densities in the nucleon and nucleus are the same, then $S_i = 1$.  Then the integration over the nuclear volume gives $F_i^A(x,Q^2,k_T) = A f_i^p(x,Q^2,k_T)$.

The nuclear density distribution $\rho_A(s)$ is described by a three-parameter Woods-Saxon form:
\begin{equation}
\rho_A(s) = \rho_0 \left(1 + \omega \frac{s^2}{R_A^2} \right) \left(1 + \exp\left(\frac{s - R_A}{d}\right) \right)^{-1},
\end{equation}
where $R_A$ is the nuclear radius, $d$ the surface thickness, $\omega$ allows for central density variation, and $\rho_0$ is the density at the center of the nucleus.

Two shadowing parameterizations were proposed in Ref.~\cite{Emelyanov:1997guf}, both giving stronger shadowing at low $b$ (central collisions) and weaker at high $b$ (peripheral collisions), normalized so that
\begin{equation}
\frac{1}{A} \int d^2b\, dz\, \rho(s) \, S_i(x,Q^2,\vec{b},z) = S_i(x,Q^2).
\end{equation}

The first assumes shadowing follows the local density:
\begin{equation}
S_i^{\rm WS}(\vec{b},z) = 1 + N_{\rm WS} \left( S_i - 1 \right) \frac{\rho(\vec{b},z)}{\rho_0},
\end{equation}
where $N_{\rm WS}$ ensures proper normalization. This form has weak spatial dependence until near the surface.

Alternatively, if shadowing depends on the nuclear thickness $T_A(b)$, the $b$ dependence is stronger. In both approaches, $S_i(\vec{b},z) < S_i$ in the shadowing and EMC regions for small $b$, and $S_i(\vec{b},z) > S_i$ in the antishadowing region. If no $b$ dependence is assumed, the same shadowing is applied to all $b$, which is unrealistic. As shown in Ref.~\cite{Emelyanov:1998yul}, a model like $S_i^{\rm WS}$ enhances shadowing at $b=0$ by a few percent, halves it around $b \sim 1.5R_A$, and eliminates it for $b > 2R_A$. A stronger $b$ dependence would both increase central shadowing and localize it more effectively.

In Ref.~\cite{McGlinchey:2012bp}, these parameterizations were found insufficient to describe the PHENIX d+Au data \cite{PHENIX:2012czk}. Two additional models were tested. EPS09s \cite{Helenius:2012wd}, similar to the $T_A$-dependent form but using powers of $T_A^n(b)$ up to $n=4$ with $A$-independent coefficients $a(n)$,
\begin{equation}
S_i(\vec{b},z) = 1 - (1 - S_i) \frac{T_A^n(b)}{a(n)},
\end{equation}
was also too weak. A step function form was also tested:
\begin{equation}
S_i(\vec{b},z) = 1 - \left( \frac{1 - S_i}{a(R,d)(1 + \exp((b - R)/d))} \right),
\end{equation}
which provided the best fit with $R = 2.4$~fm and $d = 0.12$~fm, suggesting that shadowing was localized at the center of the nucleus, like a hot spot \cite{McGlinchey:2012bp}.

Later ALICE measurements indicated a weaker $b$ dependence, implying that at higher energies, shadowing is more uniformly distributed throughout the nuclear volume rather than localized at the center~\cite{Lakomov:2014yga}. However, the $b$-dependent analysis in Ref.~\cite{Lakomov:2014yga} does not account for other cold nuclear matter effects.

\subsection{Intrinsic charm}
\label{sec:intrinsic_charm}

The existence of a nonperturbative charm quark content in the proton was predicted in the early 1980s \cite{Brodsky:1980pb,Brodsky:1981se}. There has been a renewed interest in such intrinsic states over the last several years \cite{LHCb:2021stx,SMOG,SMOGpNeJ,SMOGpNeD}. The most basic of these states is a five-quark configuration of the proton, $|uud\, c \overline c\rangle$, giving final states such as $\Lambda_c + \overline D$ and $J/\psi + p$.

These intrinsic charm states are suggested to be a fundamental property of the hadronic bound-state wavefunctions \cite{Brodsky:1980pb,Brodsky:1981se}. The intrinsic states are dominated by kinematic configurations where the constituent quarks are moving at equal rapidity, manifesting the charm quarks at large $x$. When a proton in this state interacts, the coherence of the state is broken and the fluctuations can hadronize \cite{Brodsky:1980pb,Brodsky:1981se,Brodsky:1991dj}.

The basic five-particle proton state produces leading charm hadrons, carrying valence quarks of the parent hadron, such as the $\overline D$ in the proton and $D^-$ from a similar four-particle state in the $\pi^-$, resulting in asymmetries between $D^-$ and $D^+$ in $\pi^- + p$ interactions, particularly at high $x_F$ or rapidity \cite{E769,E791,WA82,SMOGpNeD}. Intrinsic states with additional light or heavy $q \overline q$ pairs have been studied \cite{RVSJB_psipsi,ANDY}, including tetraquark production \cite{RV_tetraquark, COMPASS:2022djq}.

In the original picture of intrinsic charm by Brodsky \textit{et al.} \cite{Brodsky:1980pb,Brodsky:1981se}, the frame-independent probability distribution of an $n$-particle intrinsic heavy quark state of the proton is
\begin{widetext}
\begin{equation}
\begin{split}
dP_{{\rm ic}\, n} = P_{{\rm ic}\,n}^0\, N_n \int dx_1 \cdots dx_n \int dk_{x\,1} \cdots dk_{x\,n} \int dk_{y\,1} \cdots dk_{y\,n} 
\frac{\delta\left(1 - \sum_{i=1}^5 x_i\right)\, \delta\left(\sum_{i=1}^5 k_{x\,i}\right)\, \delta\left(\sum_{i=1}^5 k_{y\,i}\right)}{\left( m_p^2 - \sum_{i=1}^n \left(\widehat{m}_i^2 / x_i\right) \right)^2} \, ,
\end{split}
\label{icdenom}
\end{equation}
\end{widetext}
where $i = 1, 2, 3$ are the light valence quarks ($u$, $u$, $d$), and $i = 4, \dots, n$ correspond to intrinsic states. In a five-particle $|uud\, c \overline c\rangle$ state, $i = 4, 5$ are $c$ and $\overline c$, respectively.

The factor $N_n$ normalizes the integrals over $x_i$, $k_{x,i}$ and $k_{y,i}$ to unity, and $P_{{\rm ic}\,5}^0$ scales the unit-normalized probability to the assumed probability for the production of the state in the proton. The final-state hadrons, such as $\Lambda_c$ or $\overline D$, are produced by coalescence of the quarks in the state. Other models of intrinsic charm in the proton also exist, see Refs.~\cite{Paiva:1996dd,Steffens:1999hx,Hobbs:2013bia,Pumplin:2007wg,Brodsky:2015fna}.

The production cross section of intrinsic charm from the $|uud\, c \overline c\rangle$ state can be written as
\begin{equation}
\sigma_{\rm ic}(pp) = P_{{\rm ic}\, 5} \, \sigma_{pN}^{\rm in} \left( \frac{\mu^2}{4 \widehat{m}_c^2} \right) \, .
\end{equation}
The resolving factor $\mu^2 / (4 \widehat{m}_c^2)$, arising from a soft interaction that breaks the coherence of the state \cite{Brodsky:1991dj}, was introduced to calculate the intrinsic charm cross section in Ref.~\cite{Vogt:1994zf}. Here $\mu^2 = 0.1$~GeV$^2$ is assumed \cite{RV_SeaQuest}. The inelastic cross section, $\sigma_{pN}^{\rm in} = 30$~mb, is appropriate for fixed-target interactions and can be used as a conservative estimate at higher energies. The cross sections for higher states can be scaled appropriately by the masses; see the discussion in Ref.~\cite{RV_tetraquark}.

Intrinsic heavy quark production differs from standard perturbatively produced heavy quarks in the treatment of the initial state. As previously noted,  intrinsic production occurs in forward kinematic regions since the constituent quarks of the final-state hadron all emerge from the projectile hadron itself. In contrast, in perturbative production of heavy quarks, e.g., via $g g \rightarrow Q \overline Q X$, one gluon arises from the projectile and the other from the target, and the $Q$ and $\overline Q$ are produced in the center of the collision with no significant distinction between the kinematics of the final-state $Q$ and $\overline Q$.

The cross section for a final-state hadron $H$ that can be produced both perturbatively and from an intrinsic state can be expressed as the sum of the perturbative and intrinsic components, including their $A$ dependence if a nuclear target is involved \cite{RV_SeaQuest}:
\begin{equation}
\sigma_{pN}^H = A^\alpha \sigma^H_{\rm pQCD}(pN) + A^\beta \sigma_{\rm ic}^H(pN) \, .
\end{equation}
Here $N$ denotes either a $p$ or $A$ target, with $A = 1$ in the $A$-dependent coefficients for $p+p$ collisions. The exponents $\alpha$ and $\beta$ encapsulate the different nuclear dependencies of the two components. In general, CNM kinematic effects, such as nPDFs, energy loss, and momentum broadening, can be incorporated into the perturbative QCD cross section, leaving only final-state effects such as nuclear absorption or comover suppression to be included in $\alpha$.

The perturbative $A$ dependence is nearly linear, $\alpha \approx 1$, for most assumed values of the $J/\psi$ absorption cross section, an effect dependent on the nuclear volume. On the other hand, the nuclear dependence of the intrinsic states is assumed to be more a diffractive-like surface effect with $\beta = 0.71$ for the proton \cite{NA3:1983tt}. For details of such calculations, including nuclear effects, see Refs.~\cite{RV_IC_EN,RV_SMOG}.

Intrinsic charm distributions have been included in global analyses of the parton densities \cite{Pumplin:2007wg,Nadolsky:2008zw,Dulat:2013hea,Jimenez-Delgado:2014zga,Ball:2016neh,Lyonnet:2015dca} with contradictory results. The NNPDF collaboration established the existence of intrinsic charm to the $3\sigma$ level in their analysis \cite{NNPDF_2021}, consistent with both the LHCb $Z+$charm results \cite{LHCb:2021stx} and the previous EMC charm quark structure function, $F_2^c$ measurements \cite{EuropeanMuon:1981cbg}. However, other work has called the NNPDF results into question based on how the LHCb $Z+c$-jet data were analyzed \cite{Guzzi:2022rca}. 

Current and forthcoming fixed-target experiments at similar and lower energies may shed more light on the issue \cite{NA60:2022sze,Adams:2018pwt}. In addition, the lower energies of hadron beams at the EIC could facilitate a discovery measurement.

\subsection{Observables}
\label{sec:initial-state-np-observations}

Finally, this section discusses a selection of observables that can be used to study initial-state effects such as impact parameter dependent shadowing and intrinsic charm. The main challenge is to isolate these initial-state effects while minimizing the influence of other nuclear effects, as discussed throughout this report.
\begin{itemize}[left=0.1cm]
\item  The impact parameter dependence of the nPDFs has not been studied in any systematic way so far.  Most of the focus has been on studies of quarkonium production in $h+A$ collisions as a function of centrality-related variables.  The PHENIX d+Au data~\cite{PHENIX:2012czk} studied in Ref.~\cite{McGlinchey:2012bp} suggested a stronger effect than ALICE charmonium centrality-dependent measurements in $p+{\rm Pb}$ collisions \cite{ALICE:2015kgk,ALICE:2022gpu}.  The CMS collaboration \cite{CMS:2013jsu} has studied $\Upsilon$ production as a function of event activity. Differences in the RHIC and LHC results could arise from changes in the gluon density of the nucleus as higher energies probe lower $x$ regions and more central collisions could herald a transition from a more dilute system with small hot spots as suggested in Ref.~\cite{McGlinchey:2012bp} to a more fully saturated system of gluons.

However, these measurements are not unambiguous because quarkonium states are, as has been demonstrated in this report, subject to multiple CNM effects.  Other recent work has implied that hot matter suppression may also be active in these small systems, see {\it e.g.} Ref.~\cite{Strickland:2024oat}.
Thus studying other potential observables could be helpful.

The advent of the EIC could allow for centrality-dependent studies of nPDFs based on event multiplicity, preferably for light hadrons and open heavy flavors to reduce at least some model ambiguity.

\item On the other hand, a number of observables have been proposed for intrinsic charm and related intrinsic states.  The high $x$  and large $Q^2$ charm structure function, $F_2^c$, measurement by the EMC collaboration \cite{EuropeanMuon:1981cbg} have not been reproduced and higher energy measurements concentrated on lower $x$ ranges where intrinsic charm contributions are negligible.  EIC measurements may be more decisive, especially for the lower energy range.   In addition, higher than expected charm yields at future low energy fixed-target measurements, such as NA60+ \cite{NA60:2022sze}, could be due to intrinsic charm near midrapidity.    
\end{itemize}   
\section{Future Experimental Directions}
\label{sec:future_experience}

The systematic study of Drell–Yan production and other hard probes such as heavy flavor and quarkonia in $p+A$ collisions as a function of $A$, $x_F$, and $p_T$, is crucial for disentangling the mechanisms of CNM energy loss and the modification of nuclear PDFs, as well as other CNM effects. Ongoing and future experiments, such as SeaQuest/SpinQuest at Fermilab \cite{SeaQuest:2017kjt} and AMBER at CERN \cite{Adams:2018pwt}, will provide increasingly precise data to test theoretical models and improve our understanding of quark propagation and energy loss in cold nuclear matter and separate these contributions from nPDF effects. The proposed Forward Physics Facility at CERN~\cite{Feng:2022inv} could also contribute significantly to  the understanding of proton and nuclear structure, coherent power corrections, and quantum electrodynamics (QED) corrections.   

The EIC will provide a clean electromagnetic probes of nuclear structure, offering an ideal environment to disentangle cold nuclear matter effects, particularly energy loss mechanisms. The EIC has key features that will help distinguish coherent from incoherent processes.

\subsection{Fixed-Target Experiments}

Several ongoing and upcoming fixed‐target experiments will expand our understanding of cold nuclear matter effects.

At Fermilab, the SeaQuest experiment (E906) \cite{Isenhower:1999rbq}, which concluded data taking in 2017, with ongoing analyses, used a 120 GeV proton beam to study Drell–Yan production with several nuclear targets (carbon, iron, tungsten) along with reference targets (liquid hydrogen, deuterium). By mapping the Drell–Yan cross section across a range of nuclear masses, SeaQuest can separate partonic energy loss from other CNM effects, such as nuclear shadowing, while refining nPDFs in the medium-to-high $x$ region. SeaQuest also studied $J/\psi$ production and could provide insight into intrinsic charm contributions \cite{RV_SeaQuest}.  Its successor, SpinQuest (E1039) \cite{Brown:2014sea}, collects data since 2019. SpinQuest focuses on spin-dependent phenomena with polarized targets, exploring the Sivers function, which provides insights into the orbital angular momentum of sea quarks. Future phases may include nuclear targets to enhance CNM physics measurements.

At CERN, AMBER (NA66) \cite{Adams:2018pwt} will extend the legacy of COMPASS \cite{COMPASS:2010shj} by exploring hadronic structure and nuclear effects with high‐intensity pion, kaon, and proton beams from 60 to 190 GeV. AMBER’s Drell–Yan and charmonium measurements on nuclear targets will place high‐precision constraints on initial‐state energy loss and distinguish between coherent and incoherent mechanisms. Its broad kinematic coverage will also refine our understanding of the nPDFs.  

The proposed NA60+ experiment at CERN \cite{NA60:2022sze} (an extension of NA60), aims to measure dimuon production with improved precision with proton beams of 40, 80 and 120~GeV. Their high precision measurements of quarkonium and Drell–Yan at these low energies will offer a more comprehensive energy dependence of partonic interactions in a nuclear medium. The NA61/SHINE experiment \cite{Laszlo:2009vg}, taking data since 2009, although primarily focused on hadron yields, could provide additional insights into nuclear PDFs and energy loss mechanisms in dedicated runs enabled by the post-2020 upgrade proposal \cite{Larsen:2019due}.

At the LHC, the SMOG2 upgrade at LHCb \cite{2707819,BoenteGarcia:2024kba} provides a unique approach to fixed‐target physics at TeV energies. By injecting low‐pressure gas into the beam pipe, SMOG2 enables $p + A$ and ${\rm Pb} + A$ collisions at forward rapidities, providing a unique environment for probing CNM effects. Although Drell–Yan measurements may be limited by relatively low integrated luminosities, they would complement quarkonium and open heavy flavor studies, especially for energy-loss dynamics at high fixed--target energies~\cite{Hadjidakis:2018ifr,Flore:2025uhv}.  Further measurements of asymmetries between leading and nonleading $D$ mesons with SMOG2  could help separate possible intrinsic charm contributions from other hadronization mechanisms \cite{RV_SMOG}.

Together, these experiments will deliver complementary measurements under different beam conditions and energy scales, providing a comprehensive understanding of CNM phenomena.  These studies will complement and pave the way for CNM studies at the EIC. 

\subsection{Ultraperipheral Heavy-Ion Collisions at the LHC (2026–2035)}

In the coming years, new measurements of open heavy-flavor hadrons, quarkonia, and jets in ultraperipheral heavy-ion collisions will provide high accuracy constraints on quark and gluon PDFs at low $x$ that are unaffected by final-state interactions. This program will benefit from the high center-of-mass energies at the LHC. The LHC’s high-luminosity upgrade for Run~4 (HL-LHC) starting in 2030 will enhance precision and expand detector coverage, leading to stronger constraints on PDFs in a wider range of \(x\) and \(Q^2\). By combining the low \(x\) reach of UPC measurements with the precision of $e+A$ collisions at the EIC, the kinematic reach of the combined results will place tighter constraints on nPDFs.

Future proton–lead runs, and data-taking with lighter ions (such as oxygen–oxygen or neon–neon), will extend this program, providing access to high-statistics photon–proton collisions and improved constraints on  light to intermediate mass nPDFs which are still poorly known.

\subsection{Fixed-Target Program at the EIC}

The Long Range Plan for Nuclear Sience listed the EIC as the  highest priority new US facility \cite{LRP}. The EIC will collide electron beams of up to 18 GeV with ion beams of up to 100~GeV/nucleon \cite{AbdulKhalek:2022hcn}.   The RHIC facility will be shut down and the EIC will be constructed in its stead.  This transition will eliminate studies of $p$+$p$, $p+A$, and $A+A$ collisions in the US at RHIC, making the LHC the only facility worldwide for ultrarelativistic hadronic collisions.

\begin{figure}[tbp]
    \centering
    \includegraphics[width=\linewidth,clip]{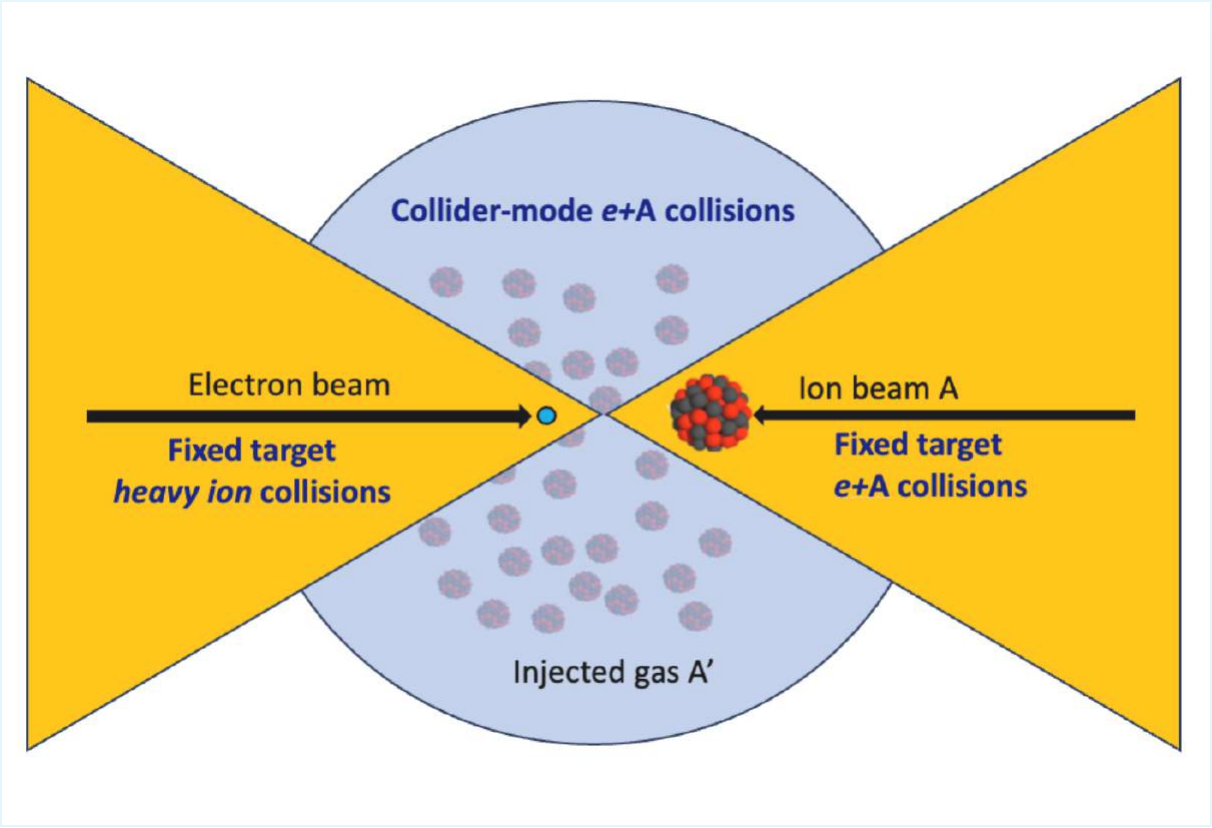}
    \caption{Conceptual diagram of the EIC collision region with a fixed target.  Collider $e+A$ collisions will produce particles across all angles, while in the $A$-going direction (left) particles from fixed target $A+A'$ collisions can be measured, and in the $e$-going direction (right) lower energy $e+A'$ collisions can be studied.  This  configuration  woud be equivalent to running three experiments simultaneously.}
    \label{fig:3for1}
\end{figure}

A gaseous fixed target at the EIC would significantly expand its physics program. By injecting a gas species $A’$ into the interaction region, low energy $e+A’$ collisions can be studied, while $A+A’$ collisions would allow continued exploration of high baryon density regions, similar to RHIC beam-energy scans \cite{Liu:2019wbq,Liu:2022cqh,Yang:2017llt} and the NA60+ proposal \cite{NA60:2022sze}. The addition of a gas target would also improve the precision of the luminosity determination, making cross section measurements more accurate \cite{BGI}. This approach has been used in HERMES \cite{HERMES:2004vsf} and SMOG at LHCb \cite{SMOG2}.  A conceptual diagram of fixed-target collisions at the EIC is shown in Fig.~\ref{fig:3for1}.

In the $e$-going direction, the beam energy of 18 GeV exceeds CEBAF's 12 GeV, while in the $A$-going direction, the 100 GeV/nucleon beam will provide fixed-target collisions at $\sqrt{s_{NN}} \approx 14$ GeV, enabling studies at high baryon chemical potential $\mu_{B}$, as in the RHIC beam-energy scan.

\subsection{The Forward Physics Facility}

The Forward Physics Facility (FPF)~\cite{Feng:2022inv, FPFWorkingGroups:2025rsc} is a proposed addition to the CERN infrastructure designed to take advantage of the physics potential of the forward region in $p+p$ collisions at the LHC. The region that the FPF proposes to cover, close to the beamline, has been  underexplored so far in high-energy collider experiments but has the potential to probe processes at lower $x$ values than those reached in the current LHC detectors. The FPF would be located roughly 620 m downstream from the ATLAS interaction point, far enough away to detect long-flying neutrinos, primarily from charm decays.   

The FPF physics program would encompass:
\begin{itemize}[left=0.1cm]
\item searches for new physics beyond the Standard Model, including light weakly-interacting particles such as dark photons, axion-like particles, and sterile neutrinos;
\item neutrino physics at the TeV scale,  including direct detection of electron, muon, and tau neutrinos from LHC collisions in a regime inaccessible to other neutrino experiments;
\item cosmic ray physics, which could benefit from measurements of high-energy particle production in the forward direction to constrain hadronic interaction models of air showers, improving the understanding of ultra-high-energy cosmic rays;
\item forward particle production in QCD, refining proton PDFs and fragmentation functions for $x \sim 10^{-7}$.
\end{itemize}

This physics program would be implemented by detectors such as FASER \cite{FASER:2022hcn}, a spectrometer to measure light, weakly-interacting states; FASER$\nu$ \cite{FASER:2020gpr}, employing emulsion and electronics to detect neutrinos in the TeV range; FLArE \cite{Batell:2021blf}, a noble liquid detector for neutrinos and light dark matter;  and FORMOSA \cite{Foroughi-Abari:2020qar}, a scintillator detector sensitive to non-integer charges and weakly-interacting particles. 

Most relevant to these studies, the FPF neutrino studies as a fixed-target experiment could be complementary to EIC program.  The energy flux of neutrinos  detected in the FPF, with a peak around 1~TeV, is very broad.  In a fixed target configuration, this energy flux would correspond to $\sqrt{s_{NN}} \approx 45^{+45}_{-23}$~GeV due to the energy spread of the neutrino flux. Such a fixed-target configuration would nicely complement the EIC physics program with $\nu + A$ DIS measurements at energies similar to those of the EIC.  For example, studies of coherent power corrections predicts a measurable  difference in the  shadowing 
pattern of the  $F_2^A$ and
$F_3^A$~\cite{Qiu:2004qk} structure functions as well as significant low- and moderate-$Q^2$ modifications 
of QCD sum rules based on these structure functions~\cite{Gross:1969jf}.  These complementary measurements could help determine the origin of shadowing.

QED radiative corrections to charged lepton and neutrino interactions in nuclei are also subject to modification in cold nuclear matter~\cite{Tomalak:2022kjd,Tomalak:2023kwl,Tomalak:2024lme}. These modifications can jointly be explored at the FPF and the EIC~\cite{Bhattacharya:2025pje}. Such complementary neutrino and electron DIS studies can provide more precise insights into the structure of matter. 

\subsection{The ideal experiment} 

Surveying the experimental landscape for future studies of CNM effects, one can envision an ideal setup that combines both collider and fixed-target modes to maximize physics reach. In collider mode, a proton beam could interact with various nuclear beams at $\sqrt{s_{NN}} \sim 100$ GeV. This configuration provides access to relatively high $p_T$ and broad $x_F$ coverage both forward and backward of the center of the collision, enabling sensitive probes of CNM dynamics.

In fixed-target mode, the same proton beam could interact with an internal gas target, as pioneered by the SMOG2 system at LHCb \cite{2707819,BoenteGarcia:2024kba}. Realistic gas targets include H, D, He, Ne, Ar, Kr, and Xe.  Solid nuclear targets such as Au or Pb could also be considered. Such  configurations would result in a center-of-mass energy of $\sqrt{s_{NN}} \sim 10$~GeV. While lower in energy, the fixed-target mode offers high luminosity and access to large $x_F$, both of which are essential for highlighting the strong suppression pattern in that region.

High-precision silicon vertex detectors would be essential to improve vertex resolution, particularly challenging with solid targets, and reconstruct heavy-flavor hadrons and quarkonia, providing complementary observables to probe both initial- and final-state effects.

By combining collider and fixed-target modes, such an experiment would offer broad kinematic coverage, enabling detailed studies of nuclear parton distribution functions, quark energy loss in nuclei, and other CNM effects. Heavy-flavor measurements, facilitated by muon or electron triggers, would enhance sensitivity to these phenomena. This approach would provide complementary insights to existing experiments at the LHC (high energy) and future fixed-target projects (e.g., AFTER@LHC \cite{Massacrier:2017oeq} or an EIC fixed-target setup).  
To further enhance the physics reach of this and upcoming facilities like the EIC, it is essential to adopt strategies that disentangle the interplay between nPDF modifications, parton saturation, and energy loss mechanisms. This could be achieved by selecting appropriate observables, as described in the previous sections of this work.    
\section{Conclusion}
\label{sec:conclusion}
The study of CNM effects in hadron-nucleus and electron-nucleus collisions is critical for understanding QCD in nuclei. This review has examined the nuclear modifications observed in hard processes such as Drell-Yan, heavy flavor, and quarkonium production, highlighting the challenges in disentangling the contributions of various CNM effects, including nuclear parton distribution functions, gluon saturation, parton energy loss, and final-state interactions. Despite decades of research, a unified theoretical framework capable of consistently describing these phenomena is still lacking.

Recently, heavy-flavor data have been incorporated into global nPDF analyses to provide a more direct constraint on the small $x$ gluon distribution. While this marks a major step forward, it also significantly impacts the extraction of leading-twist shadowing and may introduce nontrivial biases in the interpretation of a broad range of hard-scattering observables.

Key observations from experimental data reveal suppression patterns that depend on the specific process, kinematic regime, and nuclear medium. The interplay between perturbative and nonperturbative QCD dynamics complicates the interpretation of these results, necessitating the identification of promising observables to isolate individual effects. Future facilities, such as the EIC, ongoing and new fixed-target experiments, and the program of ultraperipheral collisions at the LHC, promise unprecedented precision combined with an expanded kinematic coverage to enable deeper insights into the properties of CNM.

Motivated by these developments, this document seeks to clarify the role of various cold nuclear matter effects and their impact on the hard processes under consideration. 

Although no single “golden observable” can unambiguously isolate a specific CNM effect, we have compiled a set of promising observables that collectively address the puzzles of cold nuclear effects, as discussed in Secs.~\ref{sec:saturation-observables}, \ref{sec:energy-loss-observables}, \ref{sec:final-state-observations}, and \ref{sec:initial-state-np-observations}. 

Together, these proposed observables can provide our best answers to the three fundamental questions introduced in Sec.~\ref{sec:introduction}:
\begin{enumerate}[left=0.1cm]
\item \textbf{What are the relative contributions of perturbative (gluon saturation) and nonperturbative (nPDF) QCD dynamics to the modification of nuclear structure functions and particle production spectra at low $x$?}

Disentangling leading-twist nPDFs from small-$x$ saturation effects remains a central challenge in cold QCD. Current global nPDF fits often absorb dynamic suppression mechanisms not related to genuine PDFs. As discussed in Section~\ref{sec:saturation-observables}, a dedicated program using inclusive, semi-inclusive, and diffractive observables will be essential to separate nonperturbative modifications from saturation-induced suppression.

\item \textbf{How do parton energy loss mechanisms affect particle production in $h+A$ and $e+A$ collisions?}

Energy loss in cold nuclear matter depends strongly on the process under study and its kinematic regime. Although no universal regime applies across all observables, the transport coefficient $\hat{q}$ provides a unifying parameterization of the scattering properties of the medium. Section~\ref{sec:energy-loss-observables} outlines key observables sensitive to distinct regimes of radiative energy loss and offers strategies to constrain $\hat{q}$ across quarkonia, heavy flavor and electroweak production. 

\item \textbf{What other effects beyond energy loss modify particle production and how do we distinguish among final-state interactions?}

Final-state effects such as absorption and comover interactions further suppress hadronic yields, particularly at mid and backward rapidity where formation times are short. Their contribution can be isolated through a careful analysis of rapidity and species dependence, as discussed in Section~\ref{sec:final-state-observations}, and must be clearly separated from other CNM effects to avoid misinterpretation of suppression patterns.
\end{enumerate}

Addressing the fundamental questions outlined in this review, such as the relative roles of gluon saturation and nPDFs, the mechanisms of parton energy loss, and the impact of final-state interactions, will pave the way for a more comprehensive understanding of cold nuclear matter. This knowledge is essential not only for advancing QCD theory but also for interpreting data from heavy-ion collisions, where CNM effects serve as a baseline for studying the quark-gluon plasma \cite{Arslandok:2023utm}.

In summary, the path forward requires a concerted effort combining high-precision experiments, multi-variable phenomenological analyses, and close collaboration between experimentalists and theorists. By leveraging upcoming experimental opportunities, the scientific community can bridge the gap between theory and observation, ultimately achieving a unified, universal, and coherent description of cold nuclear matter effects in high-energy collisions.    
\\
\\
{ \large \bf Acknowledgments} \\

The work of A. Deshpande by the U.S. Department of Energy through Grant No. DE‐FG02‐05ER41372. J. M. Durham is supported by DOE Office of Science Early Career Awards program. J. Jalilian-Marian is supported by the US DOE Office of Nuclear Physics through Grant No. DE-SC0002307 and the framework of the Saturated Glue (SURGE) Topical Theory Collaboration. A. Kusina is grateful for the support of National Science Centre Poland under the OPUS grant no. 2023/49/B/ST2/03862. The work of M. X. Liu is supported by DOE Office of Science Nuclear Physics Program through Los Alamos National Laboratory. The work of Y. Mehtar-Tani is supported by the U.S. Department of Energy under Contract No. DE-SC0012704 and the framework of the Saturated Glue (SURGE) Topical Theory Collaboration.. The work of C.-J. Naïm is supported by the Center for Frontiers in Nuclear Science and the Simons Foundation. The work of H. Paukkunen is supported by the Academy of Finland, the Centre of Excellence in Quark Matter, project 364194. I. Vitev is supported by the Laboratory Directed Research and Development program of Los Alamos National Laboratory under project numbers 20240127ER and 20240131ER and the HEFTY Topical Collaboration for Nuclear Theory. R. Vogt is supported by Lawrence Livermore National Laboratory under Contract DE-AC52-07NA27344, the LLNL-LDRD Program under Project No. 23-LW-036, and through the Topical Collaboration in Nuclear Theory on Heavy-Flavor Theory (HEFTY) for QCD Matter under award no. DE-SC0023547.   

\bibliographystyle{apsrev4-2}
\bibliography{biblio}

\end{document}